\AtBeginDocument{\date{24 April 2019}}

\documentclass[prd,preprintnumbers,nofootinbib,twocolumn,longbibliography]{revtex4-2}

\usepackage{graphicx}
\usepackage{amsmath}
\usepackage{amssymb}
\usepackage{braket}
\usepackage{color}
\usepackage{units}
\usepackage{metric}

\usepackage[bookmarksopen,bookmarksnumbered]{hyperref}

\newcommand\3[1]{\boldsymbol{#1}}
\newcommand\ve[1]{\3{#1}}

\newcommand\op[1]{\widehat{#1}}

\DeclareRobustCommand*\diff[2][]{%
   \mathop{ {\mathrm{d}}^{#1} {}#2}\nolimits
}

\DeclareMathOperator*{\SumInt}{%
\mathchoice%
  {\ooalign{$\displaystyle\sum$\cr\hidewidth$\displaystyle\int$\hidewidth\cr}}
  {\ooalign{\raisebox{.14\height}{\scalebox{.7}{$\textstyle\sum$}}\cr\hidewidth$\textstyle\int$\hidewidth\cr}}
  {\ooalign{\raisebox{.2\height}{\scalebox{.6}{$\scriptstyle\sum$}}\cr$\scriptstyle\int$\cr}}
  {\ooalign{\raisebox{.2\height}{\scalebox{.6}{$\scriptstyle\sum$}}\cr$\scriptstyle\int$\cr}}
}

\DeclareMathOperator{\sign}{sign}

\newcommand \In {{\rm in}}
\newcommand \Out {{\rm out}}
\newcommand \Free {{\rm free}}

\newcommand{\difftilde}[1]{\mathop{\widetilde{\diff{#1}}}\nolimits}

\newcommand{\deriv}[2][t]{\frac{\partial#2}{\partial #1}}
\newcommand{\derivB}[1][t]{\overleftrightarrow{\frac{\partial}{\partial #1}}}

\newcommand\limSub[1]{\lim_{\substack{#1}}}
\newcommand{\limTIn}{\limSub{ \rm infinite \\ \rm past \\ \rm time}}
\newcommand{\limTOut}{\limSub{ \rm infinite \\ \rm future \\ \rm time}}
\newcommand{\limT}{\limSub{\rm infinite \\ \rm time }}

\newcommand{\pc}{\3p_{\rm c}}

\newcommand{\eqdef}{\stackrel{\textrm{def}}{=}}

\newcommand{\eqbad}{\stackrel{\textrm{?}}{=}}

\let\oldleft\left
\def\xleft{\mathopen{}\oldleft}

\newcommand\note[1]{{\color{red}#1}}
\newcommand\IGNORE[1]{}
\long\def\??#1{\note{#1}}

\begin{document}

\title{A new approach to the LSZ reduction formula%
  \texorpdfstring{\ifmetricBD\else\\\color{red}{(Metric signature is
      \metricSig)}\fi}{}%
}

\author{John Collins}
\email{jcc8@psu.edu}
\affiliation{%
  Department of Physics, Penn State University, University Park PA 16802, USA}

\begin{abstract}
  Lehmann, Symanzik and Zimmermann (LSZ) proved a theorem showing how to
  obtain the S-matrix from time-ordered Green functions.  Their result,
  the reduction formula, is fundamental to practical calculations of
  scattering processes.  A known problem is that the operators that they
  use to create asymptotic states create much else besides the intended
  particles for a scattering process.  In the infinite-time limits
  appropriate to scattering, the extra contributions only disappear in
  matrix elements with normalizable states, rather than in the created
  states themselves, i.e., the infinite-time limits of the LSZ creation
  operators are weak limits.  The extra particles that are created are in
  a different region of space-time than the intended scattering process.
  To be able to work with particle creation at non-asymptotic times, e.g.,
  to give a transparent and fully deductive treatment for scattering with
  long-lived unstable particles, it is necessary to have operators for
  which the infinite-time limits are strong limits.  In this paper, I give
  an improved method of constructing such operators.  I use them to give
  an improved systematic account of scattering theory in relativistic
  quantum field theories, including a new proof of the reduction formula.
  Among the features of the new treatment are explicit Feynman rules for
  the vertices corresponding to the creation operators, both for the LSZ
  ones and for the new ones.  With these I make explicit calculations to
  illustrate the problems with the LSZ operators and their solution with
  the new operators.  Not only do these verify the existence of the extra
  particles created by the LSZ operators and indicate a physical
  interpretation, but they also show that the extra components are so
  large that their contribution to the norm of the state is ultra-violet
  divergent in renormalizable theories.  Finally, I discuss the relation
  of this work to the work of Haag and Ruelle on scattering theory.
\end{abstract}

\maketitle

\section{Introduction}

The reduction formula of Lehmann, Symanzik and Zimmermann
\cite{Lehmann:1954rq} (LSZ)\footnote{Other useful references for
  proofs following LSZ's strategy are in Refs.\
  \cite{Duncan:2012.book,Goldberger.Watson} } is very important for
applications of quantum field theory (QFT) to experiment because it
shows how to compute S-matrix elements from time-ordered Green
functions, including the correct external line factors.

Unfortunately, there are some problems, as was realized a long time
ago --- see the papers by Haag \cite{Haag:1958vt,Haag:1959ozr}, Ruelle
\cite{Ruelle:1962}, and Hepp \cite{Hepp:1965.LSZ}.  The problems do
not in fact impact the validity of the reduction formula itself, or
even the validity of LSZ's proof.  Instead, the problems manifest
themselves when one tries extending the LSZ methods to more general
situations.  As we will see, such cases occur quite dramatically when
the operators used by LSZ to create asymptotic particles are applied
in experimentally relevant situations at finite times instead of
infinite times.

More explicitly, suppose we are given a single-particle
positive-energy wave function\footnote{See Sec.\ \ref{sec:wfs} for a
  specification of what is meant here.} $f(x)$. Then LSZ define a
time-dependent operator $a^{\dagger}_f(t)$ that is intended to create a
single particle in the in-state or the out-state in the limit that
$t\to-\infty$ or $t\to+\infty$, with the particle's state corresponding to the wave
function $f$.  However, when one of these operators acts on the
vacuum, what is created is a lot more than the intended particle;
taking time to infinity does not help.  This will be illustrated in
Sec.\ \ref{sec:extra.particles} with the aid of explicit perturbative
calculations.  Moreover, we will see that, the extra contributions are
not merely nonzero, but in a renormalizable QFT are also generically
ultra-violet (UV) divergent, as measured by the norm of the state that
is created.

In the restricted context of the matrix elements used to obtain the
S-matrix, a careful application of the infinite-time limits, as in the
LSZ paper, does remove the extra contributions.  This can be
characterized \cite{Haag:1958vt,Ruelle:1962} by saying that the limits
used by LSZ are weak limits, but not strong limits.  (See App.\
\ref{sec:limits} for characterization of these concepts, together with
summaries of methods by which it can be determined which kind of limit
is applicable in particular cases.)

In contrast, for an operator $A^{\dagger}_{f}(t,\Delta t)$ that actually does
asymptotically create a single particle only, then the strong limit
exists.  As indicated by the notation, it will be useful to introduce
an extra parameter $\Delta t$ that is a range of time involved in defining
the operator; its inverse is essentially an uncertainty in energy.
The operator creates a single particle in the limit\footnote{At finite
  times $A^{\dagger}_{f}(t,\Delta t)$ does create extra contributions in addition
  to the intended single particle.  The extra contributions vanish in
  the limit that $\Delta t\to\infty$.  Therefore they are small if $\Delta t$ is large
  enough.  The smallness of the extra contributions is what can allow
  the application to unstable particles, etc. Notice that to ensure
  that the operator creates one particle to a good approximation, it
  is $\Delta t$ that needs to be large, not $t$ itself.  That is suitable
  for creating a particle in a chosen finite region of space-time.  It
  might be within an experimental apparatus instead of being
  infinitely far away.} $\Delta t\to\infty$.  Its application to creating
particles in a scattering process involves taking both $t$ and $\Delta t$
to infinity in such a way that $\Delta t/|t|\to0$.

Such an operator allows one to make an adequate treatment when the
strict limits of infinite time are not taken.  Such would be the case
for treating long-lived but unstable particles or for a fully
deductive treatment of neutrino scattering and
oscillations.\footnote{See Refs.\ \cite{Akhmedov:2019iyt,
    Akhmedov:2010ms} a systematic account that includes an analysis of
  the confusion that sometimes results when textbook results on
  scattering theory are applied to neutrino oscillations, together
  with relevant references.  It would be interesting to combine the
  account given there with the methods of the present paper.}  The
extra particles created by using the original LSZ operator instead of
the new operator would be detected in a suitably located detector.

Textbook treatments given for these situations typically start from
the strict infinite-time formalism for standard scattering.  They then
graft on something like a semiclassical analysis of isolated free
particles, with a good dose of intuition and hand-waving.\footnote{See
  also the comments by Coleman \cite{Coleman:2011xi,Coleman.2018} on
  the hand-waving used in the usual treatments of scattering.}

The primary purpose of the paper is to provide an improved proof that
overcomes the problems just described.

In the LSZ paper, the creation operator $a^{\dagger}_f(t)$ involves just an
integral over all space, with the field (and its time derivative)
being taken at one specific value of time $t$.  It is the use of
integrals at fixed time that causes the problems, essentially by a
kind of application of an uncertainty principle: A fixed time implies
infinite uncertainty in energy.

The main innovation applied in the present paper is to find a good way
of defining the operator $A^{\dagger}_{f}(t,\Delta t)$, by averaging $a^{\dagger}_f(t)$
over a range of time.  Many conceptual subtleties then robustly
disappear.

Related techniques were used by Haag and Ruelle in their formulation of
scattering theory \cite{Haag:1958vt,Ruelle:1962}.\footnote{See also Hepp's
  account \cite{Hepp:1965.LSZ} of their method, as well as the recent
  account by Duncan \cite{Duncan:2012.book}.}  But their method was
formulated somewhat differently, and in a way that calculations using
their operators difficult.  They focused heavily on mathematical aspects
of the theory as opposed to possible applications.  The different
construction given in the present paper makes its much easier to treat the
asymptotics and allows a simplification of the proof of the reduction
formula.  The new construction provides simple explicit formulas for the
new operators both in coordinate space and momentum space.  See Sec.\
\ref{sec:HR.comparison} for a comparison of the new method with the
Haag-Ruelle method.

It is important to emphasize that, as regards the LSZ reduction
formula itself, the issues just summarized concern its proof.  In the
case that we use a theory in which all particles are massive and that
we treat only scattering of exactly stable particles, the LSZ
reduction formula remains correct and unchanged.

One advantage of the new formulation is that once the new definition of
$A^{\dagger}_{f}(t,\Delta t)$ has been provided, then the derivation of the S-matrix
is essentially a straightforward calculation.  The bulk of this paper
primarily concerns motivation, examples, and derivations of the
prerequisites for performing the calculations.  As already mentioned,
another advantage of the new methods are that they allow straightforward
extensions to situations at non-asymptotic times, e.g., to treat unstable
particles.  Boyanovsky \cite{Boyanovsky:2018swn} has recently treated the
space-time properties of the decay of unstable particles, and encountered
complications that are closely related to the issues treated here.  In
particular, he provides an independent calculation of the ultra-violet
divergence in the state created by an LSZ operator.

Another possible extension is to scattering with massless particles.
As is well-known, the postulates of standard scattering theory fail in
theories with massless particles.  A fully systematic treatment
requires extensions or modifications to the versions of scattering
theory that are valid for massive particles.  There is recent
interest, e.g., \cite{Kapec:2017tkm, Hoare:2018jim}, in finding better
treatments for the massless case.\footnote{See also Ref.\
  \cite{Zamolodchikov:1992zr} for further information about the
  S-matrix in massless integrable theories.}  Off-shell or finite-time
Green functions do exist in such theories.  Therefore what is in
question is the nature of the infinite-time limits and their relation
to physically implementable scattering.  A strategy for defining good
finite-time approximations to the creation of single particles could
be very useful to finding a better formulation of the infinite-time
limits with massless particles.  The formalism presented in this paper
is suitable for use in perturbative calculational examples that can be
used to test the formulation of general abstract theorems.

\section{Overall view: starting point, motivations, strategy}
\label{sec:overview}

\subsection{Aims}

A primary technical aim is the determination of the S-matrix in a
quantum field theory (QFT) from its Green functions.\footnote{I.e.,
  vacuum expectation values of time-ordered products of field
  operators.}.  A related aim is to construct definitions of operators
that can be applied to the vacuum state to construct in and out
states, which are states of well-separated individual particles.  The
operators give a construction of the state space of the theory in
terms of field operators applied to the vacuum, with parameterizations
of the states that are suitable for experimentally relevant scattering
processes.  

We assume that a QFT exists, as specified by its set of fields and its
Lagrangian density, that it obeys the standard properties of QFTs, and
that the task is to compute S-matrix elements from the Green
functions.  In doing so, one also verifies many of the properties of
scattering processes that underlie the definition and use of the
S-matrix.  Motivations for the emphasis on Green functions will be
given next.

\subsection{Position with respect to logical framework for QFT}
\label{sec:logical.structure}

Underlying those practical aims is a deeper issue.  This concerns what
it means to solve a particular QFT, and what exactly is the logic by
which the results are derived and checked.

A QFT is specified by listing a set of basic fields, which are
operator valued functions of space-time (strictly speaking,
operator-valued distributions), and by postulating certain of their
properties, notably equal-time canonical commutation relations
(ETCCRs) and equations of motion.  Normally these are determined from
a formula for a Lagrangian density in terms of the basic fields.  A
solution entails determining what the state space is and how the
operators act on it, after which one can compute quantities of
experimental interest.  Of course, after solving for the state space
and the operators by deductions from the initial postulates that
specify the theory, it is useful to verify self-consistency by showing
that the constructed operators do obey the postulated
properties.\footnote{The complications entailed by ultra-violet
  divergences need not concern us here.  They require an
  implementation of renormalization, thereby entailing modification of
  the underlying postulates in order to get self-consistent results.}

In contrast, the situation is rather different in the case of the
non-relativistic quantum mechanics of a finite number of particles.
In the first formulation of quantum mechanics, i.e., Heisenberg's
matrix mechanics, the above procedure was followed to determine the
matrices that implement the position and momentum operators.  See the
paper by Born and Jordan \cite{Born.Jordan} for the case of the
harmonic oscillator.  In normal current terminology, we would say that
the matrices consist of the matrix elements of the corresponding
operators between energy eigenstates.

It was quickly realized, at least in effect, that in these relatively
simple theories there is a unique representation of the ETCCRs, up to
unitary equivalence.  Thus the state space and how the operators act
on it are determined uniquely.  States can then be realized as
Schr\"odinger wave functions.  That is all independent of the details of
the Hamiltonian, e.g., as to what the potential is.  Predictions of
the theory can be determined by solving the Schr\"odinger equation for
time dependence of the state or for energy eigenstates, etc.

In QFT, the situation is radically different.  Because of the infinite
number of degrees of freedom, there is no longer a unique
representation of the ETCCRs.  Moreover, it is found that the
different representations get used.  Calculations show pathologies and
inconsistencies --- e.g., \cite{vanHove1951, vanHove1952145} --- if
one assumes that the state space of an interacting theory is the same
as that of a free theory and that the operators at one fixed time are
the same in both theories, as is done to define the interaction
picture.  Moreover, Haag's theorem \cite{Haag:1955ev, Streater:1964,
  Duncan:2012.book} guarantees that this is not just a difficulty in
particular examples, but a general property of relativistic QFTs.

One way of stating this is that the Hilbert space of states for an
interacting theory is orthogonal to that for a corresponding free
theory.  However, the Hilbert spaces for the free and interacting
theories are isomorphic, so one could alternatively arrange things
such that the Hilbert spaces are the same; but in that case, Haag's
theorem shows that the free and interacting fields cannot be related
by a unitary transformation, contrary to what happens in the widely
used interaction picture.

These results considerably complicate the derivation of useful
consequences from a given QFT.  Solving the theory requires,
implicitly or explicitly, a determination of the state space and the
action of the field operators on it.  The vast majority of work on
making predictions effectively evades the issue of what the states and
operators are.  Perturbative calculations using Feynman graphs give
only matrix elements.  Non-perturbative calculations using Monte-Carlo
lattice methods provide an implementation of the functional integral
of a QFT, and have as their immediate target time-ordered Green
functions continued to Euclidean time; thus they give
vacuum-expectation values of certain operators.

Nevertheless, underlying any derivation of the methods from the
foundational postulates of a QFT is an assumption that there are
operators acting on the state space.

A useful way of handling the issues is to make the Green functions
then primary target of calculations, such as in Refs.\
\cite{Sterman:1994ce, Srednicki:2007qs, Coleman:2011xi, Coleman.2018}.
In perturbation theory, the Green functions can be obtained from the
Gell-Mann-Low formula.  This allows the calculation\footnote{Note that
  the straight application of perturbation theory is often
  supplemented by many kinds of ``resummation'' methods to extend
  calculations beyond where strict fixed-order perturbation theory
  applies. In addition, in QCD the operator product expansion and more
  general kinds of factorization are used to allow certain kinds of
  predictions to be made from perturbative calculations even in the
  presence of strong non-perturbative phenomena.}  of Green functions
in the full theory from certain matrix elements in the free theory.
The formula can be derived from the functional integral, but it is
often also derived from a use of the interaction picture.  Normally
Haag's theorem prevents the consistent use of the interaction picture.
But in deriving the Gell-Mann-Low formula, the derivation using the
interaction picture can be first applied to a regulated theory with a
finite number of degrees of freedom.  A projection onto the exact
ground state can be made with the use of the evolution operator at a
time that is somewhat rotated towards imaginary values
\cite{Peskin:1995ev}.  Then the regulators can be removed to give a
continuum theory in an infinite volume of space, with the application
of any necessary renormalization.  In a correct derivation, the
numerator and denominator of the Gell-Mann-Low formula both contain a
factor $|\langle0|0;\Free\rangle|^2$, the squared overlap of the vacuum states in
the interacting and free theories.  Haag's theorem manifests itself in
this overlap going to zero when the infinite volume limit is taken.
But since the factor cancels between numerator and denominator, the
final results for the Green function are valid and well-behaved in the
limit that the regulators are removed.  Even though the operators and
states have rather singular properties as the regulator is removed,
the Green functions have smooth limits. 

The Green functions obey equations of motion that encode both the
equations of motion for the fields and their (anti)commutation
relations on a ``surface of quantization''.  Since it is readily
proved that the perturbative expansion of Green functions obeys these
equations, we know that at least the perturbative solution for Green
functions exists independently of any qualms one might have about the
adequacy of particular textbook derivations from first principles,
e.g., concerning the existence of the functional-integral
representation of Minkowski-space Green functions, or the asymptotic
limits used in applying the interaction picture.

An approach via Green functions recognizes that the particle content
and scattering processes arise as emergent phenomena from the solution
of a QFT.  The particle concept in interacting relativistic QFTs is
essentially identical to the quasi-particle concept
\cite[Sec.\ 5.7]{Jain2007.book} in condensed matter physics, certainly if
one uses the word ``particle'' to refer not only to strictly stable
particles but also to unstable and confined particles. The primary
practical differences in condensed matter physics are that there is an
obvious preferred rest frame, and that the background medium is at
non-zero temperature, thereby giving rise to notable dissipative
effects.

Then the project initiated by LSZ of obtaining the S-matrix (and in
fact other matrix elements of time-ordered operators) is in effect a
determination of the state space of the theory in a useful basis and
of how the field operators are implemented in that basis.  (In fact,
there are two useful sets of basis states, one for incoming states in
a scattering process and one for outgoing states.)

Hence the overall logic is to start with the postulates specifying a
particular QFT.  From them one deduces methods for calculating Green
functions, with care taken to avoid invalidation of the derivations by
Haag's theory.  Finally one constructs the scattering states, and the
other consequences of the theory from the Green functions.  The LSZ
reduction formula is the core tool to get from the off-shell Green
functions to S-matrix elements and to matrix elements of any operator.

In contrast to the Green function route, many books --- e.g.,
\cite{Weinberg:1995mt} --- take the S-matrix as primary.  Such an
approach can be useful, e.g., \cite{Coleman:2011xi, Coleman.2018}, to
gain initial insight from low-order perturbation theory about
elementary experimental implications of a given QFT.  But in a
complete treatment, use of the S-matrix as primary is problematic.  In
its most natural form, such a treatment assumes that the spectra of
the free and interacting theories are the same (e.g., p.\ 110 of
\cite{Weinberg:1995mt}) and hence that the particle types are in
one-to-one correspondence with the fields.  But such an assumption is
generally very incorrect.  For example, in the Standard Model, the
only elementary fields that correspond to particles in the strict
sense of scattering theory are those for the photon, electron, and
neutrinos.  The particles, or quasiparticles, that correspond to the
other fields are either unstable (e.g., muon), or confined (e.g.,
quarks), or both.  On the other hand, there is a large collection of
stable bound states (proton, and many nuclei, atoms and molecules)
that do not correspond to the elementary fields.

Moreover, in theories with massless particles, the standard theory of
scattering and the S-matrix needs modification, as manifested by the
existence of infra-red divergences in calculations of the S-matrix and
cross sections by standard methods.  In contrast, the off-shell Green
functions do not have such problems.  So it is again useful to
separate the issue of solving the theory, as manifested in the Green
functions, from that of determining properties of scattering.

Furthermore, treatments that take the S-matrix as primary typically
use the interaction picture in a way that runs badly afoul of Haag's
theorem.  For example, the treatment in Ref.\ \cite{Weinberg:1995mt}
starts from an assertion, (3.1.12) and (3.1.13), of the large-time
asymptotics of interaction-picture states.  The assertion is intended
to capture in mathematical form the intuitive notion of states
approaching states of separated particles.  But Haag's theorem ensures
that the asserted asymptotic properties are simply wrong, and in a
sense infinitely wrong. The incorrectness of the stated properties is
readily verified by low order perturbative calculations, as was
well-known in the early 1950s, e.g., \cite{vanHove1951,
  vanHove1952145}.  

These direct derivations of the S-matrix can be regarded as
constructing a perturbative solution of a theory on the basis of
certain postulates about its properties.  Once that solution has been
constructed, it can be investigated whether the constructed solution
self-consistently has the properties attributed to the solution.  In
this case, it is readily seen from perturbative calculations that the
solution does not have these properties.  A critical question is
whether the final answer for the perturbative solution is correct
despite the false hypotheses used to derive it or whether the answer
itself is wrong.  In this case it is only the hypotheses that are
wrong, and the solution can be derived by better methods.

\subsection{Structure of presentation}

The overall structure of the presentation and derivation in this paper
is summarized by the following items:
\begin{enumerate}

\item In Sec.\ \ref{sec:scattering} a review is given of the
  formulation of scattering theory in terms of Fock-space structures
  for the in- and out-states.  This is a framework that is strongly
  motivated by an examination of what happens in scattering processes
  and in non-relativistic quantum mechanics \cite{Goldberger.Watson,
    Taylor.scattering.1972}.  From a logical point of view, it may be
  best to regard the formalism as a conjecture, to be a target of and
  then verified by subsequent derivations.

\item Then there is made an examination of the asymptotics of Green
  functions in coordinate space, for large positive and negative
  times, together with the relation to properties of the Green
  functions in momentum space, notably the poles in external lines.
  This motivates which properties of Green functions need to be
  examined to derive the S-matrix.

\item An essential part of the specification of in- and out-states
  concerns wave functions for the center-of-mass motion of each of the
  asymptotic particles.  These are used in both momentum and
  coordinate space. The coordinate-space wave functions are simply
  positive energy solutions of the Klein-Gordon equation. In Sec.\
  \ref{sec:wfs} an account of properties of these wave functions is
  given, since these properties will be used in essential ways in the
  derivation of the reduction formula.  The material is by no means
  new, but it is not always found in standard textbooks, so it is
  useful to provide a systematic exposition here.

\item In Sec.\ \ref{sec:LSZ.statement}, a statement of the reduction
  formula is given, an improved derivation of which is the aim of
  later sections.

\item In Sec.\ \ref{sec:creation.ops} it is shown how to construct
  creation operators for particles in the in- and out-states, such
  that the necessary limits of infinite time are valid as strong
  limits, rather than merely weak limits.  As a motivation for the
  definitions, the LSZ versions of the operators are stated, and their
  deficiencies are demonstrated with the aid of explicit perturbative
  calculations.  The structure of the definition of the new creation
  operators will be such as to trivially avoid the problems, as we
  will see after the proof of the reduction formula.

  Some elementary properties of the operators are obtained in Sec.\
  \ref{sec:Af.ME.simple}.

\item In Sec.\ \ref{sec:LSZ.derivation}, the new derivation of the
  reduction formula is made. The derivation starts from the vacuum
  matrix elements of products the new annihilation and creation
  operators, and then analyzes the relevant limits of large times.

\item In Sec.\ \ref{sec:verify}, verification of important properties
  of the new annihilation and creation operators is made, including
  that the infinite-time limits are strong limits. 

\item Finally, some indications of possible generalizations are
  summarized in Sec.\ \ref{sec:generalizations}, and a comparison with
  the Haag-Ruelle method is given in Sec.\ \ref{sec:HR.comparison}.

\end{enumerate}

\section{Notations and conventions}
\label{sec:notation}

All the fields are in the Heisenberg picture, so that the states are
time-independent.  If renormalization needs to be considered, then the
fields are taken to be renormalized fields; for these, the
time-ordered Green functions are finite. The space-time metric has the
signature \metricSig.

In expanding quantities like fields in integrals over modes, I use the
Lorentz invariant form of integral with the same convention and
notation as Itzykson and Zuber's book \cite{Itzykson:1980rh}.  Thus a
free Klein-Gordon field obeys 
\begin{multline}
\label{eq:KG1}
    \phi_{\Free}(t,\ve{x})
    =
          \int \frac{\diff[3]{\ve{k}}}{(2\pi)^3\,2E_{\ve{k}}}
\times\\\times
            \left[
               a_{\ve{k}, \Free} e^{-iE_{\ve{k}}t + i\ve{k}\cdot\ve{x}}
             + a_{\ve{k}, \Free}^\dagger e^{iE_{\ve{k}}t - i\ve{k}\cdot\ve{x}}
            \right],
\end{multline}
where $E_{\ve{k}} = \sqrt{\ve{k}^2+m^2}$, $m$ is the mass of the
particle, and the commutators of the annihilation and creation
operators are
\begin{subequations}
\label{eq:a.a.adag.comm}
\begin{gather}
\label{eq:a.adag.comm}
  [a_{\ve{k},\Free}, a_{\ve{l},\Free}^\dagger] = 2E_{\ve{k}}\, (2\pi)^3 \, \delta^{(3)}(\ve{k}-\ve{l}) ,
\\
\label{eq:a.a.comm}
  [a_{\ve{k},\Free}, a_{\ve{l},\Free}] = [a_{\ve{k},\Free}^\dagger , a_{\ve{l},\Free}^\dagger] = 0 .
\end{gather}
\end{subequations}
Correspondingly, the normalization condition for single-particle
momentum eigenstates $|\3k\rangle=a_{\ve{k},\Free}^\dagger|0\rangle$ is
\begin{equation}
  \langle\3k|\3l\rangle = 2E_{\ve{k}}\, (2\pi)^3 \, \delta^{(3)}(\ve{k}-\ve{l}).
\end{equation}

Following Itzykson and Zuber, I define a notation $\difftilde{k}$
by
\begin{align}
\label{eq:mode.sum}
    \int \difftilde{k} \, \dots
   & \eqdef \int \frac{ \diff[3]{\ve{k}} }{ 2E_{\ve{k}} (2\pi)^3 } \, \dots
\nonumber\\
   & = \int \frac{ \diff[4]{k} }{ (2\pi)^4 } \, 2\pi\delta(\BDpos k^2-m^2) \, \theta(k^0) \, \dots .
\end{align}
Then we can write
\begin{equation}
\label{eq:KG}
  \phi_{\Free}(x)  =
          \int \difftilde{k}
            \left[
               a_{\ve{k},\Free} e^{\BDneg{ik\cdot x}}
             + a_{\ve{k},\Free}^\dagger e^{\BDpos{ik\cdot x}}
            \right].
\end{equation}
I use the standard convention that a 4-vector like $x$ is notated in
italics, while its spatial part is in boldface: $\ve{x}$.

Note that many authors use different conventions for
the momentum eigenstates and wave functions.  Correspondingly they
have slightly different integrals in their versions of Eqs.\
\eqref{eq:KG1}--\eqref{eq:KG} and in later equations.

We will make much use of time-ordered Green functions of the quantum
field(s) of a theory.  When there is one scalar field, which is the
only case we will treat explicitly, we use the notation
\begin{equation}
\label{eq:G.N}
  G_N(x_1,\dots,x_N) 
  \eqdef \langle0| T \prod_{j=1}^N \phi(x_j) |0\rangle.
\end{equation}
Its Fourier transform is defined by
\begin{multline}
\label{eq:tilde.G.N}
  \tilde{G}_N(k_1,\dots,k_N) 
  \eqdef \int \diff[4]{x_1} \dots \diff[4]{x_N}
\times\\\times
    e^{\BDneg i k_1\cdot x_1 \BDminus \dots \BDminus i k_N\cdot x_N}
    \, G_N(x_1,\dots,x_N) ,
\end{multline}
with the convention that the $k_j$ are treated as momenta flowing into
the Green function.

\section{Scattering formalism}
\label{sec:scattering}

In this section, I review the formalism of in- and out-states that is
used to formulate scattering theory in QFT.  Although the material is
more or less standard, it is useful to present it here, so that the
necessary background and motivation for the reduction formula are
given.  It is also useful to organize the presentation to show certain
differences in relativistic QFT compared with the situation in
non-relativistic quantum mechanics.  

The essential point is to provide a quantum-mechanical formulation of
the intuitive idea of scattering, to do this in Heisenberg picture,
and to do it in such a way as to be immune to issues such as those
associated with Haag's theorem and the non-existence of the
interaction picture in QFT.  Later sections will be concerned with
relating the results to properties of operators and Green functions.
When we prove the reduction formula, we are, among other things,
effectively verifying that the formalism is indeed appropriate.

We are familiar with scattering processes, where at asymptotically
large negative times a system's state consists of two incoming free
particles each moving classically.  The particles scatter in some
essentially finite region of space and time, and then at
asymptotically large positive times, the state is a linear combination
of various states consisting of outgoing free particles propagating
classically.  Experimental apparatus makes a measurement of the final
state, with approximate localization of the detected outgoing
particles in both space-time and momentum.  In normal applications of
QFT we only examine the momenta of the incoming and outgoing
particles, and present calculational results in terms of the S-matrix
(commonly in perturbative approximations).

\subsection{Scattering in the Schr\"odinger formulation of
  non-relativistic quantum mechanics}

We first examine how the intuitive ideas about scattering are
translated into quantum-mechanical form for systems of a finite number
of non-relativistic particles with interactions mediated by
potentials.  The results can be formalized in terms of Schr\"odinger
wave functions.  Essential simplifications compared with QFT are:
\begin{itemize}

\item Haag's theorem does not apply, so that the state space and the
  action of operators on it can be specified independently of the
  interaction.  Schr\"odinger wave functions are effectively an
  expansion of states in terms of eigenstates of the position
  operators. Thus for a single particle we can write its Schr\"odinger
  picture state as
  \begin{equation}
  \label{eq:Sch.w.f}
    |\psi,t\rangle = \int \diff[3]{\3a} \, |\3a\rangle \, \psi(\3a,t),
  \end{equation}
  where $|\3a\rangle$ is an eigenstate of the position operators with
  eigenvalues $\3a$, and with the normalization condition
  \begin{equation}
    \langle\3a|\3b\rangle = \delta^{(3)}(\3a-\3b).
  \end{equation}
  Observe that the state $|\3a\rangle$ can be considered as having the wave
  function $f_{\3a}(\3x) = \delta^{(3)}(\3x-\3a)$.  This is a distribution
  but not an ordinary function of position.  So any valid use has to
  be considered as having an implicit or explicit integral with a
  smooth test function, as in Eq.\ (\ref{eq:Sch.w.f}).  Effectively,
  one can treat $|\3a\rangle$ as a state-valued distribution, i.e., a
  mapping from smooth functions to states.  (Similar conceptual issues
  will apply when we work with momentum eigenstates in QFT.)

\item Asymptotically when particles are separated by much more than
  the range of the potential, their propagation is simply that of free
  particles: the action of the potential operator on the state goes to
  zero.  In contrast, in an interacting QFT, one can never turn off
  the interactions inside a particle.  Relative to a corresponding
  free theory, even a single particle in a QFT can be thought of as
  consisting of a complicated linear combination of states in the free
  field theory.  Moreover, Haag's theorem guarantees (in the
  relativistic case) that these linear combinations are badly
  divergent.  So effectively the free and interacting theories use
  different state spaces, which are dynamically determined.  Hence
  expressing the true single particle states in terms of free-particle
  states is a manner of speaking, only suggestive of the true
  situation.

\end{itemize}

The simplest case is of one particle in an external potential that falls
off rapidly enough at large distance, and that has no bound states.  The
Hamiltonian is
\begin{equation}
  H = \frac{\3{\op{p}}^2}{2m} + V(\3{\op{x}}).
\end{equation}
Here, to avoid confusion between operators and numeric-valued variables of
the same name, I have labeled QM operators with a hat. 

Scattering is implemented by a wave function $\psi(\3x,t)$ that solves
the time-dependent Schr\"odinger equation.  As $t\to-\infty$, $\psi(\3x,t)$
approaches the wave function for a freely propagating particle:
\begin{equation}
\label{eq:psi.past}
  \psi(\3x,t) \stackrel{t\to-\infty}{\to}
  \int \frac{\diff[3]{p}}{(2\pi)^3} \, \tilde\psi_{\In}(\3p) 
    \, e^{-iE_{\3p}t + i\3p\cdot\3x},
\end{equation}
where $E_{\3p} = \3p^2/(2m)$ and $\tilde\psi_{\In}(\3p)$ is a
momentum-space wave function narrowly peaked around some value of
momentum.\footnote{More general wave functions can be considered, but
  to correspond to the natural notion of scattering, wave packet
  states with momentum-space wave functions peaked around some
  particular momentum are appropriate.  In any case, more general
  states can be made by the taking of linear combinations.}  At large
positive times, the state has a similar expansion, but with different
coefficients:
\begin{equation}
  \psi(\3x,t) \stackrel{t\to+\infty}{\to}
  \int \frac{\diff[3]{\3p}}{(2\pi)^3} \, \tilde\psi_{\Out}(\3p) 
    \, e^{-iE_{\3p}t + i\3p\cdot\3x}.
\end{equation}

\subsection{Basis in- and out-states in elementary quantum mechanics}

To obtain a formulation in Heisenberg picture, we use two sets of
eigenfunctions of the Hamiltonian with certain boundary conditions at
spatial infinity.  These give what we will call the in- and out-basis
functions.  

For the wave functions $\phi_{\3p;\,\In}(\3x)$ for the in-basis we write
\begin{equation}
\label{eq;eigen.fn.in}
  \phi_{\3p;\,\In}(\3x) = e^{i\3p\cdot\3x} + g_{\3p;\,\In}(\3x), 
\end{equation}
with a corresponding state-vector notated as $|\3p;\,\In\rangle$.  It is an
eigenfunction of the Hamiltonian,
\begin{equation}
  H |\3p;\,\In\rangle = E_{\3p} |\3p;\,\In\rangle,
\end{equation}
that obeys the boundary condition that at large $\3x$, the scattered wave
$g_{\3p;\,\In}(\3x)$ has only an outgoing part at large $|\3x|$, i.e.,
\begin{equation}
  g_{\3p;\,\In}(\3x) \stackrel{|\3x|\to\infty}{\to} \frac{e^{i|\3p| |\3x|}}{|\3x|}
                f_{\3p;\,\In}(\theta,\phi),
\end{equation}
with no incoming term.  That is, there is no term with $x$ dependence
of the form $e^{-i|\3p| |\3x|}/|\3x|$.  The factor
$f_{\3p;\,\In}(\theta,\phi)$ is a function of the polar angle of $\3x$; it is
a result of the solution.

Thus at large $\3x$, the function $\phi_{\3p;\,\In}(\3x)$ is a
combination of a plane wave and an outgoing scattered wave:
\begin{equation}
\label{eq:psi.in.p}
  \phi_{\3p;\,\In}(\3x) = e^{i\3p\cdot\3x} 
                + \frac{e^{i|\3p| |\3x|}}{|\3x|} f_{\3p;\,\In}(\theta,\phi)
                + O(1/\3x^2).
\end{equation}

The out-basis functions are defined similarly, except that the scattered
wave has only an incoming part:
\begin{equation}
  \phi_{\3p;\,\Out}(\3x) = e^{i\3p\cdot\3x} 
                + \frac{e^{-i|\3p| |\3x|}}{|\3x|} f_{\3p;\,\Out}(\theta,\phi)
                + O(1/\3x^2).
\end{equation}
The two solutions can be related by a time-reversal transformation. 

Then a general solution of the time-dependent Schr\"odinger equation is of
the form
\begin{equation}
\label{eq:psi.in.t}
  \psi(\3x,t) =
  \int \frac{\diff[3]{\3p}}{(2\pi)^3} \, \tilde\psi_{\In}(\3p) 
    \, e^{-iE_{\3p}t} \, \phi_{\3p;\,\In}(\3x),
\end{equation}
i.e.,
\begin{equation}
  |\psi;t\rangle = 
  \int \frac{\diff[3]{\3p}}{(2\pi)^3} \, |\3p;\,\In\rangle
    \, \tilde\psi_{\In}(\3p) 
    \, e^{-iE_{\3p}t}.
\end{equation}

A stationary-phase argument can be used to show that at large negative
times, only the $e^{i\3p\cdot\3x}$ term in Eq.\ (\ref{eq:psi.in.p})
contributes.  The contribution of the scattered wave is strongly
suppressed.  Then the wave function $\psi(\3x,t)$ obeys the condition of
a free incoming particle, as in Eq.\ (\ref{eq:psi.past}).  At large
positive time, the scattered wave also contributes.

A reversed set of conditions applies to an expansion in the out-basis
states. 

The Heisenberg-picture state is defined to be the Schr\"odinger-picture
state at time 0.  Thus we can expand the Heisenberg state in terms of
either set of basis states:
\begin{subequations}
\label{eq:expand}
\begin{align}
\label{eq:expand.in}
  |\psi\rangle_{\rm H}
  & =   \int \frac{\diff[3]{\3p}}{(2\pi)^3} \, |\3p;\,\In\rangle \, \tilde\psi_{\In}(\3p)
\\
\label{eq:expand.out}
  & =   \int \frac{\diff[3]{\3p}}{(2\pi)^3} \, |\3p;\,\Out\rangle \,
  \tilde\psi_{\Out}(\3p).
\end{align}
\end{subequations}

We now show that the inner product of the basis states of the same type
has the standard normalization:
\begin{equation}
\label{eq:norm.1}
  \langle\3p;\,\In|\3q;\,\In\rangle
  = (2\pi)^3 \delta^{(3)}(\3p-\3q)
  = \langle\3p;\,\Out|\3q;\,\Out\rangle.
\end{equation}
(Since we are in a non-relativistic situation, we omit the $2E_{\3p}$
factor that we use in the relativistic case.)  The derivation is by
considering the inner product $\langle\psi_1;t|\psi_2;t\rangle$ of two states of the
form given in Eq.\ (\ref{eq:psi.in.t}).  Because time-evolution is
unitary, the inner product is independent of $t$.  From the expansion
in the in-basis states we have
\begin{multline}
  \langle\psi_1;t|\psi_2;t\rangle = \int \frac{\diff[3]{\3p}}{(2\pi)^3} \int
                  \frac{\diff[3]{\3q}}{(2\pi)^3}
\times\\\times
              \tilde\psi_{1;\,\In}^*(\3p) \tilde\psi_{2;\,\In}(\3q) 
              \, \langle\3p;\,\In|\3q;\,\In\rangle.
\end{multline}
But because of the time-independence of the inner product, we can also
compute in the limit of $t\to-\infty$, when we can replace the wave functions by
plane waves, as at Eq.\ (\ref{eq:psi.past}), and then we use the usual
inner product of plane waves to give
\begin{equation}
  \langle\psi_1;-\infty|\psi_2;-\infty\rangle = \int \frac{\diff[3]{\3p}}{(2\pi)^3}
              \, \tilde\psi_{1;\,\In}^*(\3p) \tilde\psi_{2;\,\In}(\3p).
\end{equation}
A similar derivation applies to the expansion in out-basis states.  Hence
the scattering solutions obey (\ref{eq:norm.1}), and this normalization
follows directly from the normalization of the plane-wave part in
(\ref{eq;eigen.fn.in}). 

Observe that the inner product (\ref{eq:norm.1}) has to be interpreted
in a distributional sense, i.e., integrated with (smooth) test
function(s).  This is evidenced by the presence of a delta-function.
Trying to calculate the inner product directly, by an integral over
$\3x$ of $\phi_{\3q;\,\In}(\3x)^\phi_{\3p;\,\In}(\3x)$ is prevented by the
lack of convergence of the integral.

\subsection{The S-matrix in elementary quantum mechanics}

The S-matrix can be defined as a relation between the expansions in
the 2 sets of basis functions given in (\ref{eq:expand}). Let us work
in the Heisenberg picture. We have
\begin{align}
  |\psi\rangle_{\rm H}
  & =   \int \frac{\diff[3]{\3p}}{(2\pi)^3} \, |\3p;\,\In\rangle \, \tilde\psi_{\In}(\3p)
\nonumber\\
  & =   \int \frac{\diff[3]{\3p}}{(2\pi)^3} \frac{\diff[3]{\3k}}{(2\pi)^3}
        \, |\3k;\,\Out\rangle \, \langle\3k;\,\Out |\3p;\,\In\rangle \, \tilde\psi_{\In}(\3p).
\end{align}
Then the S-matrix could be defined by
\begin{equation}
\label{eq:S.QM.1.def}
  S_{\3k,\3p} = \langle\3k;\,\Out |\3p;\,\In\rangle,
  = \int \diff[3]{\3x} \phi_{\3k;\,\Out}^*(\3x) \phi_{\3p;\,\In}(\3x).
\end{equation}
Then the two sets of expansion coefficients are related by
\begin{equation}
\label{eq:S.QM.2.def}
  \tilde\psi_{\Out}(\3k) 
  = \int \frac{\diff[3]{\3p}}{(2\pi)^3} \, S_{\3k,\3p}
  \, \tilde\psi_{\In}(\3p).
\end{equation}

As with many other formulas involving basis states labeled by momenta,
the definition (\ref{eq:S.QM.1.def}) is to be interpreted
distributionally, i.e., when integrated with smooth test functions.
If nothing else the $\3x$ integral in (\ref{eq:S.QM.1.def}) would
otherwise diverge.  So we could better define the S-matrix by Eq.\
(\ref{eq:S.QM.2.def}), as a relation between expansion coefficients;
effectively this will be the definition we use in QFT, thereby
avoiding a definition directly in terms of basis states.

In working with the S-matrix, it is convenient to extract all the
delta functions, i.e., to make explicit the intrinsically
distributional part, and thus to write
\begin{equation}
  S_{\3k,\3p} = (2\pi)^3 \delta^{(3)}(\3k-\3p) + 2\pi \delta(E_{\3k}-E_{\3p}) \, \mathcal{A}(\3k,\3p).
\end{equation}
The amplitude $\mathcal{A}(\3k,\3p)$ is an ordinary function of its
arguments (but restricted to the case that the energies of the two momenta
are equal).  This amplitude is a natural target of Feynman-graph
calculations, especially in the generalization of these results to QFT.

It can be shown that it is related to the function $f$ that is the
coefficient of the $1/x$ term in (\ref{eq:psi.in.p}) by
\begin{equation}
  \mathcal{A}(\3k,\3p)
  = \frac{-2\pi i}{m} f_{\3p;\,\In}(\theta_{\3k},\phi_{\3k}).
\end{equation}
A fundamental derivation from first principles can be made by
computing the asymptotics of $\psi(\3x,t)$ for $t\to\pm\infty$; this is done
starting from the expansion in $|\3p;\,\In\rangle$, using stationary phase
methods, and then matching onto a plane-wave expansion as $t\to\infty$.  That
expansion has coefficients $\tilde{\psi}_{\Out}(\3k)$.  Somewhat shorter
derivations can be made with the aid of insights as to what happens in
the limit that the expansion function $\tilde{\psi}_{\In}(\3p)$
approaches a delta function, so that the $t\to-\infty$ wave function
approaches a plane wave cut off at very large distances.  An
appropriate function would be
\begin{equation}
\label{eq:narrow.wf}
  \tilde{\psi}_{\In}(\3p)
  = \left( \frac{4\pi}{\Delta p^2}\right)^{3/2} e^{-(\3p-\3p_0)^2/\Delta p^2},
\end{equation}
with $\Delta p\to0$.

\subsection{Generalization}
\label{sec:generalization}

When we go to relativistic QFT, we will need the multiparticle case,
where we will write basis states with arbitrarily many particles as
\begin{subequations}
\label{eq:bases}
\begin{gather}
\label{eq:in.basis}
  |\3{p}_1,\dots, \3{p}_n;\,\In\rangle,
\\
\label{eq:out.basis}
  |\3{q}_1,\dots, \3{q}_n;\,\Out\rangle,
\end{gather}
\end{subequations}
often with the natural generalizations to allow labels for particle type
and for spin states.  However, unlike the case of Schr\"odinger
wave-function theory, we will not have a direct definition\footnote{Many
  textbooks by reputable authors appear to provide constructions of the
  basis states with the aid of the interaction picture and manipulations
  inspired by those that done in elementary quantum mechanics. However,
  Haag's theorem guarantees that the interaction picture does not exist
  --- cf.\ Streater and Wightman's \cite{Streater:1964} ironic restatement
  of Haag's theorem as ``The interaction picture exists if and only if
  there is no interaction''.  So any direct construction of scattering
  states and the S-matrix by interaction-picture methods must be regarded
  as highly suspect, at the least.} of these objects, e.g., as wave
functions that are eigenfunctions of the Hamiltonian subject to certain
boundary conditions.  So we will formulate the necessary concepts in terms
of normalizable states with specified asymptotic particle content, and
arrange that further derivations use only normalizable states as starting
points.

In a QFT we construct normalizable states by applying products of
field operators to the vacuum and integrating with smooth functions of
the positions of the field operators.  The taking of linear
combinations then gives general states. A main aim of this paper is to
provide a construction of this kind for a state that has a specified
momentum content for asymptotic incoming particles, and similarly for
outgoing particles.

This implies that it is useful to formulate the methods in terms of
normalizable states only, i.e., states genuinely in the Hilbert space
of the theory, and only after that to provide a formulation involving
the momentum-dependent basis states.  That is done by the natural
generalization of the construction given above (\ref{eq:narrow.wf}).
To implement the idea of states with specified incoming particle
content, we will first specify the relevant properties of a Fock space
decomposition of the state space.  This simply matches the
corresponding structures in wave-function theory.  Later sections will
provide constructions of states that implement the Fock space.  The
physical interpretation as free incoming particles of definite
momentum content will be determined by the localization of the
particles as determined by locations of the fields used to construct
the states, and by a computation of the effect of applying the
momentum operator on the states.  In Sec.\
\ref{sec:localization.momentum}, we will find that indeed the
particles propagate asymptotically along the appropriate classical
trajectories and have the expected momenta.

The states for which we actually give a construction have product wave
functions.  This will be sufficient, because the taking of linear
combinations gives general states in the Hilbert space, i.e., the
states with product wave functions form a complete set in the sense
used in Hilbert space.

After the construction of states with specified content in the initial
or final states, various quantities of interest can be computed from
the Green functions of the theory; these include S-matrix elements,
and matrix elements of operators between specified states.

In Schr\"odinger wave-function theory, basis states for incoming
particles, as in (\ref{eq:bases}), are defined to be the sum of a
multidimensional plane wave and what is asymptotically an out-going
wave.  Whenever a bound state is one of the asymptotic particles, then
the proper generalization of the plane wave idea is that there is a
plane wave factor for the center-of-mass coordinate of the bound
state, and this is multiplied by the wave function depending on the
relative coordinates of the elementary constituents. 

We now abstract from the above discussion the Fock-space formalism that
applies in QFT to states with specified asymptotic particle content.  For
this presentation following, we assume that there is one type of particle,
and that it is a boson of nonzero mass $m$.  Generalizations to multiple
types of particle, including fermions, are elementary, and can be worked
out from material in standard textbooks.

Given the basis in-states (\ref{eq:in.basis}), a general normalizable
state is specified by an infinite array of momentum-space wave
functions, $\underline{f}=(f_0,f_1,\ldots)$, and has the form
\begin{equation}
\label{eq:f.gen.in}
  |\underline{f};\,\In\rangle
  = \sum_{n=0}^\infty \frac{1}{n!}
    \int \prod_{j=1}^n \difftilde{p_j}  f_n(\3p_1,\ldots,\3p_n) \,
    |\3{p}_1,\dots, \3{p}_n;\,\In\rangle.
\end{equation}
The wave functions are assumed to be symmetric in their arguments, and
the $1/n!$ factor is a choice of normalization to reflect the multiple
counting of identical states in the integral over all momenta.  We
have now restored the relativistic normalization for integrals.  Note
that the label ``$\In$'' in $|\underline{f};\In\rangle$ does not refer to a
particular type of state. Rather it refers to the specification of the
state in terms of a given array of momentum-space functions
$\underline{f}$, which specify the state in terms of its particle
content at asymptotically large negative times.  Equation
(\ref{eq:f.gen.in}) is an expansion of one particular Heisenberg-picture
state, so the coefficients have no time dependence.

Exactly similar considerations apply to treating states with given
momentum content in the asymptotic future, i.e., states denoted
$|\underline{f};\Out\rangle$.

The Hilbert-space structure can now be specified without mention of
the basis states themselves, by referring everything to the
normalizable states $|\underline{f};\In\rangle$.  The inner product is then
\begin{multline}
  \label{eq:inner.product}
  \langle\underline{f};\,\In | \underline{g};\,\In\rangle
\\
  = \sum_{n=0}^\infty \frac{1}{n!}
    \int \left( \prod_{j=1}^n \difftilde{p_j} \right)
   g_n^*(\3p_1,\ldots,\3p_n) f_n(\3p_1,\ldots,\3p_n).
\end{multline}

\subsection{Product states}

For our derivations of the S-matrix etc, it will be sufficient to
restrict to product states.  Thus given momentum-space wave functions
$\tilde{f}_1(\3p)$ and $\tilde{f}_2(\3p)$ for single particles, we
will define a two-particle state $|f_1,f_2;\,\In\rangle$ to have the wave
function
\begin{equation}
  \tilde{f}_1(\3p_1) \tilde{f}_2(\3p_2) + \tilde{f}_2(\3p_1) \tilde{f}_1(\3p_2).
\end{equation}
Since we will often work with related functions in coordinate space, I
now use the over-tilde to denote momentum-space quantities.  

More generally an $n$-particle initial state $|f_1,\dots,f_n;\,\In\rangle$
is defined to have the wave function
\begin{equation}
  \prod_{j=1}^n \tilde{f}_j(\3p_j) + \mbox{permutations},
\end{equation}
with a total of $n!$ terms.  A product state $|g_1,\dots,g_n;\,\Out\rangle$
with specified content in the far future is defined similarly.

In the notation using basis states, we write, for example, a two-particle
initial product state as
\begin{equation}
\label{eq:i}
  |f_1,f_2;\,\In\rangle = \int \difftilde{p_1} \, \difftilde{p_2}
                \, |\3p_1,\3p_2;\,\In\rangle
                \, \tilde{f}_1(\3p_1) \tilde{f}_2(\3p_2).
\end{equation}
The symmetrization appropriate to bosons is enforced by the basis
states, and no separate symmetrization of the wave function is needed.

\begin{widetext}
With the conventions of Sec.\ \ref{sec:notation}, the normalization of
this state is given by
\begin{align}
\label{eq:in-norm}
  \langle f_1,f_2;\,\In|f_1,f_2;\,\In\rangle
  & = \int \difftilde{p} | \tilde{f}_1(\3p) |^2 
      \times \int \difftilde{p} | \tilde{f}_2(\3p) |^2 
    + \int \difftilde{p} \tilde{f}_1(\3p)^*\tilde{f}_2(\3p)
      \times \int \difftilde{p} \tilde{f}_2(\3p)^*\tilde{f}_1(\3p)
\nonumber\\
  & \simeq \int \difftilde{p} | \tilde{f}_1(\3p) |^2 
      \times \int \difftilde{p} | \tilde{f}_2(\3p) |^2 .
\end{align}
In normal applications, the two wave functions are chosen to describe
two very distinct incoming particles and therefore have negligible
overlap, or even zero overlap.  Then only the first term in Eq.\
(\ref{eq:in-norm}) needs to be retained, as indicated on the second
line.  It is generally sensible to normalize each wave function
separately to
\begin{equation}
\label{eq:one-norm}
  \int \difftilde{p} | \tilde{f}_1(\3p) |^2
  = \int \difftilde{p} | \tilde{f}_2(\3p) |^2 
  = 1,
\end{equation}
which gives a normalized state $|f_1,f_2;\,\In\rangle$ to a very good
approximation.

One can apply the formalism of in-states to have more than 2 incoming
particles, and this is done in the general theory for the S-matrix and
the LSZ reduction formula, etc.  But such states are not normally used
for describing standard experiments.

As regards out-states, the situation concerning multiple particles is
of course different. In the same way as we did for in-states, we write
an expansion of normalizable out-states in terms of basis states:
\begin{equation}
\label{eq:f}
  |g_1,\dots,g_n;\,\Out\rangle 
  = \int \difftilde{q_1} \, \dots \, \difftilde{q_{n'}}
                \, |\3q_1,\dots,\3q_{n'};\,\Out\rangle
                \, \prod_{k=1}^n \tilde{g}_k(\3q_k) .
\end{equation}

Generally, in QFT we do not have an adequate direct definition of the
basis states.  Instead we will see how to construct normalizable
states $|f_1,\dots,f_n;\,\In\rangle$ and $|g_1,\dots,g_n;\,\Out\rangle$ by
applying suitable explicitly defined operators to the vacuum.  After
that we can construct basis states by the use of a limit of wave
functions that approach delta functions, effectively a distributional
construction.

\subsection{The S-matrix in general}

Probabilities relevant for scattering are constructed from (the
absolute value squared) of overlap amplitudes such as $\langle
g_1,\dots,g_n;\,\Out|f_1,f_2;\,\In\rangle$.  These can be expressed as integrals
over the wave functions.  Thus we write
\begin{equation}
\label{eq:g.f}
  \langle g_1,\dots,g_n;\,\Out|f_1,f_2;\,\In\rangle
  = \int \prod_{k=1}^n \left( \difftilde{q_k} \right)
      \, \prod_{j=1}^{2} \left( \difftilde{p_j} \right)
      \, \prod_{k=1}^n \tilde{g}_k^*(\3q_k)
      \, \prod_{j=1}^{2} \tilde{f}_j(\3p_j)
      \, S_{\3q_1,\dots,\3q_n;\,\3{p}_1,\3{p}_2} .
\end{equation}
\end{widetext}
The quantity $S_{\3q_1,\dots,\3q_n;\,\3{p}_1,\3{p}_2}$ is called the
S-matrix. It can be considered as the overlap
$\langle\3q_1,\dots,\3q_n;\,\Out|\3{p}_1,\3{p}_2;\,\In\rangle$ of basis states.
More compactly, if $\alpha$ and $\beta$ are arrays of momentum labels for in-
and out-basis state, of the form given in Eq.\ (\ref{eq:bases}), then
we write
\begin{equation}
\label{eq:S.beta.alpha}
  S_{\beta;\alpha} = \langle\beta;\,\Out|\alpha;\,\In\rangle.
\end{equation}
The S-matrix has two components.  One is a unit matrix term, which can
be symbolized by $\delta_{\beta\alpha}$, and that is the expression of a situation with
no scattering.  The other is the term with scattering.  One therefore
can write:
\begin{equation}
\label{eq:S.1+iT}
  S_{\beta\alpha} = \delta_{\beta\alpha} + iT_{\beta\alpha}, 
\end{equation}
where the T-matrix contains the contribution of actual scattering.
When we restrict to 2-body initial states, as is normal, then the
reduction formula, to be discussed below, gives the T-matrix in terms
of connected Feynman graphs only.

However, overlaps such as those on the right-hand side of
\eqref{eq:S.beta.alpha}, involving some kind of generalized plane-wave
states, are hard, if not impossible, to define directly.  One can
already see this in elementary quantum mechanics in Eq.\
(\ref{eq:S.QM.1.def}), where the basis wave functions do not fall off
for large $\3x$ and so the integral over all $\3x$ is not defined as
an ordinary integral.  Distributional methods give an appropriate
definition, as explained around that equation.  Then implicitly or
explicitly there is an integral with momentum-space wave functions.
These considerations apply equally to QFT.

Now the left-hand side of \eqref{eq:g.f} is properly defined in
itself.  Then the S-matrix is a kind of master function with the aid
of which all of $\langle g_1,\dots,g_n;\,\Out|f_1,\dots,f_{n'};\,\In\rangle$ can
be computed by integrating the S-matrix multiplied by wave functions.
One could equally say that the S-matrix gives a basis for constructing
all cases of $\langle g_1,\dots,g_n;\,\Out|f_1,\dots,f_{n'};\,\In\rangle$.

The significance of the reduction formula, to be discussed below, is
that it shows how to express the S-matrix in terms of quantities that
can be computed (e.g., from Feynman graphs) in momentum space with
perfectly definite values of external momentum.  The primary results
of many calculations are for values of particular S-matrix elements.
Then, by a well-known formula, scattering cross sections are expressed
in terms of these, thereby giving experimentally testable predictions.

Generally, we conceive of the state of a system involving scattering
as being specified by the contents of the initial state, e.g., by Eq.\
\eqref{eq:i}.  This is a Heisenberg-picture state, which is
independent of time.  Measurements involve the determination of
momenta of the outgoing particles after the scattering.  It is
therefore useful to express the state $|f_1,f_2;\,\In\rangle$ as a combination
of the (momentum-space-basis) out-states:
\begin{widetext}
\begin{equation}
\label{eq:state.expansion}
  |f_1,f_2;\,\In\rangle
  = \sum_{n=2}^\infty \frac{1}{n!} \int \prod_{k=1}^n \left( \difftilde{q_k} \right)
      \, \int \prod_{j=1}^{2} \left( \difftilde{p_j} \right)
      \, |\3{q}_1,\dots, \3{q}_n;\,\Out\rangle
      \, S_{\3q_1,\dots,\3q_n;\,\3{p}_1,\3{p}_2}
      \, \prod_{j=1}^{2} \tilde{f}_j(\3p_j) ,
\end{equation}
where the $g$ functions no longer appear, and the $1/n!$ factor takes
care of the effect of the indistinguishability of the $n$ final-state
particles.  This formula follows easily from the previous ones.  It
has a sum over all possible numbers of particles in the final state.
Nonzero terms will, of course, be restricted in any particular
application to those permitted by momentum conservation (implemented
by a delta function in the S-matrix).  Equation
\eqref{eq:state.expansion} can be conceived of as
\begin{subequations}
\begin{align}
\label{eq:state.expansion.2}
  |f_1,f_2;\,\In\rangle
  &= \sum_{n=2}^\infty \int \prod_{k=1}^n \left( \difftilde{q_k} \right)
      \, \int \prod_{j=1}^{2} \left( \difftilde{p_j} \right)
      \, |\3{q}_1,\dots, \3{q}_n;\,\Out\rangle
      \, \langle\3q_1,\dots,\3q_n;\,\Out|\3{p}_1,\3{p}_2;\,\In\rangle
      \, \prod_{j=1}^{2} \tilde{f}_j(\3p_j) 
\\
  &= \sum_\beta |\beta;\,\Out\rangle \langle\beta;\,\Out|f_1,f_2;\,\In\rangle,
\end{align}
\end{subequations}
where the sum over $\beta$ is a very symbolic notation for the sum and
integral over all possible final states (including identical-particle
effects, where needed).
\end{widetext}

\subsection{Further comments on the need to use normalizable states}
\label{sec:why.wf}

Actual scattering events are approximately localized in space and
time.  Thus, if a physical state of a system is represented by a
Heisenberg-picture state $|\psi\rangle$ in which there is a standard
scattering, then we are able to say that to a good approximation the
state is composed of two incoming particles for time $t$ less than
some value $t_-$.  For large enough times, $t>t_+$, the state is of
some number of outgoing particles, or rather is a superposition of
such configurations.  The times $t_-$ and $t_+$ can be determined from
the state, to some approximation.  Similarly an approximate spatial
location of the scattering can be determined.  The state is certainly
not invariant under translations in space and time.

Suppose we considered a state $|P\rangle$ of particles of exactly definite
momenta.  Then the state is an eigenstate of total 4-momentum.  Since
the operators for 4-momentum generate translations, the effect of a
translation is to multiply the state by a phase.  The phase is
irrelevant to the physical content and there is no way to generate
preferred values of time and position from the state $|P\rangle$.  

Now it could be argued that over the scale of the scattering, the
particles are governed by wave functions that are plane waves to a
good approximation, and that therefore the wave functions are not
particularly relevant.  Moreover it is true that actual Feynman graph
calculations for scattering use mathematically exact values of momenta
on the external lines, and the methods of calculation are indeed
justified by the reduction formula.  But the actual derivation of the
calculational methods does need the wave packets if it is to be valid.
What is shown is that the details of the wave packets drop out
provided that their size is simultaneously much larger than the
spatial size of the scattering event itself and much smaller than the
distance to the experimental apparatus for detecting outgoing
particles.  This condition is obviously satisfied by many orders of
magnitude in typical experiments.

See also Coleman's lecture notes on QFT
\cite{Coleman:2011xi,Coleman.2018} for another explanation of why full
scattering theory in QFT cannot be formulated directly in terms of
plane-wave states.

\section{Large-time asymptotics of Green function}
\label{sec:asy.Green.fn}

Even before seeing an actual complete demonstration, it is natural to
expect that the $n'\to n$ S-matrix is related to the asymptotics for an
$n+n'$-point Green function where the times of $n'$ of the fields are
taken to $-\infty$, and the times of the other $n$ fields are taken to
$+\infty$, to correspond to the initial and final particles.  It is the LSZ
reduction formula that realizes this expectation and gives the exact
quantitative relation.

Since the details of the proof of the reduction formula are rather
abstract, it is useful to start with a direct examination of the
coordinate-space asymptotics of Green functions, in simple
generalizable examples.  This motivates and illuminates the technical
steps of the proof of the reduction formula.

\begin{figure}
  \centering

    \includegraphics[scale=0.5]{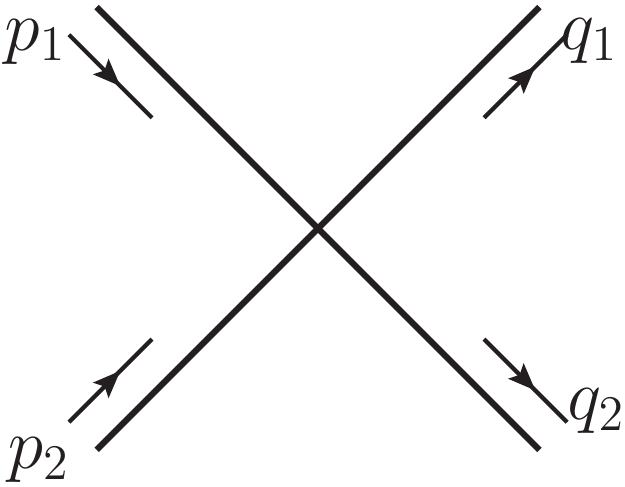}
    \caption{Lowest-order graph for 4-point Green function in $\phi^4$
      theory, with momentum labeling appropriate for $2\to2$
      scattering.}
\label{fig:2-to-2-LO}
\end{figure}

We consider first the connected 4-point Green function in lowest-order
in $\phi^4$ theory, Fig.\ \ref{fig:2-to-2-LO}, whose value in coordinate
space is expressed in terms of the well-known momentum-space formula
by
\begin{widetext}
\begin{align}
  G_4(x_1,x_2,y_1,y_2)
  = {}&
   -i\lambda \int \frac{\diff[4]{p_1}}{(2\pi)^4} \frac{\diff[4]{p_2}}{(2\pi)^4}
     \frac{\diff[4]{q_1}}{(2\pi)^4} \frac{\diff[4]{q_2}}{(2\pi)^4}
    \, e^{\BDpos i p_1\cdot x_1 \BDplus i p_2\cdot x_2 \BDminus i q_1\cdot y_1 \BDminus i q_2\cdot y_2}
\times \nonumber\\& \times 
     \, \frac{i}{\BDpos p_1^2-m^2+i\epsilon}
     \, \frac{i}{\BDpos p_2^2-m^2+i\epsilon}
     \, \frac{i}{\BDpos q_1^2-m^2+i\epsilon}
     \, \frac{i}{\BDpos q_2^2-m^2+i\epsilon}
     \, (2\pi)^4 \delta^{(4)}(p_1+p_2-q_1-q_2).
\end{align}
We analyze this in the limit that each $x_j^0\to-\infty$ and each
$y_j^0\to+\infty$.  Since the asymptotics are ultimately governed by the
separation between each external vertex and the interaction vertex, it
is useful to use the formula
\begin{equation}
  (2\pi)^4 \delta^{(4)}(p_1+p_2-q_1-q_2) 
   = \int \diff[4]{z} e^{\BDneg i(p_1+p_2-q_1-q_2)\cdot z} 
\end{equation}
to express the Green function as an integral over the position of the
interaction together with independent integrals over the momentum of
each line:
\begin{align}
  G_4(x_1,x_2,y_1,y_3)
  = {}&
   -i\lambda \int \diff[4]{z} \int \frac{\diff[4]{p_1}}{(2\pi)^4} \frac{\diff[4]{p_2}}{(2\pi)^4}
     \frac{\diff[4]{q_1}}{(2\pi)^4} \frac{\diff[4]{q_2}}{(2\pi)^4}
    \, e^{ \BDpos i p_1\cdot (x_1-z) \BDplus i p_2\cdot (x_2-z) 
          \BDminus i q_1\cdot (y_1-z) \BDminus i q_2\cdot (y_2-z) }
\times \nonumber\\& \times 
     \, \frac{i}{\BDpos p_1^2-m^2+i\epsilon}
     \, \frac{i}{\BDpos p_2^2-m^2+i\epsilon}
     \, \frac{i}{\BDpos q_1^2-m^2+i\epsilon}
     \, \frac{i}{\BDpos q_2^2-m^2+i\epsilon} .
\end{align}
\end{widetext}

For each propagator we have an integral of the form
\begin{equation}
  \int \frac{\diff[4]{k}}{(2\pi)^4} \, e^{ \BDpos i k\cdot (x-z) }
     \, \frac{i}{\BDpos k^2-m^2+i\epsilon}.
\end{equation}
Now, in the limit that we are interested in, the position difference
$x-z$ between ends of the propagator is scaled to be large.  Then for
almost all values of $k$, we can deform the integration of $k$ off the
real axis and get a strong suppression from the effect of the
imaginary part of $k$ in the exponential.  We need to determine where
the deformation is not possible, and what the consequences are.

To formalize the analysis, we notate the contour deformation as
\begin{equation}
\label{eq:deform}
  k = k_R + i\kappa k_I(k_R).
\end{equation}
Here $k_R$ is real, $k_I(k_R)$ is a real-valued function, and $\kappa$ is a
parameter ranging from 0 to 1.  The variable of integration is $k_R$.
Then, given a function $k_I(k_R)$ and a value of $\kappa$, Eq.\
(\ref{eq:deform}) determines a contour of 4 real dimensions in a
complex space of 8 real dimensions.  Varying $\kappa$ from 0 to 1 gives a
continuous family of contours, starting from an integration over all
real $k$.  Cauchy's theorem tells us that the integral is independent
of $\kappa$, provided that no singularities are encountered as the contour
is deformed.

To get a suppression we need
\begin{equation}
\label{eq:suppression}
  \BDpos k_I \cdot (x-z) > 0,
\quad \text{for suppression by exponential}.
\end{equation}
The suppression of the integral is exponential in the large scaling of
$x-z$.  

However at certain points the contour deformation is obstructed by the
propagator pole.  At the pole, $k^2=\BDpos m^2$.  Suppose at some
point the pole \emph{fails} to obstruct the contour deformation, then
at $\kappa=0$, $k_I$ times the derivative of $\BDpos k^2-m^2$ is positive,
compatible with the $i\epsilon$ prescription:
\begin{multline}
\label{eq:avoid.pole}
  k_I^\mu \frac{\partial}{\partial k_R^\mu} (\BDpos k_R^2-m^2) = \BDpos 2 k_I\cdot k_R > 0,
\\ \text{for non-obstruction by pole}.
\end{multline}
If there exists a $k_I$ obeying both of conditions
(\ref{eq:suppression}) and (\ref{eq:avoid.pole}), then we can deform
the contour and get an exponential suppression.  

To get asymptotics of the Green function, we are interested in
unsuppressed contributions, and these arise where such a $k_I$ fails
to exist.  The failure occurs when the vectors $x-z$ and $k_R$ are in
opposite directions, i.e., when $x-z + \alpha k_R=0$ for some positive $\alpha$.
This immediately implies that $x-z$ is time-like, since $k_R$ must be
on-shell in order that there is a pole to obstruct the deformation.

Consider the case that $k$ is one of the incoming momenta $p_1$ or
$p_2$.  We already know $x-z$ is time-like for an unsuppressed
contribution.  Since $x$ has large negative time, $x-z$ is a time-like
past-pointing vector.  The non-suppression condition then states that
$k_R$ has positive energy and is on-shell.  Thus the non-suppressed
contributions come from near the configuration where $k_R$ corresponds
to propagation of a classical particle from $x$ to $z$, which is
exactly what we expect for incoming asymptotic particles in a
scattering process.

Similarly the asymptotic behavior when the times of $y_1$ and $y_2$
get large and positive corresponds to momenta $q_1$ and $q_2$ for
outgoing classical particles.

Hence the asymptotic large-time behavior of the Green function is
controlled by the poles of the propagators on the external lines, with
the momenta involved being those of the appropriate classical
particles.  The mass of a particle is determined by the position of
the pole in the propagator.

\begin{figure}
  \centering

    \includegraphics[scale=0.45]{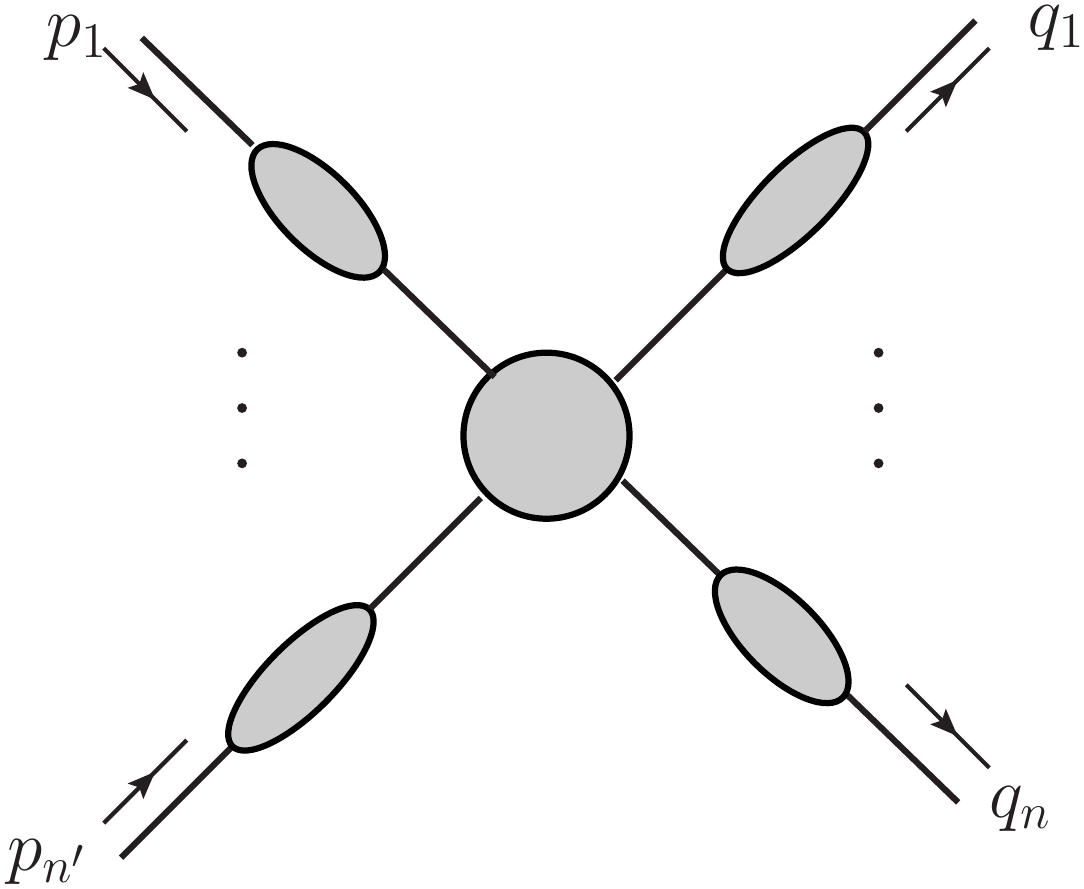}
    \caption{Graphical structure of general connected Green function
      with factorization of full external propagators, with momentum
      labeling appropriate for $n'\to n$ scattering.}
\label{fig:n'-to-n}
\end{figure}

This idea generalizes readily.  In Fig.\ \ref{fig:n'-to-n} the
connected part of an $n+n'$-point Green function is decomposed into
the product of an amputated part and full propagators for each
external line.  Generally a full propagator with momentum $k$ has not
only a particle pole, but a set of other, weaker, singularities at
higher values of $\BDpos k^2$ that are thresholds for $k$ to make
multiple particles.  The dominant large time behavior is governed by
the strongest singularity, i.e., the particle pole.

\section{K\"allen-Lehmann representation}
\label{sec:KL}

The K\"allen-Lehmann representation \cite{Kallen:1952zz,Lehmann:1954xi}
(or ``spectral representation'') is important to the general analysis
of the 2-point function, and to the generality of the correspondence
between single-particle states and the positions of poles of 2-point
functions in momentum space.  More detailed treatments can be found in
many textbooks on QFT, and I will only summarize here the results
needed for this paper.

The K\"allen-Lehmann representation concerns the 2-field correlator:
\begin{equation}
\label{eq:2.fixed}
  \langle0| \phi(x) \phi(y) |0\rangle,
\end{equation}
and it is obtained by inserting a complete set of in- or out-basis
states between the two fields and by using the fact that for a state
$|p\rangle$ of 4-momentum $p$, the $x$ dependence of $\langle0| \phi(x) |p\rangle$ obeys
\begin{equation}
  \langle0| \phi(x) |p\rangle = \langle0| \phi(0) |p\rangle e^{\BDneg i p\cdot x} .
\end{equation}
Both of the time-ordered propagator and the vacuum expectation value
of equal-time commutators (and non-equal-time commutators) can be
obtained from (\ref{eq:2.fixed}).  All of these are expressed in terms
of a non-negative spectral function $\rho(s)$ which measures the size of
$\left|\langle0| \phi |p\rangle\right|^2$ for states of invariant squared mass $s$,
and is defined by
\begin{equation}
  \label{eq:rho.s}
  \rho(s) \eqdef (2\pi)^3 \sum_X \delta^{(4)}(p_X-p_s) |\langle0| \phi(0) |X\rangle|^2,
\end{equation}
where $\sum_X$ denotes a sum/integral over a complete set of 4-momentum
eigenstates (which can be chosen to be the out-basis states or the
in-basis states), $p_X$ is the 4-momentum of $|X\rangle$, and $p_s$ is a
vector obeying $p_s^2=\BDpos s$ and having a positive energy
component.

We can now express the 2-field correlator in terms of a free field
correlator:
\begin{align}
\label{eq:prod.prop}
  \langle0| \phi(x) \phi(y) |0\rangle &= \int_0^\infty \diff{s} \rho(s) \,
                       \Delta(\BDpos(x-y)^2;s)
\nonumber\\
  &\hspace*{-1.3cm}=
  \int_0^\infty \diff{s} \rho(s) \,
  \int \frac{\diff[3]{\3k}}{(2\pi)^3 2 \sqrt{\3k^2+s}}
  e^{\BDneg ik\cdot (x-y) },
\nonumber\\
  &\hspace*{-1.3cm}=
  \int \frac{\diff[4]{k}}{(2\pi)^3} \rho(k^2)\theta(k^0)
  e^{\BDneg ik\cdot (x-y) },
\end{align}
where in the exponent on the second line $k^0=\sqrt{\3k^2+s}$.  The
quantity $\Delta(\BDpos(x-y)^2;s)$ is the correlator (\ref{eq:2.fixed}) for
the case of a free field of mass $\sqrt{s}$, which for space-like
$x-y$ is in terms of a particular Bessel function (e.g., Ref.\
\cite{Bogoliubov.Shirkov}):
\begin{multline}
  \Delta(\BDpos(x-y)^2;s)
  = \frac{\sqrt{s}}{4\pi^2\sqrt{\BDneg (x-y)^2}}
    K_1\xleft( \sqrt{\BDneg s(x-y)^2} \right)
\\
  ~\mbox{for space-like $x-y$}.
\end{multline}

It then follows that the propagator, i.e., the Fourier transform of
the time-ordered version of the correlator, is
\begin{equation}
\label{eq:KL.prop}
   \hat{G}_2(p^2) = \int_0^\infty \diff{s} \rho(s) \,
               \frac{i}{ \BDpos p^2-s + i\epsilon }.
\end{equation}

Suppose that the field has a non-zero matrix element between the
vacuum and a single particle state:
\begin{equation}
\label{eq:1-0-ME}
  \langle 0 | \phi(x) | \ve{p} \rangle = c \, e^{\BDneg ip\cdot x},
\end{equation}
where $p^0 = E_{\ve{p}} = \sqrt{\ve{p}^2 + m_{\rm phys}^2}$, and
$m_{\rm phys}$ is the physical mass of the particle.\footnote{The form
  of the dependence on $x$ follows from an application of the
  translation operator to the field.} Then there is a contribution to
$\rho(s)$ of the form $|c|^2\delta(s-m_{\rm phys}^2)$.  All other
contributions are from continuum parts of the allowed energies, and
start at higher particle thresholds; they give no further delta
functions.  Hence the propagator has a pole with residue $|c|^2$ at
$p^2 = \BDpos m_{\rm phys}^2$:
\begin{equation}
\label{eq:prop.residue}
  \hat{G}_2(p^2) = \frac{ i |c|^2 }{ \BDpos p^2-m_{\rm phys}^2 + i\epsilon }
           + \mbox{non-pole term}.
\end{equation}
We write the residue as $R=|c|^2$.  

Commonly one normalizes the single-particle states to make $c$ real
and positive.  But this is only possible in one matrix element like
(\ref{eq:1-0-ME}).  If we examine 2-point functions for this and other
fields that have nonzero coupling between the vacuum and the
single-particle state, then $c$ is normally different in each case and
can only be normalized to be real and positive for one of them.

\section{Space-time wave functions}
\label{sec:wfs}

Since space-time localization is important, let us define a
coordinate-space wave function for each particle in states such as
those in Eqs.\ (\ref{eq:i}) and (\ref{eq:f}), by writing
\begin{align}
\label{eq:wf.coord.f}
  f(x) = f(t,\3x)
   & \eqdef \int \difftilde{p} \tilde{f}(\ve{p}) \, e^{\BDneg ip\cdot x}
\nonumber\\
   & = \int \difftilde{p} \tilde{f}(\ve{p}) \, e^{-iE_{\3p}t + i \3p\cdot\3x},
\end{align}
where $f$ corresponds to any of the $f_j$ or $g(x)$.  Here $p$ is on
shell, of course, at the \emph{physical} particle mass. Each of these
wave functions is a function of time and spatial position.  It is thus
like an ordinary Schr\"odinger wave function in the non-relativistic
quantum mechanics of a single particle.

Although, in general, there are great difficulties in using wave
functions\footnote{For the purposes of the discussion here, a wave
  function in the non-relativistic quantum mechanics of a finite
  number of particles can be treated as an expansion of a general
  state in a basis of states obtained in a corresponding theory of
  free particles.  Ordinary Schr\"odinger wave functions are such an
  expansion in a basis of position eigenstates.  When one tries to
  follow the same approach in relativistic QFTs, severe difficulties
  and impossibilities arise.  See, for example, \cite{vanHove1951,
    vanHove1952145, Haag:1955ev}, and references therein.} in
relativistic theories with interactions, the concept of a wave
function is valid in a \emph{free}-field theory, for the state of one
particle.  Our use of wave functions is as useful auxiliary
quantities, in the analysis of a state in terms of its particle
content in the infinite past or future, where the state corresponds to
a set of isolated particles.  That is, we use the concept of wave
function only for the center-of-mass motion of a single particle and
then only when the particle is being correctly approximated as a free
particle.  (Since it is the center-of-mass motion that is relevant
here, these ideas apply equally when the particle is a bound state of
more elementary constituents.)

Unlike the case of multiparticle wave functions in non-relativistic
quantum mechanics, we do not assign a common time variable to all the
single-particle wave functions for a state of multiple particles.  The
purpose of our wave functions is not the one used in non-relativistic
quantum mechanics, where time-dependent wave functions implement
time-dependent states in the Schr\"odinger picture.  Here their purpose
is to simply give a useful quantity with which to analyze the relation
between the states in Eqs.\ (\ref{eq:i}) and (\ref{eq:f}), the
S-matrix, and certain matrix elements involving the Heisenberg-picture
field operators.

In basic applications to scattering, we assume that the momentum-space
wave functions are sharply peaked about one momentum.  Then the
coordinate-space wave functions describe propagating wave packets, as
will now verify.  Thus they correspond to propagation of free
particles with approximately definite momenta.  For analyzing
scattering processes, we will need information about the asymptotic
behavior of wave functions in coordinate space for large positive and
negative times.

Consider a momentum-space wave function that is sharply peaked around
one value of momentum, $\3p_0$, and that has a width characterized by
one\footnote{One can generalize to the case that there are very
  different widths in different directions. But that will only add
  notational complexity without changing the principles.} number $\Delta
p$.  We will initially choose the wave function also to be real and
non-negative. One simply possibility would be a Gaussian
\begin{equation}
  \label{eq:Gaussian}
  f(\3p;\3p_0,\Delta p) \eqbad D e^{-(\3p-\3p_0)^2/\Delta p^2},
\end{equation}
with $D$ adjusted to give unit normalization:
\begin{equation}
\label{eq:wf.norm}
  \int \difftilde{p} |f(\3p)|^2 = 1.
\end{equation}
However, this has a (small) tail extending out to infinite momentum.
For reasons to be reviewed below, this gives large-time behavior that
is not quite optimal for constructing a proof of the reduction
formula.  Therefore \cite{Ruelle:1962, Hepp:1965.LSZ} it is better to
choose a wave function of compact support, i.e., one that vanishes
outside a finite range of $\3p$.  One simple possibility would be the
infinitely differentiable function
\begin{equation}
  \label{eq:comp.supp.wf}
  f(\3p;\3p_0,\Delta p)
  = 
  \begin{cases}
     D e^{-1/(1-|\3p-\3p_0|^2/\Delta p^2)} 
     & \hspace*{-2mm}\mbox{if $|\3p_0-\3p|<\Delta p$},
\\
     0   & \hspace*{-2mm}\mbox{if $|\3p_0-\3p|\geq\Delta p$},
  \end{cases}
\end{equation}
with $D$ being again adjusted to give a normalized wave function,
(\ref{eq:wf.norm}).

The corresponding coordinate-space wave function, from Eq.\
(\ref{eq:wf.coord.f}), is of maximum size at the origin, i.e., when
$x=0$.  This is because at that point, the integrand is strictly
positive.  At all other values of $x$, the phase factor $e^{-iE_{\3p}t
  + i \3p\cdot\3x}$ is not always unity, and therefore there are some
cancellations.

\subsection{Properties needed}

For a treatment of scattering theory and a derivation of the S-matrix,
we need to know roughly the behavior of single-particle wave functions
in coordinate space.  In the derivation, we will encounter integrals
of coordinate-space wave functions multiplied by Green functions, as
in Eq.\ (\ref{eq:f.i.coord.T}), in which certain ranges of time are
selected.  We will need to understand which regions of the integrals
over spatial coordinates give non-zero contributions in a limit of
infinite time and which regions give asymptotically vanishing
contributions, and to see that the asymptotic space-time structure
does in fact corresponding to the expected scattering phenomena.

To this end, suppose we are given a wave function obeying the
properties given just above. Then we need estimates of the following:
\begin{enumerate}
\item The location in $\3x$ of the peak of the wave function, for a
  given value of $t$.
\item The corresponding width in $\3x$.
\item The asymptotic behavior when $t$ goes to infinity (or negative
  infinity), as a function of $\3x$. This is especially needed for the
  asymptotic behavior far from the peak of the wave function. 
\end{enumerate}
However, only rough estimates will be needed.

When related to our use of $f(x)$ in QFT, the location of the peak
verifies that the particle propagates classically.  The width of the
peak quantifies the inaccuracy of a purely classical view; in a
scattering situation, the width determines when and where the
different particles in the initial or final states can start to be
regarded as separate non-interacting particles.  The asymptotic
behavior is needed to ensure that asymptotically the particles are
cleanly separated; it is here that the motivation will arise to use
wave functions of compact support in momentum space.

If we use a momentum-space wave function of a form such as is defined
in (\ref{eq:Gaussian}) or (\ref{eq:comp.supp.wf}), we will find that
the classical trajectory of the particle goes through the origin of
spatial coordinates at time zero, and also has minimum width there.
Of course, we need to allow more general possibilities for the
trajectory; this is easily done by applying a space-time translation
--- see Sec.\ \ref{sec:wf.shift} below.

\subsection{Stationary phase}

The integrand in (\ref{eq:wf.coord.f}) for the coordinate-space wave
function is a real, positive factor multiplied by the phase
$e^{-iE_{\3p}t+i\3p\cdot\3x}$.  Suppose that $t$ and/or $\3x$ is made
large.  Then the integrand, as a function of $\3p$, generally has
rapid oscillations, which result in a small contribution to $f(x)$.
Given a particular value of time $t$, if $\3x$ is increased
sufficiently, then for most values of $\3p$ the oscillations become
arbitrarily rapid, and the corresponding contribution to $f(x)$
decreases rapidly to zero, i.e., faster than any power of $\3x$, by a
standard theorem.

The exception to these statements occurs where there is a lack of
oscillations, i.e., at and close to the point of stationary phase.
Given a value of $t$ and $\3x$, the stationary-phase point is the
value $\pc$ where
\begin{equation}
  \label{eq:stat.phase.cond}
  0 = \frac{ \partial(-E_{\pc}t + \pc\cdot\3x) }{ \partial\pc }
    = -\frac{ \pc t }{ E_{\pc} } + \3x,
\end{equation}
i.e.,
\begin{equation}
  \label{eq:stat.phase}
  \pc = \frac{ m \3x \sign{t} }{ \sqrt{t^2-\3x^2} },
\end{equation}
so that
\begin{equation}
  E_{\pc} = \frac{ m |t| }{ \sqrt{t^2-\3x^2} },
\end{equation}
Notice that the stationary phase condition only has a solution when
$x=(t,\3x)$ is time-like or zero.\footnote{We have assumed that the
  mass $m$ is non-zero, as is the case throughout this paper.
  Modifications to the treatment are needed if $m=0$. Furthermore, if
  the momentum-space wave function does not fall off sufficiently
  rapidly as $|\3p|\to\infty$, then a degenerate case of Eq.\
  (\ref{eq:stat.phase.cond}) is relevant in the limit of infinite
  $\pc$ with light-like $x$. A compact-support condition on
  $\tilde{f}(\3p)$ avoids that issue, among others.}

Now let use restrict to the case that $\tilde{f}(\3p)$ is like the
examples in (\ref{eq:Gaussian}) or (\ref{eq:comp.supp.wf}), i.e.,
real, non-negative, and strongly peaked at one value $\3p=\3p_0$.
Then, given $t$, the coordinate-space wave function is largest when
the stationary phase point is close to the maximum of the function
$\tilde{f}(\3p)$, i.e., when $\pc = \3p_0$, and thus when
\begin{equation}
  \label{eq:peak.t}
  \3x \simeq \frac{ \3p_0 t }{ E_{\3p_0} }
      = \3v_0 t.
\end{equation}
This is the trajectory of a classical relativistic particle, which
therefore, as expected, matches the overall propagation of the wave
packet.  The 3-velocity is $\3v_0=\3p_0/E_{\3p_0}$.

\subsection{Width}

For the analysis of the width of the coordinate-space wave function in
coordinate space for a given $t$, we still consider cases with
momentum-space wave functions like those in (\ref{eq:Gaussian}) or
(\ref{eq:comp.supp.wf}).  First consider $t=0$.  The peak of the wave
function is at the origin, $\3x=0$.  As $\3x$ moves away from this
position, we reach a situation when about one oscillation of the
$e^{i\3p\cdot\3x}$ factor as a function of $\3p$ fits inside the peak of
the momentum-space wave function.  Hence the width of the wave
function is of order $1/\Delta p$.  We can call this the
uncertainty-principle value or the quantum mechanical uncertainty.

When $|t|$ is increased sufficiently much away from zero, the
oscillations in $e^{-iE_{\3p}t}$ are important.  For determining the
width of the wave function, these first become significant when there
is a change of order unity in the phase when one moves from
$\3p=\3p_0$ to a value differing by $\Delta p$.  We estimate where this
happens by using the derivative of $E_{\3p}$ at $\3p_0$ and our
general assumption that $\Delta p$ is small.  Then the
uncertainty-principle estimate continues to apply when
\begin{equation}
  |t| \lesssim  \frac{ E_{\3p_0} }{ \Delta p (|\3p_0|+\Delta p) }.
\end{equation}
Notice that if $\Delta p$ is very small, this is a large range of times. But
always, for a given wave function, once $t$ gets large enough in size,
positive or negative, the uncertainty principle uncertainty is
insufficient.

At large enough values of time, it is the stationary phase point given
in (\ref{eq:stat.phase}) that is relevant.  The size of the
coordinate-space wave function corresponds to the size of the
$\tilde{f}(\3p)$ at $\3p=\pc$.  This is multiplied by an
overall $t$-independent factor from the integration measure, and by a
factor with power-law $t$ dependence.  The second factor can be
estimated asymptotically by expanding the exponent of the phase factor
to second order in momentum about $\3p=\pc$, and doing a
saddle point expansion.

Hence, for large enough $|t|$, the width of the wave function is
determined by where the stationary phase point deviates from the
central value of momentum by $\Delta p$.  That is it is determined by where
$|\pc(t,\3x)-\3p_0|=O(\Delta p)$.  To understand what this implies for
where in $\3x$, the coordinate-space wave function starts to fall off,
we start with a value of $\pc$ obeying the last condition, and compute
the corresponding value of $\3x=\3p_{\rm c}t/E_{\pc}$.  This
corresponds to the propagation of a classical particle of momentum
$\pc$.  In this situation of large enough $|t|$, the width of the wave
function therefore corresponds to classical dispersion, i.e., to the
different velocities of classical particles of different
momenta.\footnote{Note that the width in $\3x$ is then quite different
  in a direction perpendicular to $\3p_0$ than parallel.  This is
  simply a consequence of the form of the dependence of velocity on
  momentum.}  The dependence of this contribution on the width, as a
function of $\Delta p$ and $t$, is completely different to that of the
quantum uncertainty-principle width.  It increases proportionally to
$t$, whereas the uncertainty-principle is independent of time.
Moreover it is proportional to $\Delta p$, decreasing to zero when $\Delta p\to0$.
In contrast, the uncertainty principle width is of order $1/\Delta p$ and
becomes infinite when $\Delta p\to0$.

An appropriate estimate for the order of magnitude of the width is
simply to add the uncertainty-principle and classical-dispersion
widths or to add them in quadrature.

\subsection{Asymptote}
\label{sec:wf.asy}

We now evaluate the asymptotics of the wave function as $t\to\pm\infty$.  If
$(t,\3x)$ is time like, then the wave function's value is dominated by
an integral near the corresponding stationary phase point $\pc$.  This
is multiplied by a power law in $t$.  When $(t,\3x)$ is space-like,
there is no stationary-phase point and hence the coordinate-space wave
function falls off faster than any power of $t$ (for a fixed ratio
$\3x/t$), as follows from standard properties of Fourier transforms of
infinitely differentiable functions.

For time-like $(t,\3x)$, the leading asymptote is given by a Gaussian
approximation around the stationary-phase point, as I now show.  Let
$\delta\3p = \3p-\pc$, with $\pc$ given by Eq.\ (\ref{eq:stat.phase}), and
let $\delta p_{\|}$ and $\delta\3p_\perp$ be the components of $\delta\3p$ parallel and
perpendicular to $\pc$.  Then the exponent in Eq.\
(\ref{eq:wf.coord.f}) is
\begin{align}
    -iE_{\3p}t + i \3p\cdot\3x
  = & -im\sqrt{t^2-\3x^2}
\nonumber\\&
      -i \frac{\delta p_\|^2 (t^2-\3x^2)^{3/2} }{ 2mt^2 }
      -i \frac{\delta p_\perp^2 (t^2-\3x^2)^{1/2} }{ 2m }
\nonumber\\&
      + \dots ,
\end{align}
where the cubic and higher terms indicated by $\dots$ are suppressed
by a power of $|\delta\3p|/E_{\pc}$ relative to the quadratic term.

We can deform the integrals over $\delta p_\|$ and $\delta\3p_\perp$ into the complex
plane to go down the directions of steepest descent.  For large $|t|$,
the integral is dominated by $\delta\3p$ of order $1/t$.  More exactly, the
dominance is by $\delta p_\|$ of order $\sqrt{m}t(t^2-\3x^2)^{-3/4}$ and
$\delta p_\perp$ of order $\sqrt{m}(t^2-\3x^2)^{-1/4}$.  We can use a Gaussian
approximation to estimate $f(x)$:
\begin{align}
\label{eq:wf.asy}
  f(x) \sim &
  \frac{\tilde{f}(\pc)e^{-3\pi i/4}m^{3/2}}{(2\pi)^32E_{\pc}}
  e^{-i\sqrt{t^2-\3x^2}}
  \frac{t}{(t^2-\3x^2)^{5/4}}
\nonumber\\=&
  \frac{\tilde{f}(\pc)e^{-3\pi i/4}E_{\pc}^{3/2}}{(2\pi)^32m}
  \, \frac{ e^{-i\sqrt{t^2-\3x^2}} \sign t }{ |t|^{3/2} },
\end{align}
with errors suppressed by a power of $1/t$.  There is implicit
dependence on $\3x$ and $t$, in the dependence on $\pc$, since $\pc =
m (\3x/t)/\sqrt{1-\3x^2/t^2}$, but this is only a dependence on the
ratio $\3x/t$, i.e., a velocity.  By inverting the relationship of
$\pc$ to $\3x/t$, we see that the momentum-space wave function
$\tilde{f}(\3p)$ at a particular value of momentum gives a
contribution to the asymptote of $f(x)$ at a corresponding value of
velocity
\begin{equation}
  \frac{\3x}{t} = \frac{\3p}{E_{\3p}} = \frac{\3p}{\sqrt{m^2+\3p^2}},
\end{equation}
which is of course the well-known value for the relativistic
propagation of a classical particle.  

At fixed velocity $\3x/t$, Eq.\ (\ref{eq:wf.asy}) shows that $f(x)$
decreases like $1/|t|^{3/2}$ at large $t$.  When we construct the
S-matrix, it will be the $1/|t|^{3/2}$ asymptote that gives the
S-matrix; weaker contributions to $f(x)$ are irrelevant to the
S-matrix. 

\begin{figure}
  \centering
  \includegraphics[width=5cm]{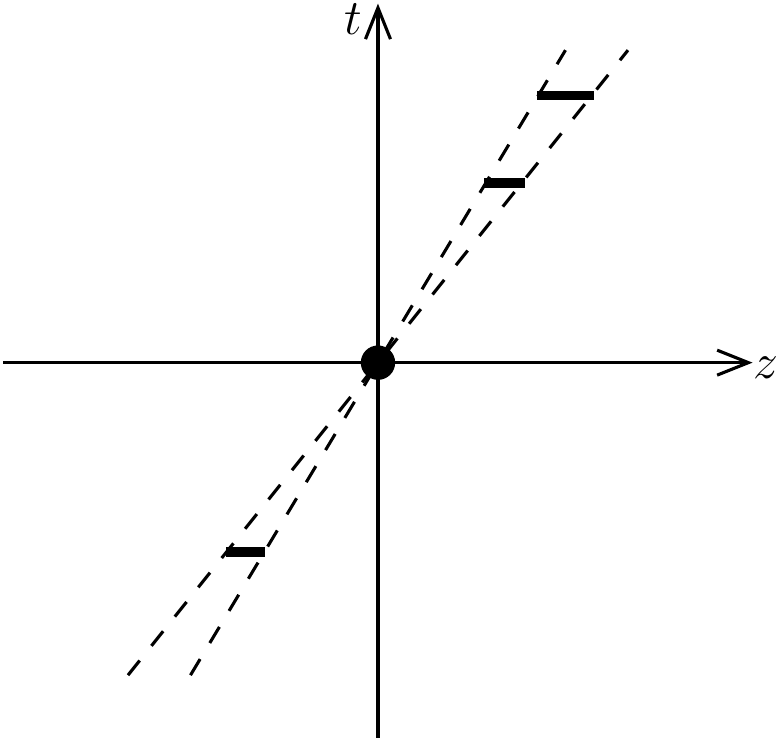}
  \caption{Illustrating space-time evolution of wave function
    $f(t,\3x)$.  The horizontal lines indicate at particular times the
    width of the wave packet.}
  \label{fig:wf.evol}
\end{figure}

\subsection{Asymptote for compact support momentum-space wave
  function}
\label{sec:compact.supp}

Some simplifications occur if the momentum-space wave function has
compact support, i.e., if it vanishes outside a finite range of $\3p$.
Then, for the coordinate-space wave function, the range of velocities
$\3x/t$ for which applies the $1/|t|^{3/2}$ decrease is equally
compact, i.e., bounded.  Outside this range, $f(x)$ decreases more
rapidly with $t$.  This provides a useful visualizable localization,
as in Fig.\ \ref{fig:wf.evol}.  For a given large value of $|t|$, the
range of $\3x$ in which the $1/|t|^{3/2}$ decrease applies has a
volume of order $|t|^3$.

The space of functions of compact support is dense in the space of
normalizable wave functions, so we have no loss of generality if we
restrict attention to wave functions of compact support in momentum
space.

In contrast it is not particularly useful to impose a condition of
compact support in spatial position on the coordinate-space wave
function.  To see this, observe that if a function of $\3x$ has
compact support, then its Fourier transform into $\3p$ space can be
continued to all complex $\3p$ and is analytic everywhere.
Conversely, if there is a singularity of the Fourier transform for
some value of $\3p$, then the original function of $\3x$ cannot be of
compact support.  Now suppose, the wave function $f(x)$ in
(\ref{eq:wf.coord.f}) is of compact support in $\3x$ for one value of
$t$, say $t=t_0$. Then $\tilde{f}(\ve{p}) \, e^{-iE_{\3p}t_0} /
E_{\3p}$ is analytic for all complex $\3p$.  Going to another value of
$t$ gives an extra factor $e^{-iE_{\3p}(t-t_0)}$ on the right-hand
side of (\ref{eq:wf.coord.f}).  This is not analytic, because it has a
singularity where $\3p^2=-m_{\rm phys}^2$, and hence $f(t,\3x)$ is not
of compact support in $\3x$.  Thus a condition of compact support in
space can be maintained for at most one instant in time.

\subsection{Shift of wave function in coordinate space}
\label{sec:wf.shift}

We now ask how to shift the wave, so that the classical trajectory is
\begin{equation}
  \3x \simeq \3x_a + \3v_0 (t-t_a),
\end{equation}
so that instead of occurring at the origin, the minimum width occurs
at time $t_a$, with the position of the particle at that time being
$\3x_a$.  This is done by multiplying the momentum-space wave function
by a suitable phase.  We replace $f(\3p;\3p_0,\Delta p)$ by
\begin{equation}
  f(\3p;\3p_0,\Delta p) \mapsto f(\3p;\3p_0,\Delta p) e^{iE_{\3p}t_a-\3p\cdot\3x_a}.
\end{equation}
Then the coordinate-space wave function gets changed by
\begin{equation}
  f(t,\3x) \mapsto f(t-t_a,\3x-\3x_a),
\end{equation}
as can be verified by substituting the modified momentum-space wave
function in the defining formula (\ref{eq:wf.coord.f}) for the
coordinate-space wave function.

\begin{widetext}
\section{The reduction formula}
\label{sec:LSZ.statement}

In this section, I state the reduction formula, of which an improved
proof will be in Sec.\ \ref{sec:LSZ.derivation}.  It is useful to
focus attention separately on the connected components of the Green
functions (and hence of the S-matrix).  It is the fully connected term
that is relevant to standard calculations, and we will focus
exclusively on that in this section.

Let $\tilde{G}_N^{\text{conn}}(p_1,p_2, \dots p_N)$ be a
\emph{connected} $N$-point Green function, and let $\Gamma_N(p_1,p_2, \dots
p_{N-1})(2\pi)^4\delta^{(4)}(p_1+\dots+p_N)$ be the corresponding
\emph{amputated} Green function.  It is convenient to choose $\Gamma_N$ to
be defined without the momentum conservation delta function.  So it
only has $N-1$ independent momentum arguments, and the last momentum
obeys $p_N = -\sum_{j=1}^{N-1} p_j$.  We choose the convention for the
arguments of Green functions that all the momenta flow in.

The unamputated and amputated Green functions are related by
\begin{equation}
\label{eq:G.Gamma}
  \tilde{G}_N^{\text{conn}}(p_1,p_2, \dots p_N) 
=
  \prod_{j=1}^N \hat{G}_2(p_j^2) \,\,
  \Gamma_N(p_1,p_2, \dots p_{N-1}) 
  (2\pi)^4\delta^{(4)}(p_1+\dots+p_N) ,
\end{equation}
where $\hat{G}_2(p^2)$ is the propagator, i.e., the 2-point function
without its momentum-conservation delta function.  Let $c$ be the
coefficient in the normalization of the vacuum-to-one-particle matrix
element, as in Eq.\ (\ref{eq:1-0-ME}). Then, as seen at Eq.\
(\ref{eq:prop.residue}), the propagator has a pole at the physical
mass of the particle and the propagator residue is $R=|c|^2$.

The LSZ theorem \cite{Lehmann:1954rq} both states that the wave packet
states $\langle g_1,\dots,g_n;\,\Out|f_1,\dots,f_{n'};\,\In\rangle$ have the form
\eqref{eq:g.f} and states how the S-matrix is given in terms of Green
functions.  The LSZ result for the connected part\footnote{Similar
  formulas can also be worked out for the disconnected parts, but it
  is the connected part that is relevant for computing cross sections.
  Moreover, for typical practical applications, one only has two
  incoming particles, $n'=2$.}  of the S-matrix element for $n'\to n$
scattering is
\begin{equation}
\label{eq:LSZ.1}
  S^{\text{conn}}_{\ve{q}_1, \dots \ve{q}_n; \, \ve{p}_1, \dots \ve{p}_{n'} }
= 
  \lim_{\rm on-shell}
  \frac{1}{ c^n (c^*)^{n'} } \, 
   \prod_{k=1}^n \frac{ \BDpos q_j^2-m_{\rm phys}^2 }{ i } \,\,
   \prod_{j=1}^{n'} \frac{ \BDpos p_j^2-m_{\rm phys}^2 }{ i } \,\,
   \tilde{G}_{n+n'}^{\text{conn}}(p_1, \dots, p_{n'}, -q_1, \dots -q_{n-1}  ).
\end{equation}
Observe that the \emph{full} Green function $\tilde{G}_{n+n'}$
diverges when the external momenta are put on-shell.  This formula
asserts that the S-matrix is obtained by first multiplying the Green
function by the factors of $\BDpos p_j^2-m_{\rm phys}^2$, etc, to
cancel the poles, and by then taking the limit of on-shell momenta,
and finally by inserting the factor $1/(c^n (c^*)^{n'})$. An important
and non-trivial part of \eqref{eq:LSZ.1} is this last factor, which
involves the normalization of the vacuum-to-one-particle of the field.
In the usual case that $c$ is real and positive, the factor equals
$1/R^{(n+n')/2}$, where $R$ is the residue of the pole in the full
propagator.

A convenient version of the reduction formula is in terms of the
amputated Green function:
\begin{equation}
\label{eq:LSZ.2}
  S^{\text{conn}}_{\ve{q}_1, \dots \ve{q}_n; \, \ve{p}_1, \dots, \ve{p}_{n'} }
= 
  (2\pi)^4 \delta^{(4)}\left({\textstyle\sum}_{j=1}^{n'}p_j-{\textstyle\sum}_{k=1}^n q_k \right) 
  \,
 (c^*)^nc^{n'} \, 
  \Gamma_{n+n'}(p_1, \dots, p_{n'}, -q_1, \dots -q_{n-1}  ) \,
  .
\end{equation}
\end{widetext}
Here the external momenta of the amputated Green function are set on
shell from the beginning.  This formula gives the following procedure
for computing the S-matrix:
\begin{enumerate}
\item Replace each full external propagator of the full Green function
  by the factor $c$ for incoming lines and $c^*$ for outgoing lines.
\item Set the external momenta on-shell.
\end{enumerate}

The actual statement of the theorem by LSZ was for the case that the
vacuum-field-particle matrix element had unit normalization, i.e.,
$c=1$.  But it is elementary to extend the theorem to where $c\neq1$, as
can be seen in many textbooks, e.g., \cite{Bjorken:1965zz,
  Sterman:1994ce, Peskin:1995ev, Srednicki:2007qs}.  LSZ presented
their result in a different but equivalent form to that given here.
Their formula is obtained from Eq.\ \eqref{eq:g.f} by substituting
Eq.\ \eqref{eq:LSZ.1} for the S-matrix and then expressing the
momentum-space Green function in terms of the corresponding
coordinate-space Green function. It is quite elementary to reverse the
procedure, i.e., to perform the Fourier transforms in LSZ's actual
formula to obtain the combination of Eqs.\ \eqref{eq:g.f} and
\eqref{eq:LSZ.1}.  Many, but not all, authors do present the
momentum-space form that is relevant for actual calculations.

A simple extension to matrix elements of operators between in and out
states is also elementary --- see p.\ 53 of \cite{Sterman:1994ce}.
Such matrix elements have the form
\begin{equation}
\label{eq:OME}
  \langle\alpha;\,\Out|\,T\,\text{operator(s)}\,|\beta;\,\In\rangle.  
\end{equation}
Cases in regular use are where operators are currents in QCD and the
matrix elements are for the hadronic part of scattering amplitudes or
cross sections that involve both QCD and the electroweak interactions,
e.g., deeply inelastic scattering and the Drell-Yan process.

\begin{figure}
  \centering
  \includegraphics[scale=0.7]{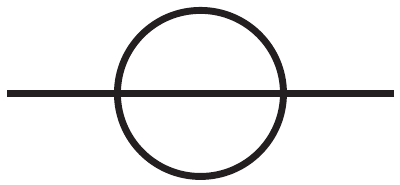}
  \qquad
  \includegraphics[scale=0.7]{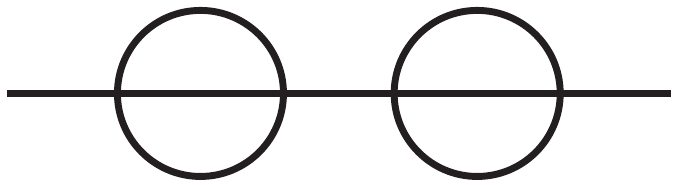}
  \caption{Propagator corrections.}
  \label{fig:prop.corr}
\end{figure}

Practical calculations involve on-shell amputated Green functions and
a separate calculation of the physical mass and the propagator
residue.  Direct calculations of perturbative corrections to the
propagator are not useful in themselves because they give terms with
higher-order poles, from two or more free propagators in series, e.g.,
Fig.\ \ref{fig:prop.corr}.  This results in non-convergence of the sum
when the momentum is near the particle pole.

This problem is evaded by a resummation in terms of the self-energy
function $\Sigma(p^2,m_R,\lambda_R)$, which is defined to be $i$ times a 2-point
Green function that is irreducible in the external line; it includes
any necessary renormalization counterterms.  It is analytic when $p^2$
is in a neighborhood of its on-shell value. I have written
$\Sigma(p^2,m_R,\lambda_R)$ with explicit arguments for renormalized mass and
coupling.  In many schemes there is also a renormalization mass $\mu$,
but I have left that argument implicit.  There is no requirement for
the renormalized mass to equal the physical mass.

In terms of the self-energy, the propagator is
\begin{equation}
  \hat{G}_2(p^2) = \frac{ i }{ \BDpos p^2-m_R^2 -\Sigma(p^2,m_R,\lambda_R) + i\epsilon } .
\end{equation}
The physical mass can be determined in terms of the parameters of the
theory, i.e., the renormalized mass and coupling, by solving
\begin{equation}
  m_{\rm phys}^2-m_R^2 -\Sigma(\BDpos m_{\rm phys}^2,m_R,\lambda_R) = 0,
\end{equation}
which can be done order-by-order in perturbation theory.  The
propagator's residue is then
\begin{equation}
  R = \left. \frac{1}{1 \BDminus \partial \Sigma(p^2) / \partial p^2 }
      \right|_{p^2=\BDpos m_{\rm phys}^2} ,
\end{equation}
which is also susceptible to perturbative calculation.

\section{Creation of in- and out-states by fields}
\label{sec:creation.ops}

In Eqs.\ (\ref{eq:i}) and (\ref{eq:f}), we proposed states
$|f_1,f_2;\,\In\rangle$ and $|g_1,\dots;\,\Out\rangle$ that have simple expressions
when parameterized by what we termed momentum-space wave functions:
$\tilde{f}_j(\3p)$, etc.

For all our considerations concerning scattering, we assume that each
momentum-space wave function is peaked around a particular value of
momentum.  Since the space spanned by such functions is the whole
space of wave functions, this condition will not give any loss of
generality.  But it will enable arguments concerning scattering to be
visualizable. 

The state $|f_1,f_2;\,\In\rangle$ is intended to be a state which at very
large negative times approaches a state of free separated individual
particles with wave function $\tilde{f}_1(\ve{p}_1) \,
\tilde{f}_2(\ve{p}_2)$.  The simplest case to understand is when,
following Sec.\ \ref{sec:compact.supp}, we choose $\tilde{f}_1$ and
$\tilde{f}_2$ to be of compact support, and also choose them to have
non-overlapping supports.  In this case, we have non-overlap of the
region of asymptotic $1/|t|^{3/2}$ decrease of the corresponding
coordinate space wave functions.  It is the region of $1/|t|^{3/2}$
decrease that matters, because that is the part that is relevant for
obtaining the S-matrix.  In the non-overlapping case, the regions of
$1/|t|^{3/2}$ decrease for $f_1$ and $f_2$ are space-like separated
for a given large enough value of $t$ (positive or negative), and then
the particles are causally separated.

When we go to the more general overlapping case, a somewhat less trivial
argument is needed to show that the two particles do not causally
influence each other.  Our proof of the reduction formula will be
organized in such a way that the issue of separation of the different
particles at infinite time is handled rather indirectly.

Similar statements apply to the state $|g_1,\dots;\,\Out\rangle$, except that
they are applied in the far future.  Exactly how far one has to go to
the past and future to have a good free-particle approximation depends
on the state, i.e., on the wave functions.

The aim of this section is to construct creation operators that
genuinely create single asymptotic particles in the far past and
future.  The complication we will need to overcome is that, as we will
see, the simplest and natural candidate definition for these
operators, i.e., the one used by LSZ, gives operators that create much
more than the intended single particles.  The new operators will then
be used in a proof of the reduction formula.  Separately, after the
proof is completed, we will be able to verify not only the property
inherent in the definition that one creation operator applied to the
vacuum creates the intended single particle, but also that multiple
applications of the operator create the intended multiparticle state
and nothing else.

The starting point consists of the following assumptions (a) that we
have a QFT that exists and that obeys standard principles; (b) that it
has a scattering theory that obeys the principles given in Sec.\
\ref{sec:scattering}; and (c) that there is a nonzero
vacuum-to-one-particle matrix element of the field $\phi(x)$.  Poincar\'e
invariance is assumed, as usual, but that assumption can be relaxed,
normally with a penalty only in notational complexity.

I will first demonstrate by explicit perturbative calculations the
already-summarized problem that the LSZ creation operators $a_f^\dagger(t)$
create much more than the desired single particles, and that the
limits as $t\to\pm\infty$ do not help.  Because the field operators in the
defining formula for $a_f^\dagger(t)$ are at fixed time, there is
effectively an infinite uncertainty in energy, with the corresponding
possibility of creating arbitrarily many particles.

With the results of those calculations as motivation, I will then give
a suitable definition of creation operators.  It modifies the LSZ
definition by applying a time average.

\subsection{Free fields}

We start with a review of the case of a \emph{free} field theory,
without interactions.  With the standard normalization, the Lagrangian
density is
\begin{equation}
  \label{eq:L.free}
  \mathcal{L} = \BDpos \frac{1}{2}(\partial\phi_{\Free})^2 - \frac{m^2}{2}\phi_{\Free}^2.
\end{equation}
The standard expansion of the field in terms of time-independent
creation and annihilation operators was given in Eq.\ (\ref{eq:KG}).
These operators can be found in terms of the field $\phi_{\Free}(t,\3x)$
and the canonical momentum field $\pi_{\Free}(t,\3x)=\partial\phi_{\Free}/\partial t$ at
fixed time:
\begin{subequations}
\begin{align}
\label{eq:ak.dag.free}
  a_{\ve{k},\Free}^\dagger &= \int \diff[3]{\3x} e^{\BDneg i k\cdot x}
            \left[ E_{\3k}\phi_{\Free}(x) - i\pi_{\Free}(x) \right],
\\
\label{eq:ak.free}
  a_{\ve{k},\Free} &= \int \diff[3]{\3x} e^{\BDpos i k\cdot x}
            \left[ E_{\3k}\phi_{\Free}(x) + i\pi_{\Free}(x) \right],
\end{align}
\end{subequations}
where $k^\mu = \left( \sqrt{\ve{k}^2+m^2}, \ve{k} \right)$. The $\phi$ and
$\pi$ fields are important because they are the independent fields that
appear in the Hamiltonian.  From the equal-time commutation relations,
and specifically
$[\phi_{\Free}(t,\3x),\pi_{\Free}(t,\3y)]=i\delta^{(3)}(\3x-\3y)$, follow the
standard commutation relations, (\ref{eq:a.a.adag.comm}), for the annihilation and creation operators.
Given that $a_{\ve{k},\Free}$ annihilates the vacuum, i.e.,
$a_{\ve{k},\Free}|0,\Free\rangle=0$, all the usual consequences follow.

A generic normalizable one-particle state is
\begin{align}
\label{eq:free.state.start}
   |f; \Free\rangle 
={}&
   \int \difftilde{p} \, |\ve{p}; \Free\rangle  \, \tilde{f}(\ve{p})
\nonumber\\
= {}&
   \int \difftilde{p} \, a_{\ve{p},\Free}^{\dag} \, |0; \Free\rangle  \, \tilde{f}(\ve{p}).
\end{align}
It is useful to define creation and annihilation operators
corresponding to this state by
\begin{equation}
 a_{f,\Free}^\dagger \eqdef  \int \difftilde{k} \tilde{f}(\ve{k}) \, a_{\ve{k}}^\dagger,
\quad
 a_{f,\Free} =  \int \difftilde{k} \tilde{f}^*(\ve{k}) \, a_{\ve{k}}.
\end{equation}
In terms of the coordinate-space wave function defined by Eq.\
(\ref{eq:wf.coord.f}), these operators are
\begin{subequations}
\begin{align}
  a_{f,\Free}^\dagger & = -i \int \diff[3]{\3x} f(x)\derivB \phi_{\Free}(x) ,
\\
  a_{f,\Free} & = i \int \diff[3]{\3x} f^*(x) \derivB \phi_{\Free}(x) .
\end{align}
\end{subequations}
The operators are time-independent operators, a property which will
no longer hold in an interacting theory.

Then the generic one-particle state can be written as
\begin{equation}
\label{eq:free.state}
  |f; \Free\rangle 
  =
  a_{f,\Free}^\dagger \, |0; \Free\rangle,
\end{equation}
and a multi-particle state with a product wave function is
\begin{equation}
\label{eq:free.state.multi}
   |f_1,\dots,f_n; \Free\rangle =  \prod_{j=1}^n a_{f_j,\Free}^\dagger \, |0; \Free\rangle. 
\end{equation}

\subsection{Analogs of creation and annihilation operators for
  interacting fields: first attempt}

We now wish to extend the results to an interacting theory.  The
overall aim is, if possible, to find operators that in suitable limits
of infinitely large negative and positive times create the individual
incoming and outgoing particles in (\ref{eq:i}) and (\ref{eq:f}),
i.e., such that, for example, something like the following equation is
valid:
\begin{equation}
  |f_1,f_2;\,\In\rangle = \lim_{t\to-\infty} A^{\dagger}_{f_1}(t) A^{\dagger}_{f_2}(t) |0\rangle.
\end{equation}
(Note that our actual definitions will contain a second parameter $\Delta t$
and that a non-trivial infinite-time limit is applied to both
parameters.)

A natural proposal, used in the LSZ paper, is to define the operators
in the same way as in the free theory The operators now become time
dependent. Thus operators of definite momentum are defined by
\begin{subequations}
\begin{align}
\label{eq:ak.dag}
  a_{\ve{k}}^\dagger(t) &\eqdef \int \diff[3]{\3x} e^{\BDneg i k\cdot x}
            \left[ E_{\3k}\phi(x) - i \frac{\partial\phi(x)}{\partial t} \right],
\\
\label{eq:ak}
  a_{\ve{k}}(t) &= \int \diff[3]{\3x} e^{\BDpos i k\cdot x}
            \left[ E_{\3k}\phi(x) + i\frac{\partial\phi(x)}{\partial t} \right],
\end{align}
\end{subequations}
where $k^\mu = \left( \sqrt{\ve{k}^2+m_{\rm phys}^2}, \ve{k} \right)$,
with the physical particle mass appearing in the formula for the
energy in terms of 3-momentum.

Compared with the free-field formulas (\ref{eq:ak.dag.free}) and
(\ref{eq:ak.free}), the canonical momentum field $\pi(x)$ has been
changed to the time derivative of the field.  For the free-field case
with the standard normalization, corresponding to (\ref{eq:L.free}),
the two formulas are the same, of course.  There are several reasons
for the change in the interacting case.  One is that it is exactly
what LSZ do.  The second is that when we take account of
renormalization in an interacting theory, we will typically change the
normalization of the field.  In that case the relative coefficient
between the $\phi$ and $\pi$ terms changes in the definition of the
creation operators, whereas no change in relative normalization is
needed between $\phi$ and the time derivative term in (\ref{eq:ak.dag})
and (\ref{eq:ak}).  The change in normalization does imply that the
commutation relations of the annihilation and creation operators no
longer have the standard normalization that was given in Eq.\
(\ref{eq:a.adag.comm}).  However, that will turn out to be irrelevant
to scattering physics because of non-trivial complications in an
interacting theory.  Finally, the key point is that it is with the
definitions in the form (\ref{eq:ak.dag}) and (\ref{eq:ak}), and with
our later modification of them, that the LSZ reduction formula is
proved.

Corresponding to a wave function and its Fourier transform, as in
(\ref{eq:wf.coord.f}), we define
\begin{subequations}
\begin{align}
\label{eq:af.dag}
 a_f^\dagger(t) &\eqdef   \int \difftilde{k} \tilde{f}(\ve{k}) \, a_{\ve{k}}^\dagger(t)
        = -i \int \diff[3]{\3x} f(x) \derivB \phi(x) ,
\\
\label{eq:af}
 a_f(t) &=   \int \difftilde{k} \tilde{f}^*(\ve{k}) \, a_{\ve{k}}(t)
       = i \int \diff[3]{\3x} f^*(x) \derivB \phi(x) .
\end{align}
\end{subequations}
Although these operators have time-dependence, their time-dependence
is \emph{not} given by an application of the Heisenberg equation of
motion. This is simply because their definition in terms of the field
$\phi(x)$ involves a time-dependent numerical-valued function $f(x)$. The
field $\phi(x)$ itself does obey the Heisenberg equation, as part of the
definition of the theory.

Some properties of the operators correspond to those of a free field.  For
example, from Eqs.\ (\ref{eq:ak.dag}) and (\ref{eq:ak}) it follows
that the field is expressed in terms of the $a$ and $a^\dagger$ operators in
the same form as Eq.\ (\ref{eq:KG}) for the free theory:
\begin{equation}
\label{eq:phi.ak}
  \phi(x)  =
          \int \difftilde{k}
            \left[
               a_{\ve{k}}(t) e^{\BDneg{ik\cdot x}}
             + a_{\ve{k}}^\dagger(t) e^{\BDpos{ik\cdot x}}
            \right].
\end{equation}
It also follows by straightforward calculations from the definitions
that if $\phi$ and $\partial\phi/\partial t$ were to obey equal-time commutation relations
of the same form and normalization as in the standard free theory,
then the $a$s and $a^\dagger$s would have their standard commutation
relations: these are just like (\ref{eq:a.adag.comm}) and
(\ref{eq:a.a.comm}), but with each $a_{\Free}$ and $a_{\Free}^\dagger$
replaced by its time-dependent counterpart in the interacting theory.

However, if it were simply asserted that an expansion of the form
(\ref{eq:phi.ak}) exists, then the $a$ and $a^\dagger$ cannot be uniquely
deduced from $\phi$ and $\partial\phi/\partial t$, unlike the case in the free theory.
The problem is that when one derives from Eq.\ (\ref{eq:phi.ak}) an
expression for $\partial\phi/\partial t$, the result contains terms with
time-derivatives of $a_{\ve{k}}(t)$ and $a_{\ve{k}}^\dagger(t)$.  So Fourier
transformation of $\phi$ and $\partial\phi/\partial t$ is not sufficient to determine
$a_{\ve{k}}(t)$ and $a_{\ve{k}}^\dagger(t)$ uniquely in an interacting
theory. 

We did not have this complication in a free field theory, because then
the creation and annihilation operators are time-independent; they
provide a way of presenting the general solution of the equation of
motion.

In contrast, at this point in our investigation of an interacting
theory, we are simply assuming that there exists a solution for the
state space and for the $x$-dependent field operator, without yet
specifying what they are.  Motivated by the formulas in the free
theory, we defined $a$ and $a^\dagger$ operators, anticipating that they
will be useful, but now they necessarily have time dependence.  Since
at this stage of the argument we have not determined whether or not
the operators genuinely destroy and create particles, it is best to
avoid referring to them as creation and annihilation operators.

\subsection{The operators \texorpdfstring{$a_f^\dagger(t)$}{afdag(t)} and
  \texorpdfstring{$a_f(t)$}{af(t)} create and destroy much more than single
  particles}
\label{sec:extra.particles}

The operators $a_f^\dagger(t)$ and $a_f(t)$ defined in Eqs.\
\eqref{eq:af.dag} and \eqref{eq:af} involve integrals with a single
field operator.  Therefore, momentum-space matrix elements, like
$\langle\alpha;\,\Out|a_f^\dagger(t)|\beta;\,\In\rangle$, with in- and out-states can be computed
from $\langle\alpha;\,\Out|\phi(x)|\beta;\,\In\rangle$, which is a matrix element of the field
between the same states.  This can be computed by applying the
reduction method to a Green function that has one more field than
needed for the states $\langle\alpha;\,\Out|$ and $|\beta;\,\In\rangle$.

In this section, we compute some examples elements in low-order
perturbation theory.

Since the calculations involve a theorem which is only proved later in
this paper, one should worry whether the logic is circular.  In the
first place, this section is purely motivational: it pinpoints an
inadequacy in the definition of $a_f^\dagger(t)$ and $a_f(t)$ relative to
their purposes.  It therefore indicates ways in which the definition
can be modified to be satisfactory.  From the point of view of the
logic of the proofs, the section can be safely omitted.  A second
point is that independently of the exact form of the reduction
formula, an elementary examination of the asymptotics of Green
functions, such as was done in Sec.\ \ref{sec:asy.Green.fn}, leads to
a natural conjecture that something like the reduction formula is
valid, even without adequately showing all the details.  This is
sufficient to allow motivational calculations that indicate
appropriate definitions for creation and annihilation operators.

A computation of $\langle\alpha;\,\Out|a_f^\dagger(t)|\beta;\,\In\rangle$ in perturbation theory
involves a Fourier transform of momentum-space Feynman graphs.  The
rules for computation of Feynman graphs for $\langle\alpha;\,\Out|a_f^\dagger(t)|\beta;\,\In\rangle$
need a special vertex for $a_f^\dagger(t)$ which has one line connected to
the rest of a Feynman graph.  Combining the Fourier transform with the
integral in \eqref{eq:af.dag} gives the following rule for the vertex
for $a_f^\dagger(t)$ in momentum-space:
\begin{equation}
  \label{eq:af.dag.Feyn}
  \text{Vertex}(a_f^\dagger(t))
  = \int \frac{\diff[4]{p}}{(2\pi)^4}
      \tilde{f}(\3p) 
      \, \frac{E_{\3p}+p^0}{2 E_{\3p}}
      \, e^{-i(E_{\3p}-p^0)t}. 
\end{equation}
Here the convention is that momentum $p$ is flowing \emph{into} the
rest of the graph, as is appropriate for an operator that is intended
to create a particle.  As in Eqs.\ (\ref{eq:ak.dag}) and
(\ref{eq:ak}), $E_{\3p}$ is the on-shell energy of a physical particle
in the interacting theory, i.e., $E_{\3p}=\sqrt{\ve{p}^2+m_{\rm
    phys}^2}$.  In the definition of the vertex factor, there is an
integral over the momentum of the external line, unlike the rules for
normal Green functions.  The integral is over all 4-momenta $p$, not
just over on-shell values.  The complete derivation of Eq.\
(\ref{eq:af.dag.Feyn}) is made by using the usual textbook methods for
deriving Feynman rules for Green functions, extended simply to deal
with the factor of $f(x)$ and the time-derivatives and the integral
over position.

Similarly, the vertex for $a_f(t)$ has the rule
\begin{equation}
  \label{eq:af.Feyn}
  \text{Vertex}(a_f(t))
  = \int \frac{\diff[4]{q}}{(2\pi)^4}
      \tilde{f}^*(\3q) 
      \, \frac{E_{\3q}+q^0}{2 E_{\3q}}
      \, e^{i(E_{\3q}-q^0)t}. 
\end{equation}
But here the momentum $q$ is defined to flow \emph{out} of the rest of
the graph, which is an appropriate convention for an operator intended
to destroy something.

\begin{figure}
  \centering
  \includegraphics[scale=0.5]{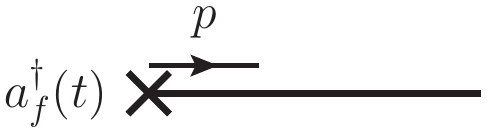}
  \qquad
  \includegraphics[scale=0.6]{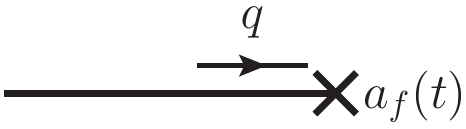}
  \caption{Notation for vertices for $a_f^\dagger(t)$ and $a_f(t)$.}
  \label{fig:af.vert}
\end{figure}

The vertices are notated as in Fig.\ \ref{fig:af.vert}.

\subsubsection{Example: Vacuum to one particle}
\begin{figure}
  \centering
  \includegraphics[scale=0.55]{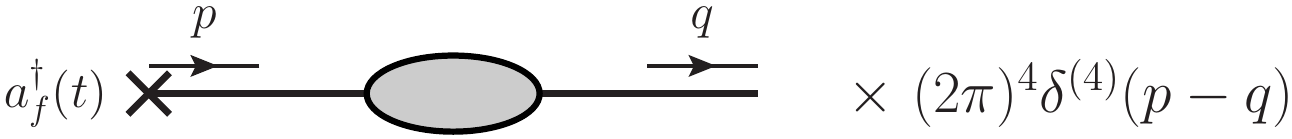}
  \caption{Green function that gives $\langle\3q|a_f^\dagger(t)|0\rangle$ after
    application of reduction method. Since the graphical 2-point
    function is often treated with the momentum-conservation delta
    function omitted, its presence has been indicated explicitly.}
  \label{fig:afdag.1.0}
\end{figure}

First consider the vacuum-to-one-particle matrix element of
$a_f^\dagger(t)$, i.e., $\langle\3q;\,\Out|a_f^\dagger(t)|0\rangle=\langle\3q|a_f^\dagger(t)|0\rangle$.  We obtain
this from the full 2-point Green function by applying the vertex rule
for $a_f^\dagger(t)$ at one end, Fig.\ \ref{fig:afdag.1.0}, and the LSZ
reduction method at the other.  This gives
\begin{equation}
  \label{eq:afdag.1.0}
  \langle\3q|a_f^\dagger(t)|0\rangle = \tilde{f}(\3q) c^*,
\end{equation}
which is the expected result, given the normalization of the matrix
element $\langle\3q|\phi(x)|0\rangle$ of the field.  In fact, the result can easily
be obtained simply by taking the one-particle-to-vacuum matrix element
of the definition \eqref{eq:af.dag} of $a_f^\dagger(t)$, and using
(\ref{eq:1-0-ME}) for the vacuum-to-one-particle matrix element of
$\phi$.  The derivation using the Feynman graph method merely checks
self-consistency of the Feynman rules and the LSZ method.

It can also be checked that the same matrix element of the conjugated
operator, intended to annihilate particles, is zero:
\begin{equation}
  \label{eq:afdag}
  \langle\3q|a_f(t)|0\rangle = 0.
\end{equation}
This follows from the factor $E_{\3q}+q^0$ in the rule
\eqref{eq:af.Feyn} for the vertex for $a_f(t)$, since with an
on-shell final-state particle, $q^0=-E_{-\3q}=-E_{\3q}$.

\subsubsection{Example: Vacuum to three particles}
\label{sec:3.afdag.0}

\begin{figure}
  \centering
  \includegraphics[scale=0.55]{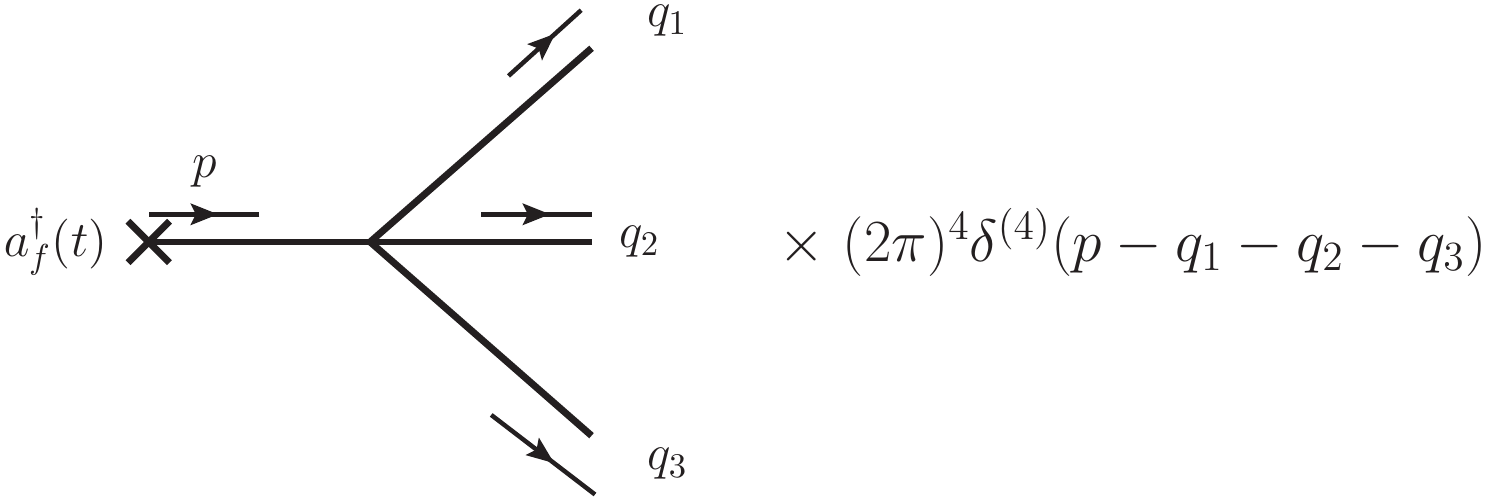}
  \caption{Lowest-order graph for
    $\langle\3q_1,\3q_2,\3q_3;\,\Out|a_f^\dagger(t)|0\rangle$, with explicit
    delta-function for momentum conservation.}
  \label{fig:afdag.3.0}
\end{figure}

In $\phi^4$ theory with coupling $\lambda$, the lowest order graph for
$a_f^\dagger(t)$ to create something other than a single particle is shown
in Fig.\ \ref{fig:afdag.3.0}, where the final state has 3 (on-shell)
particles of momenta $q_1$, $q_2$, and $q_3$.  From the Feynman rules,
we get
\begin{widetext}
\begin{align}
  \label{eq:afdag.1.3}
  \langle\3q_1,\3q_2,\3q_3;\,\Out| \, a_f^\dagger(t) \, |0\rangle
  &= \lambda\tilde{f}(\3q) \, \frac{E_{\3q} + q^0}{2E_{\3q}}
    \frac{1}{(q^0)^2-\3q^2-m^2+i\epsilon}
    \,e^{-i(E_{\3q}-q^0)t}
    \,+\, O(\lambda^2)
\nonumber\\
  &= \lambda\tilde{f}(\3q) \, \frac{1}{2E_{\3q} (q^0-E_{\3q}+i\epsilon) }
    \,e^{-i(E_{\3q}-q^0)t}
    \,+\, O(\lambda^2) .
\end{align}
Here $\3q=\sum_{j=1}^3\3q_j$ and $q^0=\sum_{j=1}^3E_{\3q_j}$.  Note that at
the order of perturbation theory to which we are working, the mass
value in the free propagator and the physical particle mass are equal:
$m_{\rm phys} = m + O(\lambda)$, so here we do not need to be careful
whether $m$ or $m_{\rm phys}$ is used in the calculation.

The matrix element in (\ref{eq:afdag.1.3}) is non-zero. So the
calculation shows unambiguously that the operator $a_f^\dagger(t)$, when
applied to the vacuum, creates more than a single particle.  We can
quantify by how much, by computing the corresponding contribution to
the squared norm of the state $a_f^\dagger(t)|0\rangle$, i.e., to $\left\|
  a_f^\dagger(t)|0\rangle \right\|^2$:
\begin{align}
\label{eq:size.state}
   \frac{1}{3!} \int \difftilde{q_1} \difftilde{q_2} \difftilde{q_3}
   \, \left| \langle\3q_1,\3q_2,\3q_3;\,\Out| \, a_f^\dagger(t) \, |0\rangle \right|^2
={}&
  \frac{\lambda^2}{3!} \int \difftilde{q_1} \difftilde{q_2} \difftilde{q_3}
  \, |\tilde{f}(\3q)|^2  \left| \frac{E_{\3q} + q^0}{2E_{\3q}} \right|^2
    \frac{1}{ \left[ (q^0)^2-\3q^2-m^2 \right]^2}
\\\nonumber
& + \text{higher order}.
\end{align}
\end{widetext}
(The overall factor of $1/3!$ is to compensate double counting of
indistinguishable states of identical particles.)  This is linearly
divergent in the UV.  To see this, we notice that the wave function
$\tilde{f}(\3q)$ strongly restricts the total 3-momentum $\3q$ to
finite values.  We can integrate freely over two of the final-state
momenta, say $\3q_1$ and $\3q_2$.  When these values are large and in
different directions, so is the third.  Let $\Lambda$ be the order of
magnitude of these momenta.  Then the propagator denominator is of
order $\Lambda^2$, and the factor $(E_{\3q}+q^0)/2E_{\3q}$ is of order $\Lambda$
divided by a finite mass scale.  Thus the power at large $\Lambda$ is $\Lambda^1$,
and the integral is linearly divergent as claimed.

As we know from renormalization theory, such UV power counting
corresponds to simple dimensional counting, and therefore it applies
equally to higher order graphs.  At best the UV divergences can be
modified by renormalization-group resummation.

The fact that calculation gives an infinite norm for $a_f^\dagger(t)|0\rangle$
shows that $a_f^\dagger(t)|0\rangle$ is not in the Hilbert space of the theory.
At best it can be said to be in some bigger space (not a Hilbert
space), which would allow the manipulations given above to be valid.
In any case it is not a physical state, and vectors in the bigger
space could well have pathological properties compared with those of
physical states in the Hilbert space.

Notice that the extra multiparticle contributions to the state
$a_f^\dagger(t)|0\rangle$ do not disappear when a limit of infinite time is taken.
This is because changing $t$ simply changes the phase in
\eqref{eq:afdag.1.3} but not the magnitude, and the same applies to
all higher order contributions.  The contribution to the norm of the
state is time-independent.  Thus the limits as $t$ goes to $-\infty$ or
$+\infty$ of $a_f^\dagger(t)|0\rangle$ do not exist.  Such limits would be called
strong limits, if they existed.

It is important that the existence of a divergence in the state is
distinct from the existence of the extra multiparticle contributions.
This can be seen be going to theory in a different space-time
dimension.  In a space-time dimension $n$ the degree of divergence of
the natural generalization of (\ref{eq:size.state}) is $2n-7$.  We get
convergence if $n<3.5$.  Thus in the integer dimensions 3 and 2 in
which the theory is super-renormalizable, there is no UV divergence in
(\ref{eq:size.state}), but the result remains non-zero.  The existence
of divergence in the physical case $n=4$ merely dramatizes the issue
that the putative creation operator applied to the vacuum creates more
than the intended single particle.  That issue is unchanged when there
is no UV divergence.

\subsubsection{Annihilation operator creates particles}

Similarly, although the one-particle-to-vacuum matrix element of the
would-be annihilation operator $a_f(t)$ is zero, the matrix elements
to multi-particle states are non-zero.  For example, we can calculate
$\langle\3q_1,\3q_2,\3q_3;\,\Out|a_f(t)|0\rangle$. The only differences compared
with the corresponding matrix element of $a_f^\dagger(t)$ are: A change of
$\tilde{f}(\3q)$ to its complex conjugate $\tilde{f}^*(\3q)$; a
complex conjugation of the $t$-dependent phase; a replacement of $q^0$
by $-\sum_{j=1}^3E_{\3q_j}$ and of $\3q$ by $-\sum_{j=1}^3\3q_j$.  The
result remains non-zero.

\subsubsection{Generalizations}

Many more examples can readily constructed.

Similar considerations also apply if more than one $a_f^\dagger(t)$ operator
is applied to the vacuum, with the intent of creating an initial state
consisting of two or more particles.  However, note that if more than
one operator is applied, as in $a_{f_1}^\dagger(t_2)a_{f_2}^\dagger(t_1)|0\rangle$, the
fields in the operator product $a_{f_1}^\dagger(t_2)a_{f_2}^\dagger(t_1)$ are
simply multiplied, without a time-ordering operations.  The use of
standard Feynman diagram methods applies to
$a_{f_1}^\dagger(t_2)a_{f_2}^\dagger(t_1)|0\rangle$ if $t_2>t_1$.

\subsubsection{Weak limit v.\ strong limit}
\label{sec:w.v.s}

Suppose instead of a state of three final-state particles of definite
momentum, we used a normalizable state,
\begin{equation}
  |G; \Out\rangle \eqdef
  \int \difftilde{q_1} \difftilde{q_2} \difftilde{q_3}
   \, |\3q_1,\3q_2,\3q_3;\,\Out\rangle \, \tilde{G}(\3q_1,\3q_2,\3q_3).
\end{equation}
Then Eq.\ (\ref{eq:afdag.1.3}) would be replaced by
\begin{multline}
  \label{eq:afdag.1.3.norm}
  \langle G;\,\Out| \, a_f^\dagger(t) \, |0\rangle
  =
    \lambda
  \int \difftilde{q_1} \difftilde{q_2} \difftilde{q_3}
   \, \tilde{G}^*(\3q_1,\3q_2,\3q_3)
\times \\ \times 
    \tilde{f}(\3q) \, \frac{1}{2E_{\3q} (q^0-E_{\3q}+i\epsilon) }
    \,e^{-i(E_{\3q}-q^0)t}
    \,+\, O(\lambda^2),
\end{multline}
still with $\3q=\sum_{j=1}^3\3q_j$ and $q^0=\sum_{j=1}^3E_{\3q_j}$.  The
oscillations in the integrand imply that when $t\to-\infty$ with $G$ fixed,
the matrix element rapidly goes to zero.  This is an example of the
property that $a_f^\dagger(t)$ converges weakly to a creation operator of a
single particle. But we have already seen that the strong limit, i.e.,
the limit of the state $a_f^\dagger(t)|0\rangle$, does not exist.

With a normalizable state instead of particles of definite momenta,
how negative $t$ needs to be for the matrix element to be close to the
asymptote depends on the wave functions $\tilde{G}$ and $\tilde{f}$.

As a physical illustration, suppose $\tilde{G}$ is real, positive, and
tightly peaked around a particular momentum for each particle.  Then
by the results of Sec.\ \ref{sec:wfs}, it corresponds to a state of
three particles of almost definite momenta that are localized quite
close to the origin of spatial coordinates at time $t=0$.  (We could
even insert a $\3q_j$-dependent phase to separate the particles a
bit.)  We also choose the wave function in $a_f^\dagger$ to be similarly
localized.

For this case, the matrix element (\ref{eq:afdag.1.3.norm}) rapidly
decreases to zero as $t$ is moved away from zero time.

Now change the wave function by a phase as follows:
\begin{equation}
   \tilde{G}(\3q_1,\3q_2,\3q_3)
\mapsto
   \tilde{G}(\3q_1,\3q_2,\3q_3)
    \,e^{-i(E_{\3q}-q^0)t_1},
\end{equation}
with $t_1$ being some very large negative value.  This corresponds to
a state that at time $t_1$ has the three particles spatially localized
to the same region as where the wave function $f(t_1,\3x)$ is
localized at the same time.

At $t=0$ the matrix element in (\ref{eq:afdag.1.3.norm}) is very small
because of the rapid oscillations in the integrand that are now in the
wave function.  However, when the time in the intended creation
operator is set to $t_1$, i.e., $t=t_1$, the oscillations are
canceled, and we get a large value.  It is only when $t$ is
significantly more negative than $t_1$ that the limiting behavior for
a creation of an one-particle initial state is approached.

Since $t_1$ is arbitrary, we cannot provide a single time below which
the state $a_f^\dagger(t)|0\rangle$ itself is close to zero.  We have thus seen
the non-existence of the strong limit with calculations that only use
normalizable states.  Obviously the calculation with particles of
definite momenta shows the same result more simply.  But the analysis
with normalizable states works purely within the Hilbert space of
states and lends itself to physical interpretation.

\subsubsection{Remarks}

\begin{figure*}
  \centering
  \setlength\tabcolsep{10mm}
  \begin{tabular}{cc}
  \includegraphics[width=3.5cm]{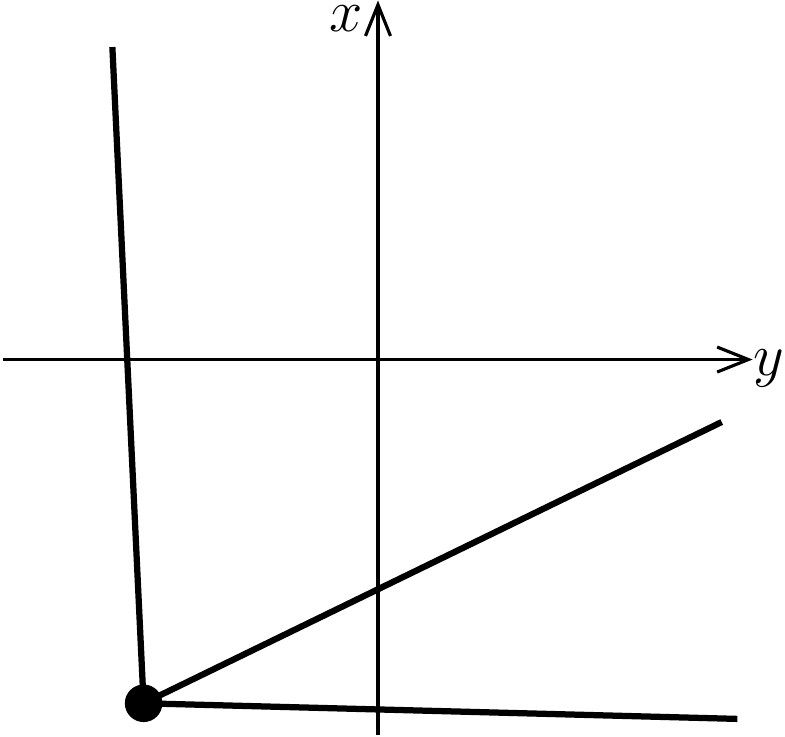}
  & \includegraphics[width=3.5cm]{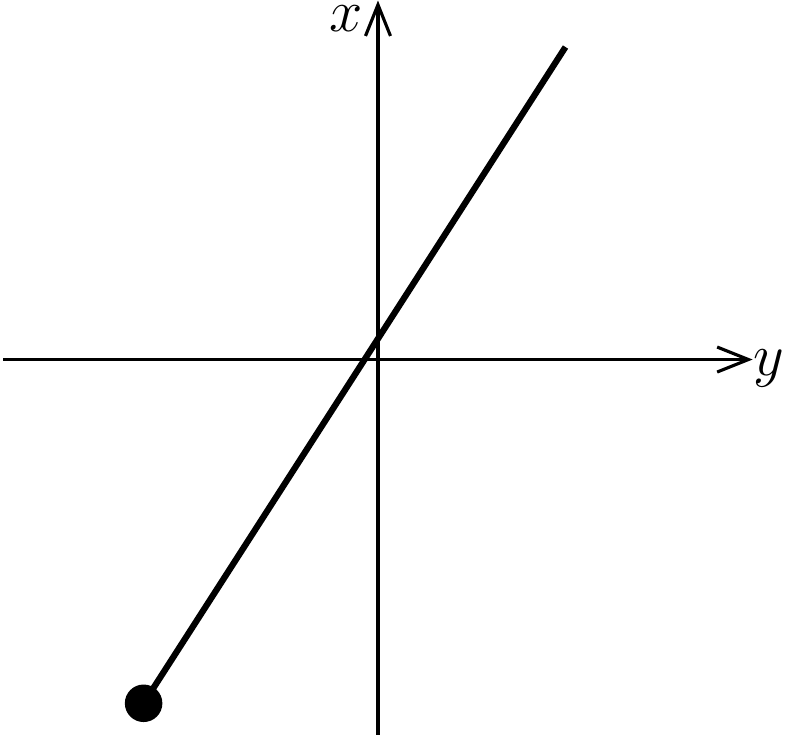}
  \\
  (a) &  (b) 
  \end{tabular}
  \caption{(a) Illustrating trajectories and hence spatial location of
    state components with extra particles that are created by
    $a_f^\dag(t)$ when $|t|$ is large and negative.  (b) The trajectories
    of the single particle corresponding to the intended
    single-particle state component created by the same operator.  
    In both cases, the dot indicates where the particle(s) were
    created. }
  \label{fig:extra.particles}
\end{figure*}

The line of argument just given shows that even though the operator
$a_f^\dagger(t)$ creates a lot of extra particles, they are asymptotically
in a different region of space time than the intended single particle.
More explicitly, observe that the coordinate-space wave function
$f(x)$ is concentrated near the classical trajectory of the particle.
When $t\to-\infty$, its position is infinitely far away from the origin. Now,
for graphs with multi-particle final states, the propagator attached
to the vertex for $a_f^\dagger(t)$ is never close to its pole.  Thus the
invariant distances involved between the ends of the propagator are of
order $1/m$ or smaller.  Hence when $t\to-\infty$, the production region of
the particles is infinitely far away, as illustrated in Fig.\
\ref{fig:extra.particles}.  Thus they essentially all avoid hitting a
finite sized detector surrounding the intended collision region in a
scattering experiment.  This is a fairly physical reason why the large
extra contributions to the initial state created by $a_f^\dagger(t)$ do not
contribute in the LSZ proof to calculations of the S-matrix itself,
despite the delicate mathematical grounding caused by the UV
divergence in the state vector.

A more abstract mathematical way of analyzing this situation is to
observe that as $t\to\pm\infty$, the phase factors in Eqs.\
(\ref{eq:af.dag.Feyn}) and (\ref{eq:af.Feyn}) oscillate infinitely
rapidly except when the momentum $p$ or $q$ corresponds to the
intended single-particle state.  If the vertices appear inside a
quantity that has a suitable integral over momentum, then the
infinitely rapid oscillations cause a corresponding vanishing of the
corresponding contribution in the integral. The oscillations
mathematically implement the physical statement that the extra
particles created or destroyed by the operators are infinitely far
from the scattering region.

However, we may not always want to use a limit of infinite time.
Notably, if we wish to calculate scattering with an initial particle
that is unstable (but perhaps long lived), then it is natural to use
an operator like $a_f^\dagger(t)$ to create the particle at a finite time
that corresponds to the experimental source of the particle.  Then the
extra particles created by $a_f^\dagger(t)$ can easily be in an
experimentally accessible location.

\begin{widetext}
\subsection{General analysis of divergence from spectral
  representation}
\label{sec:div.spectral}

The UV divergence in $\|a_f^\dagger(t)|0\rangle\|^2$ was found in an illustrative
calculation in Sec.\ \ref{sec:extra.particles} with the use of the
reduction formula.  We will now obtain a general result for this
quantity from the spectral representation, whose properties were
summarized in Sec.\ \ref{sec:KL}.  From the representation Eq.\
(\ref{eq:prod.prop}) for the 2-field correlator, and the definition
(\ref{eq:af.dag}) of $a_f^\dagger(t)$, it follows that
\begin{align}
\label{eq:size.state.full.1}
    \|a_f^\dagger(t)|0\rangle\|^2
    ={}&
    \langle0| a_f(t)a_f^\dagger(t)|0\rangle
\nonumber\\
   ={}&
   \int_0^\infty \diff{s} \rho(s) \int \difftilde{p} \difftilde{q} 
   \int \frac{\diff[3]{\3k}}{(2\pi)^3 2 \sqrt{\3k+s}} 
   \tilde{f}(\3p)^* \tilde{f}(\3q)
   \int \diff[3]{\3x} \diff[3]{\3y} 
\nonumber\\\hspace*{10mm}&
    \left[ e^{iE_{\3p}x^0 - i\3p\cdot\3x} \derivB[x^0] e^{-i\sqrt{\3k+s}x^0 + i\3k\cdot\3x} \right]
    \left[ e^{-iE_{\3q}y^0 + i\3q\cdot\3x} \derivB[y^0] e^{i\sqrt{\3k+s}y^0 - i\3k\cdot\3y} \right],
\end{align}
with $x^0$ and $y^0$ being set to $t$ after the derivatives are
taken. Straightforward manipulations lead to
\begin{align}
\label{eq:size.state.full}
    \|a_f^\dagger(t)|0\rangle\|^2
   ={}&
   \int \difftilde{p} |\tilde{f}(\3p)|^2
   \int_0^\infty \diff{s} \rho(s)
   \frac{ \left( \sqrt{\strut \3p^2+s} + \sqrt{\3p^2+m_{\rm phys}^2} \right)^2 }
      {4 \sqrt{\strut \3p^2+s} \, \sqrt{\3p^2+m_{\rm phys}^2} }.
\end{align}
\end{widetext}

Now the spectral function can be found calculationally from analyzing
the propagator and self-energy graphs.  At large $p^2$, the propagator
$\hat{G}_2(p^2)$ is known to behave like $1/p^2$ times logarithms,
order-by-order in perturbation theory.  Hence from the spectral
representation (\ref{eq:KL.prop}) for $\hat{G}_2(p^2)$, the spectral
function $\rho(s)$ behaves like $1/s$ times logarithms.

In Eq.\ (\ref{eq:size.state.full}), the wave function limits $\3p$ to
finite values, while at large $s$ the last factor grows like
$\sqrt{s}$.  It follows that order-by-order in perturbation theory,
$\|a_f^\dagger(t)|0\rangle\|^2$ has the same kind of power divergence that we found
in an explicit calculation.  The state's norm can only be finite if
the true non-perturbative $\rho(s)$ falls sufficiently more rapidly than
$1/s^{3/2}$.

The operator $a_f^\dagger(t)$ was defined by an integral over the field and
its time derivative at fixed time. Therefore the UV divergence in
$\|a_f^\dagger(t)|0\rangle\|^2$ shows that there is a failure of localization on a
quantization surface of fixed time.  The presence or absence of the
divergence depends on the dynamics of the theory.  Localization does
work in a theory that is sufficiently convergent in the UV, e.g., in a
free-field theory, or in a sufficiently super-renormalizable theory.
The culprit is not the field itself, but its time derivative.  The
time derivative gives two factors of $q^0$ in the integrand in
(\ref{eq:size.state}), and without those factors the integral would be
convergent.  See Ref.\ \cite{Schlieder:1972qr} for a further
examination of the localization properties of fields on a quantization
surface, in the free-field case, and for an examination of interesting
differences between equal-time and light-front quantization.

In a theory where there is no UV divergence in $a_f^\dagger(t)|0\rangle$, Eq.\
(\ref{eq:size.state.full}) nonperturbatively quantifies the
multi-particle contribution to the state.

\subsection{Modified creation and annihilation operators}
\label{sec:Afdag.def}

What allowed the operators $a_f^\dagger(t)$ and $a_f(t)$ to create and
annihilate states other than the intended single-particle states was
the use of the field operator at fixed time.  This gave an integral
over all energy in the momentum-space version, as exhibited in the
Feynman rules \eqref{eq:af.dag.Feyn} and \eqref{eq:af.Feyn}.

A simple way to enforce the single-particle condition is to restrict
the energy explicitly.  Let us implement this by smoothly averaging
over a range $\Delta t$ of time with an averaging function $F(t'-t, \Delta t)$,
and then taking $\Delta t$ to infinity.  We define the averaged operator by
\begin{align}
  \label{eq:Af.dag.def}
 A_f^\dagger(t;\Delta t) &\eqdef  \frac{1}{c^*} \int \diff{t'} F(t'-t, \Delta t) \, a_f^\dagger(t').
\end{align}
The averaging function is required to be real, non-negative, and to
integrate to unity:
\begin{equation}
  \int \diff{t'} F(t'-t, \Delta t) = 1.
\end{equation}
In the definition (\ref{eq:Af.dag.def}), the factor $1/c^*$ is to
normalize $A_f^\dagger(t;\Delta t)$ so that, as we will see, the one-particle
state that it creates has the standard normalization.  A suitable
averaging function is a Gaussian:
\begin{equation}
\label{eq:f.def}
  F(t'-t, \Delta t) 
= 
  \frac{ e^{ -(t'-t)^2/(\Delta t)^2 } }{ \sqrt{\pi} \Delta t } .
\end{equation}

A number of alternative forms may be written for the operator
$A_f^\dagger(t;\Delta t)$.  Let ${x'}^\mu=(t',\3x')$.  Then
\begin{widetext}
\begin{align}
\label{eq:Af.dag.def.alt}
 A_f^\dagger(t;\Delta t) &= \frac{-i}{c^*} \int \diff[4]{x'} F(t'-t, \Delta t) 
                 \left[ f(x') \derivB[t'] \phi(x') \right]
\nonumber\\
        &= \frac{i}{c^*} \int \diff[4]{x'} \phi(x')
            \left[ \frac{\partial F(t'-t, \Delta t)}{\partial t'} f(x')
                   +2 F(t'-t, \Delta t) \frac{\partial f(x')}{\partial t'} \right]
\nonumber\\
        &= \frac{1}{c^*} \int \frac{\diff[4]{p}}{(2\pi)^4} \tilde{\phi}(p)
            \, \tilde{f}(\3p)
            \, \frac{E_{\3p}+p^0}{2 E_{\3p}}
            \, e^{-i(E_{\3p}-p^0)t}
            \, \tilde{F}(p^0-E_{\3p}, \Delta t).
\end{align}
In the second line, $A_f^\dagger(t;\Delta t)$ is written as a simple integral of
the field with a function.  It is probably best to treat that second
line as the actual definition of $A_f^\dagger(t;\Delta t)$, since it avoids any
difficulties about localization of fields on a quantization surface.
The previous manipulations can then be regarded as simply
motivational, to indicate why that particular formula is used. Our
earlier results showing that the operators $a_f^\dag(t)$ fail to exist as
operators on the space of physical normalizable states show that the
steps leading to the second line of Eq.\ (\ref{eq:Af.dag.def.alt}) are
on dangerous mathematical ground.

In the last line we used the Fourier transform of the field,
\begin{equation}
\label{eq:phi.FT}
  \phi(x) = \int \frac{\diff[4]{p}}{(2\pi)^4} \tilde{\phi}(p) \, e^{\BDpos i p\cdot x},
\end{equation}
to write a momentum-space expression, with the Fourier transform of
$F$ defined by
\begin{equation}
  \tilde{F}(\delta E, \Delta t)
  = \int \diff{t'} e^{i\delta E(t'-t)} F(t'-t, \Delta t).
\end{equation}
In the case of the Gaussian, the Fourier-transformed averaging
function is
\begin{equation}
  \tilde{F}(\delta E, \Delta t)
  = e^{-\delta E^2\Delta t^2/4}.
\quad \text{($F$ Gaussian)}
\end{equation}

From the last line of Eq.\ (\ref{eq:Af.dag.def.alt}) follows the Feynman
rule for the vertex corresponding to the operator $A_f^\dagger(t;\Delta t)$.  It
is a simple generalization of (\ref{eq:af.dag.Feyn}), containing an
extra factor $\tilde{F}(p^0-E_{\3p}, \Delta t)$:
\begin{equation}
  \label{eq:Af.dag.Feyn}
  \text{Vertex}(A_f^\dagger(t;\Delta t))
  = \int \frac{\diff[4]{p}}{(2\pi)^4}
      \tilde{f}(\3p) 
      \, \frac{E_{\3p}+p^0}{2 E_{\3p}}
      \, e^{-i(E_{\3p}-p^0)t}
      \, \tilde{F}(p^0-E_{\3p}, \Delta t) .
\end{equation}
It is the new factor $\tilde{F}$ that enforces the restriction that in
the $\Delta t\to\infty$ limit, $A_f^\dagger(t;\Delta t)$ creates a single particle and
nothing else.  That is, when we take the limit of infinite averaging
time, $\Delta t\to\infty$, $\tilde{F}$ goes to zero unless $\delta E=0$, i.e., unless
the on-shell condition $p^0=E_{\3p}$ is obeyed.  

Formulas for the conjugate operator $A_f(t;\Delta t)$ are:
\begin{align}
\label{eq:Af.def.alt}
 A_f(t;\Delta t)
        &= \frac{-i}{c} \int \diff[4]{x'} \phi(x')
            \left[ \frac{\partial F(t'-t, \Delta t)}{\partial t'} f^*(x')
                   +2 F(t'-t, \Delta t) \frac{\partial f^*(x')}{\partial t'} \right]
\nonumber\\
        &= \frac{1}{c} \int \frac{\diff[4]{p}}{(2\pi)^4} \tilde{\phi}(-p)
            \, \tilde{f}^*(\3p)
            \, \frac{E_{\3p}+p^0}{2 E_{\3p}}
            \, e^{i(E_{\3p}-p^0)t}
            \, \tilde{F}^*(p^0-E_{\3p}, \Delta t),
\end{align}
and the corresponding vertex is
\begin{equation}
  \label{eq:Af.Feyn}
  \text{Vertex}(A_f(t;\Delta t))
  = \int \frac{\diff[4]{q}}{(2\pi)^4}
      \tilde{f}^*(\3q) 
      \, \frac{E_{\3q}+q^0}{2 E_{\3q}}
      \, e^{i(E_{\3q}-q^0)t}
      \, \tilde{F}^*(q^0-E_{\3q}, \Delta t) .
\end{equation}
\end{widetext}

When these formulas are used in calculations, it should be remembered
that for the $A_f^\dagger$ vertex the momentum $p$ flows \emph{into} the
rest of the graph from $A_f^\dagger$ vertex, whereas the opposite is true
for the momentum $q$ at the $A_f$ vertex, as explained after Eq.\
(\ref{eq:af.Feyn}).

Furthermore, a Green function is defined using a time-ordered product
of fields.  So except in asymptotic situations, e.g., in matrix
elements with in- and out-states, there is a mismatch between a Green
function with the vertex implied by Eq.\ (\ref{eq:Af.dag.def.alt}) and
an actual matrix element of the operator.

Finally, we need operators that create and destroy particles in the
in- and out-states, as the relevant limits.  The creation operators
are defined as
\begin{align}
\label{eq:Afdag.In.def}
  A_{f;\,\In}^\dagger & \eqdef  \lim_{\substack{ t\to-\infty\\ 
                   \Delta t \to \infty\\ 
                   \Delta t/|t| \to 0 }}
              A_f^\dagger(t;\Delta t) ,
\\
\label{eq:Afdag.Out.def}
  A_{f;\,\Out}^\dagger & \eqdef  \lim_{\substack{ t\to+\infty\\ 
                   \Delta t \to \infty\\ 
                   \Delta t/|t| \to 0 }}
              A_f^\dagger(t;\Delta t) ,
\end{align}
with the annihilation operators defined as the hermitian conjugates.
The limit $\Delta t\to\infty$ ensures that the created and destroyed particles are
just the intended single particles, because of the factor
$\tilde{F}(p^0-E_{\3p}, \Delta t)$ in the vertex for $A_f^\dagger(t;\Delta t)$.  The
limit $t\to\pm\infty$ is needed to correspond to the application to the
asymptotics of scattering processes, as usual.  To ensure that the
range of time involved in creating or destroying incoming and outgoing
particles is much less than the time between the region of scattering
and the creation and destruction of asymptotic particles, we require
that in the infinite-time limits $t$ and $\Delta t$ are taken such that 
$\Delta t/|t|\to0$.

These time scales correspond to
experimental reality, where the times involved are short of achieved
the strict mathematical limits.

The definitions (\ref{eq:Afdag.In.def}) and (\ref{eq:Afdag.Out.def})
are as \emph{strong} limits.  Given the known result that the
corresponding definitions for the elementary creation operators
$a_f^\dagger(t)$ are only \emph{weak} limits, it is important to verify
explicitly that the strong limits exist. 

Having defined the operators $A_f^\dagger(t;\Delta t)$ and $A_f(t;\Delta t)$, and
their limits as $t\to\pm\infty$, it is necessary to verify that they have the
properties that their definitions were intended to have and that they
do not have the deficiencies illustrated in Sec.\
\ref{sec:extra.particles} for the LSZ versions of the operators.  To
avoid a circularity of the logic, we will postpone this analysis until
after the proof of the reduction formula is completed.  The results
are found in Sec.\ \ref{sec:verify} below.

\begin{widetext}
\section{Simple matrix elements of annihilation and creation
  operators}
\label{sec:Af.ME.simple}

In this section we obtain some of the most basic properties of the
annihilation and creation operators.  These involve the action of one
operator on the vacuum or the vacuum expectation value of a product of
two operators.  These can be analyzed simply and non-perturbatively in
the exact theory with the aid of the spectral representation; the
analysis does not need the reduction formula.  Not only are the
results of interest in their own right, but they will be useful
ingredients for work in later sections.

\subsection{One operator on vacuum}

From the vertex formula (\ref{eq:Af.dag.Feyn}) and translation
invariance of $\langle0|\phi(x)|0\rangle$, it follows that the vacuum expectation
value of one operator is
\begin{equation}
  \langle0| A_f^\dagger(t;\Delta t) |0\rangle 
  = 
  \frac{ 1}{2c^*} 
  \langle0|\phi(0)|0\rangle \tilde{f}(\30)
            \, e^{-im_{\rm phys}t}
            \, \tilde{F}(-m_{\rm phys}, \Delta t).
\end{equation}
This goes to zero as $\Delta t\to\infty$, because $\tilde{F}(-m_{\rm phys})$ does,
and this happens independently of $t$.  In particular, we can take $t$
to $\pm\infty$ more rapidly than $\Delta t$, as is needed for the in- and
out-operators.  Thus
\begin{equation}
  \lim_{\Delta t\to\infty} \langle0| A_f^\dagger(t;\Delta t) |0\rangle = \langle0| A_{f;\,\In}^\dagger |0\rangle = \langle0| A_{f;\,\Out}^\dagger |0\rangle = 0.
\end{equation}

From the vertex formula (\ref{eq:Af.dag.Feyn}) and the known
one-particle-to-vacuum matrix element of $\langle\3p|\phi(x)|0\rangle$ it follows
that the $A_f^\dag$ operators have the desired one-particle-to-vacuum
matrix elements, independently of $t$ and $\Delta t$.  Hence
\begin{equation}
  \label{eq:Afdag.1.0}
  \langle\3p| A_f^\dagger(t;\Delta t) |0\rangle 
  = \langle\3p| A_{f;\,\In}^\dagger |0\rangle
  = \langle\3p| A_{f;\,\Out}^\dagger |0\rangle
  = \tilde{f}(\3p).
\end{equation}
Similar we have the following matrix elements of an annihilation
operator with a normalizable one-particle state:
\begin{equation}
  \label{eq:Af.0.1}
  \langle0| A_f(t;\Delta t) |g\rangle 
  = \langle0| A_{f;\,\In} |g\rangle
  = \langle0| A_{f;\,\Out} |g\rangle
  = \int \difftilde{p} \tilde{f}^*(\3p) \tilde{g}(\3p).
\end{equation}

Matrix elements of a creation operator between a multi-particle state
and the vacuum vanish in the $\Delta t\to\infty$ limit.  To see this, we follow
the pattern used in the example calculations, with basis momentum
states:
\begin{equation}
   \langle \3q_1,\dots,\3q_n;\,\In| \, A_f^\dagger(t;\Delta t) \, |0\rangle 
   = (\mbox{time-independent})
     \, e^{-i(E_{\3q}-q^0)t}
       \, \tilde{F}(q^0-E_{\3q}, \Delta t),
\end{equation}
where $q^0$ and $\3q$ are the total energy and 3-momentum of the
$n$-particle state.  As $\Delta t\to\infty$, this vanishes, because the $\tilde{F}$
factor vanishes.  We now wish to show that the strong limit of the
state $A_f^\dagger(t;\Delta t) |0\rangle$ as $\Delta t\to\infty$ is the single-particle state
$|f\rangle$.  From the already-calculated one-particle component in
$A_f^\dagger(t;\Delta t) |0\rangle$ it follows that the size of the rest of the state
is given by
\begin{equation}
  \left\| A_f^\dagger(t;\Delta t) |0\rangle - |f\rangle \right\|^2
  = \SumInt_{\substack{\rm except\\\rm single\\\rm particle}} \!\!\diff{X}
      \left| \langle X;\,\Out| A_f^\dagger(t;\Delta t) |0\rangle \right|^2
  =
  (\mbox{time-independent})
   \times \, \left| \tilde{F}(q^0-E_{\3q}, \Delta t) \right|^2,
\end{equation}
where the integral over $X$ is a sum and integral over basis states
for all but the single-particle states.  The desired strong limit
follows, plus a result for $\Delta t\to\infty$ independently of $t$:
\begin{equation}
  \label{eq:Afdag.0}
    \lim_{\Delta t\to\infty}A_f^\dagger(t;\Delta t) |0\rangle
    = A^\dag_{f;\,\In} |0\rangle = A^\dag_{f;\,\Out} |0\rangle 
    = |f\rangle .
\end{equation}
Here, as usual $|f\rangle = |f;\,\In\rangle = |f;\,\Out\rangle = \int \difftilde{p} \, |\ve{p}\rangle
\, \tilde{f}(\ve{p})$.

In textbooks it is very common to use the interaction picture, and to
employ three kinds of basis state: a free basis, and in-basis, and an
out-basis. Therefore it is necessary to emphasize that in
(\ref{eq:Afdag.0}) and elsewhere in this paper, the basis states are
strictly in the in- or out-basis, unless specifically indicated
otherwise.  Haag's theorem guarantees that the state space spanned by
the in- and out-bases in an interacting theory is orthogonal to the
corresponding state space of the free theory.

Very similarly, we find that the operator $\lim_{\Delta t\to\infty}A_f(t;\Delta t)$
annihilates the vacuum:
\begin{align}
  \label{eq:Af.0}
    \lim_{\Delta t\to\infty} A_f(t;\Delta t) |0\rangle = 0,
\end{align}
and equally for the in- and out-versions of the annihilation
operators.

None of the results above should be a surprise.  The definition of
$A_f^\dag$ was specifically designed to make them valid.  Nevertheless,
it is useful to show directly from the definitions that in the most
elementary cases the properties are valid.

\subsection{Vacuum matrix elements of two operators}
\label{sec:2op.vac.ME}

Consider now the vacuum expectation value of the product of two
$A^\dag_f(t,\Delta t)$ operators: $\langle0| A_g^\dagger(t_2;\Delta t_2) A_g^\dagger(t_1;\Delta t_1) |0\rangle$.
In view of later uses of results from this section, we will start with
independently chosen values of the parameters $t_j$ and $\Delta t_j$.  From
the definition (\ref{eq:Af.dag.def.alt}) of the operators and the
spectral representation (\ref{eq:prod.prop}), standard manipulations
give
\begin{align}
\label{eq:Adag.Adag.0.0.Deltat}
  \langle0| A_f^\dagger(t_2;\Delta t_2) A_g^\dagger(t_1;\Delta t_1) |0\rangle
  = {}&
  \frac{1}{{c^*}^2}
  \int \frac{\diff[4]{p}}{(2\pi)^3} \rho(p^2)\theta(p^0)
  \, \tilde{g}(-\3p)\, \tilde{f}(\3p)
  \, \frac{E_{-\3p}-p^0}{2 E_{-\3p}}
  \, \frac{E_{\3p}+p^0}{2 E_{\3p}}
\times\nonumber\\&\hspace*{7mm}\times
  \, e^{-i(E_{-\3p}+p^0)t_2}  \, e^{-i(E_{\3p}-p^0)t_1}
  \, \tilde{F}(-p^0-E_{-\3p}, \Delta t_2)\, \tilde{F}(p^0-E_{\3p}, \Delta t_1)
\nonumber\\
  = {}&
  \frac{1}{{c^*}^2}
  \int \frac{\diff[4]{p}}{(2\pi)^3} \rho(p^2)\theta(p^0)
  \, \tilde{g}(-\3p)\, \tilde{f}(\3p)
  \, \frac{p^2-m_{\rm phys}^2}{4 E_{\3p}^2}
\times\nonumber\\&\hspace*{7mm}\times
  \, e^{-i(E_{\3p}+p^0)t_2}  \, e^{-i(E_{\3p}-p^0)t_1}
  \, \tilde{F}(-p^0-E_{\3p}, \Delta t_2)\, \tilde{F}(p^0-E_{\3p}, \Delta t_1)
\end{align}
In a limit that $\Delta t_2\to\infty$, the first $\tilde{F}$ factor
vanishes, since $p^0$ is always positive.  Hence
\begin{align}
\label{eq:Adag.Adag.0.0}
  \lim_{\Delta t_2\to\infty}\langle0| A_g^\dagger(t_2;\Delta t_2) A_f^\dagger(t_1;\Delta t_1) |0\rangle = 0,
\end{align}
independently of whether any of the other parameters ($t_j$, $\Delta t_1$)
are held fixed or are taken to infinity.  Hence.
\begin{equation}
  \langle0| A_{g;\,\In}^\dagger A_{f;\,\In}^\dagger |0\rangle =
  \langle0| A_{g;\,\Out}^\dagger A_{f;\,\Out}^\dagger |0\rangle = 0.
\end{equation}
Note the lack of time dependence in (\ref{eq:Adag.Adag.0.0}), which is
specific to the case where the vacuum expectation value is of two
operators (or fewer).

The case of one creation and one annihilation operator is nonzero:
\begin{multline}
  \langle0| A_g(t_2;\Delta t_2) A_f^\dag(t_1;\Delta t_1) |0\rangle
\\
  =
  \frac{1}{|c|^2}
  \int \frac{\diff[4]{p}}{(2\pi)^3} \rho(p^2)\theta(p^0)
  \, \tilde{g}^*(\3p)\, \tilde{f}(\3p)
  \, \left( \frac{E_{\3p}+p^0}{2 E_{\3p}} \right)^2
  \, \, e^{i(E_{\3p}-p^0)(t_2-t_1)}
  \, \tilde{F}^*(p^0-E_{\3p}, \Delta t_2)
  \, \tilde{F}(p^0-E_{\3p}, \Delta t_1) .
\end{multline}
When one or both $\Delta t_j\to\infty$, the $\tilde{F}$ factors give zero for the
continuum contribution from $\rho(p^2)$, i.e., from $p^2>m_{\rm phys}^2$,
and leave only the contribution from the single particle intermediate
state, with its delta-function in $\rho(p^2)$.  This is again independent
of what is done with $t_j$.  So
\begin{equation}
\label{eq:A.Adag.VEV}
  \limSub{\Delta t_1 \\{\rm and/or}\\ \Delta t_2\to\infty}
   \langle0| A_g(t_2;\Delta t_2) A_f^\dag(t_1;\Delta t_1) |0\rangle
  =
  \int \difftilde{p}
  \, \tilde{g}^*(\3p)\, \tilde{f}(\3p),
\end{equation}
which is the same time-independent formula as in the free theory.  The
result is saturated by one-particle states between the two operators.
Exchanging the two operators gives zero:
\begin{equation}
  \limSub{\Delta t_1 \\{\rm and/or}\\ \Delta t_2\to\infty}
 \langle0| A_f^\dag(t_1;\Delta t_1) A_g(t_2;\Delta t_2) |0\rangle = 0,
\end{equation}
and hence the commutator's vacuum-expectation value is the standard
one
\begin{equation}
  \limSub{\Delta t_1 \\{\rm and/or}\\ \Delta t_2\to\infty}
  \langle0| \, \left[ A_g(t_2;\Delta t_2), A_f^\dag(t_1;\Delta t_1) \right] \, |0\rangle
  = \int \difftilde{p}
  \, \tilde{g}^*(\3p)\, \tilde{f}(\3p).
\end{equation}
This implies the same equations for the in and out versions of these
operators.  They are then compatible with the standard commutation
relations for the in-operators, i.e.,
\begin{align}
  \left[ A_{g;\,\In}^\dagger, A_{f;\,\In}^\dagger \right]
  = \left[ A_{g;\,\In}, A_{f;\,\In} \right]
  = 0,
\qquad  \left[ A_{g;\,\In}, A_{f;\,\In}^\dagger \right]
  = \int \difftilde{p}
  \, \tilde{g}^*(\3p)\, \tilde{f}(\3p),
\end{align}
and the same for the out-operators.  But an actual proof needs more
techniques than just the spectral representation, and will have to
wait until we derive the reduction formula.

\section{Derivation of reduction formula}
\label{sec:LSZ.derivation}

In this section, we find how to express $\langle g_1,\dots,g_n;\,\Out |
f_1,\dots,f_{n'};\,\In \rangle$ in terms of momentum-space Green functions and
thence derive the reduction formula for the S-matrix.  To make a
simplification in the notation and the analysis of connected components,
we restrict to the standard experimentally relevant situation of two
incoming particles, i.e., $n'=2$.  The generalization to other values of
$n'$ is elementary.

The overall starting point consists of the assertions that we have a
relativistic QFT obeying standard properties, that the Green functions
exist, and that the full propagator has a pole at a nonzero mass.  We
also assume that the momentum-space Green functions have the
analyticity properties attributed to them on the basis of Feynman
perturbation theory. All of these properties primarily concern the
off-shell Green functions.

We will use the operators defined in Sec.\ \ref{sec:Afdag.def} for
creating in- and out-states of specified particle content.

The derivation of the reduction formula has the following steps:
\begin{enumerate}

\item Given a particular set of momentum-space wave functions $f_1,\dots$,
  define a corresponding ``in state'' by applying the relevant product of
  $A_{f;\,\In}^\dagger$ operators to the vacuum, thereby constructing a state
  with a specified momentum content in the far past.  Similarly, construct
  an out state with the specified content in the far future.  The gives a
  construction of the states used in $\langle g_1,\dots,g_n;\,\Out |
  f_1,f_2;\,\In \rangle$.  The aim of the following manipulations is to obtain a
  useful formula for the S-matrix, which appears on the right-hand side of
  Eq.\ (\ref{eq:g.f}).

\item Apply the definitions of the $A_{f;\,\In}^\dagger$ and
  $A_{f;\,\Out}^\dagger$ operators.  The result is a limit of the
  vacuum-expectation value of a product of field operators integrated
  over space-time, multiplied by wave functions and averaging
  functions.

\item Show that in the relevant infinite-time limit, the product of field
  operators can be replaced by a time-ordered product, i.e., an ordinary
  Green function.  

\item By Fourier transformation, express the result in terms of the
  momentum-space Green functions.

\item Show that in the specified limits of infinite time, the result is of
  the usual form of momentum-space wave functions integrated with a
  quantity $S_{\3q_1,\dots,\3q_n;\,\3{p}_1,\3{p}_2}$, as on the right-hand
  side of Eq.\ (\ref{eq:g.f}), and that
  $S_{\3q_1,\dots,\3q_n;\,\3{p}_1,\3{p}_2}$ has the form given in the
  reduction formula, (\ref{eq:LSZ.1}) or (\ref{eq:LSZ.2}), for the
  connected term and related formulas for disconnected terms.

\end{enumerate}
The first two steps are simply an application of our defined creation
and annihilation operators, with the use of the assertion that the
limits involved in their definition are strong limits.  The remaining
steps consist of essentially mechanical steps to calculate the inner
product in terms of momentum-space Green functions, and hence in terms
of quantities accessible to Feynman-graph calculations.  They also
verify that the inner product has the expected structure that
corresponds to the existence of an S-matrix.

In addition, we will verify some properties of the states and operators
that are needed to show that the expected properties used in scattering
theory actually do hold.  

\subsection{Construction of asymptotic states and inner product of in
  and out states}

We have defined operators $A_{f;\,\In}^\dagger$ and $A_{f;\,\Out}^\dagger$ that are
intended to create asymptotic particles when applied to the vacuum.
For the in-state we therefore \emph{define}
a generic in-state by
\begin{equation}
\label{eq:i.state}
  |f_1,\dots,f_{n'};\,\In \rangle 
  \eqdef A_{f_1;\,\In}^\dagger \, \dots \, A_{f_{n'};\,\In}^\dagger \, |0\rangle
  = \prod_{j=1}^{n'} A_{f_j;\,\In}^\dagger \, |0\rangle .
\end{equation}
For each wave function, define the following quantity
\begin{equation}
  \label{eq:f.F}
  \hat{f}_j(x_j;t_{j,-},\Delta t_{j,-})
  =
  \frac{\partial F(x_j^0-t_{j,-}, \Delta t_{j,-})}{\partial x_j^0} f(x_j)
  +2 F(x_j^0-t_{j,-}, \Delta t_{j,-}) \frac{\partial f(x_j)}{\partial x_j^0}.
\end{equation}
With modified variable names, this is the combination of wave function
and averaging functions that appears in the second line of Eq.\
(\ref{eq:Af.dag.def.alt}) which gives formulas for what with our
modified variable names is $A_f^\dagger(t_{j,-},\Delta t_{j,-})$.  In view of how
we will use these formula, each of the $t$ and $\Delta t$ parameters is
treated as distinct.  Moreover, they are given a label $-$ to
distinguish them from corresponding parameters for final-state
particles.

Then the infinite-time limits for the $A_{f_j;\,\In}^\dagger$ operators give
\begin{align}
\label{eq:i.coord.U}
 | f_1,\dots,f_{n'};\,\In \rangle
 &=
  \frac{1}{{c^*}^{n'}}
  \limTIn
  \int \prod_{j=1}^{n'} \diff[4]{x_j}
  \prod_{j=1}^{n'} \hat{f}_j(x_j; t_{j,-}, \Delta t_{j,-}) 
  \, 
  {\textstyle  \prod_{j=1}^{n'} \phi(x_j)} \, 
  |0\rangle .
\end{align}
The exact specification of the product of fields and the limits so
that they correspond to the right-hand side of Eq.\ (\ref{eq:i.state}) is
as follows: The order of the product of the field operators is specified to
be $\phi(x_1)\dots\phi(x_{n'})$. We have $n'$ instances of the application
of the definition (\ref{eq:Afdag.In.def}), each needing its own value
of $t$ and $\Delta t$.  The infinite-time limit is be such that for each
$j$, $t_{j,-}\to-\infty$, $\Delta t_{j,-}\to\infty$, and $\Delta t_{j,-}/|t_{j,-}|\to0$.
Moreover, given the order in which the creation operators are applied,
the individual times are ordered: $t_{1,-}>\dots>t_{n',-}$ and the
separation of neighboring $t_{j,-}$ is much bigger than the $\Delta
t_{j,-}$.  
In fact, in the infinite-time limit the operators commute, and the order
in which they are applied is irrelevant, but we will only establish that
later.  

Now the averaging functions $F$ restrict the range of integration over
the time component $x_j^0$ of each $x_j$ to be dominantly within $\Delta
t_{j,-}$ of the corresponding central value of time, i.e., $t_{j,-}$.
Therefore, except for contributions that vanish in the infinite-time
limit, the product of operators can be replaced by the time-ordered
product, i.e.,
\begin{align}
\label{eq:i.coord.T}
 | f_1,\dots,f_{n'};\,\In \rangle
 &=
  \frac{1}{{c^*}^{n'}}
  \limTIn
  \int \prod_{j=1}^{n'} \diff[4]{x_j}
  \prod_{j=1}^{n'} \hat{f}_j(x_j; t_{j,-}, \Delta t_{j,-}) 
  \, 
  {\textstyle  T \prod_{j=1}^{n'} \phi(x_j)} \, 
  |0\rangle.
\end{align}

For a generic out-state, exactly the same formula applies except that
it is in the limit of infinite time is in the future and that the
ordering is anti-time ordering.  Taking the Hermitian conjugate, as
needed for the overlap of an out- and an in-state, again results in a
time-ordered product:
\begin{align}
\label{eq:f.coord.T}
 \langle g_1,\dots,g_n;\,\Out |
 &=
  \frac{1}{c^n}
  \limTOut
  \int \prod_{k=1}^n \diff[4]{y_k}
  \prod_{k=1}^n \hat{f}_j(y_k; t_{k,+}, \Delta t_{k,+}) 
  \, 
  \langle0| \, {\textstyle  T \prod_{k=1}^n \phi(y_k)} .
\end{align}
Therefore
\begin{multline}
\label{eq:f.i.coord.T}
 \langle g_1,\dots,g_n;\,\Out | f_1,\dots,f_{n'};\,\In \rangle
 =
  \frac{1}{c^n{c^*}^{n'}}
  \limT
  \int \prod_{j=1}^{n'} \diff[4]{x_j}
  \prod_{k=1}^n \diff[4]{y_k}
\times\\\times
  \left[ \prod_{k=1}^n \hat{g}_k(y_k; t_{k,+}, \Delta t_{k,+}) 
  \right]^* \,
  \prod_{j=1}^{n'} \hat{f}_j(x_j; t_{j,-}, \Delta t_{j,-}) \, 
  \langle 0 |\, T \,
  {\textstyle \prod_{k=1}^n \phi(y_k) \, \prod_{j=1}^{n'} \phi(x_j)} \, 
  |0\rangle
.  
\end{multline}
The limit of infinite time is in the future for the $t_{k,+}$ and in
the past for the $t_{j,-}$.  Since the times of $y_k$ are localized to
future times and the $x_j$ to past times, we were able to replace the
separate time-orderings of the $\phi(y_k)$ and of the $\phi(x_j)$ by
time-ordering of the whole operator product.  Thus the last factor is
a normal Green function of the theory.

Now the results of Sec.\ \ref{sec:wf.asy} show that for a
coordinate-space wave function the regions of slowest decrease in
time, $1/|t|^{3/2}$, correspond to the velocities for the region where
the momentum-space wave function is non-zero.  These regions give the
dominant contributions in the integrations over positions in
(\ref{eq:f.i.coord.T}).  These therefore correspond to the
trajectories of the classical particles corresponding to the wave
functions, as expected.  Other regions are at least power suppressed.

\subsection{Localization; momentum}
\label{sec:localization.momentum}

The states created by the $A_{f;\,\In}^\dagger$ operator are localized around
the asymptotic classical trajectory, because of the locations where the
field operator $\phi(t,\3x)$ is dominantly weighted by the wave function
factor.  Similar remarks apply for $A_{f;\,\Out}^\dagger$.  

Furthermore the construction of the operators was designed to produce
particle states with specified momentum content, with the momenta
on-shell.  It is useful to verify this explicitly.  We wish to capture in
properties of normalizable states the idea that the states
$|\3{p}_1,\dots, \3{p}_n;\,\In\rangle$ and $|\3{q}_1,\dots, \3{q}_n;\,\Out\rangle$ are
eigenstates of 4-momentum with the eigenvalues being the sum over particle
momenta.

The momentum operators $P^\mu$ are defined as usual from an integral
over the appropriate components of the Noether currents for
translations, which currents are the energy-momentum tensor $T^{\mu\nu}$.
Thus
\begin{equation}
  P^\mu = \int \diff[3]{\3x}  T^{0\mu}(t,\3x).
\end{equation}
To avoid an infinite-volume divergence due to a uniform energy density in
the vacuum, we define the energy-momentum tensor to have its
vacuum-expectation value subtracted, and then the 4-momentum of the vacuum
is zero.  

The result to be proved is that
\begin{equation}
\label{eq:P.mu.in}
  P^\mu |f_1,\dots,f_n;\,\In\rangle
  = |f^{P,\mu}_1,\dots,f_n;\,\In\rangle
    + |f_1,f^{P,\mu}_2,\dots,f_n;\,\In\rangle
    + \dots ,
\end{equation}
where $f^{P,\mu}_j$ are functions defined by
\begin{equation}
  f^{P,\mu}_j(\3p) = p^\mu f_j(\3p).
\end{equation}
For more general states, as in Sec.\ \ref{sec:generalization}, if
$|\underline{f};\,\In\rangle$ has momentum-space wave functions
$f_n(\3p_1,\dots,\3p_n)$, then $P^\mu|\underline{f};\,\In\rangle$ has wave
functions obtained by the replacement
\begin{equation}
  f_n(\3p_1,\dots,\3p_n)\mapsto f_n(\3p_1,\dots,\3p_n) \sum_{j=1}^n p_j^\mu.
\end{equation}
Similar results hold for $|g_1,\dots,g_n;\,\Out\rangle$ and
$|\underline{g};\Out\rangle$. 

To derive (\ref{eq:P.mu.in}), we first obtain the commutator of $P^\mu$
with $A_{f;\,\In}^\dagger$, by applying the commutator with the field,
\begin{equation}
  [P^\mu, \phi(x)] = \BDneg i \deriv[x_\mu]{\phi},
\end{equation}
in the formula for $A_f^\dagger(t;\Delta t)$ that is in the second line of
(\ref{eq:Af.dag.def.alt}).  An integration by parts gives
\begin{align}
    [P^\mu, A_f^\dagger(t;\Delta t)] ={}&
        \BDneg \frac{1}{c^*} \int \diff[4]{x'} \, \phi(x') \,
            \left\{ \deriv[t']{F(t'-t, \Delta t)} \, \deriv[x'_\mu]{f(x')}
                   +2 F(t'-t, \Delta t) \, \frac{\partial^2 f(x')}{\partial x'_\mu \partial t'}
\right.\nonumber\\&\hspace*{3cm}\left.
            +\delta^0_\mu\left[ \frac{\partial^2 F(t'-t, \Delta t)}{{\partial t'}^2} \, f(x')
                   +2 \deriv[t']{F(t'-t, \Delta t)} \, \deriv[t']{\partial f(x')}
               \right]
         \right\}.
\end{align}
Each derivative of $F$ with respect to $t'$ gives an extra factor of
$1/\Delta t$.  Thus in the limit $\Delta t\to\infty$, all the contributions with derivatives
of $F$ vanish, leaving
\begin{equation}
    [P^\mu, A_f^\dagger(t;\infty)] =
        \BDneg \frac{2}{c^*} \int \diff[4]{x'} \phi(x')
         F(t'-t, \Delta t) \frac{\partial^2 f(x')}{\partial x'_\mu \partial t'}.
\end{equation}
From the expression (\ref{eq:wf.coord.f}) of a coordinate-space wave
function in terms of a momentum-space wave function, we get
\begin{align}
  \deriv[x_\mu]{f(x)}
   & = \BDneg i \int \difftilde{p} p^\mu \tilde{f}(\ve{p}) \, e^{\BDneg ip\cdot x}
\nonumber\\
   & = \BDneg i \int \difftilde{p} \tilde{f}^{P,\mu}(\ve{p}) \, e^{-iE_{\3p}t + i \3p\cdot\3x}.
\end{align}
Hence 
\begin{equation}
    [P^\mu, A_f^\dagger(t;\infty)] = A_{f^{P,\mu}}^\dagger(t;\infty).
\end{equation}
That is, the commutator of $P^\mu$ with $A_f^\dagger(t;\infty)$ gives a creation
operator with the wave function $\tilde{f}(\3p)$ replaced by
$p^\mu\tilde{f}(\3p)$.

The same result applies when $t$ is taken to $-\infty$, i.e., to
$A_{f;\,\In}^\dagger$.

Given the definition (\ref{eq:i.state}) of $|f_1,\dots,f_n;\,\In \rangle$ in
terms of products of $A_{f_j;\,\In}^\dagger$ operators applied to the vacuum,
the desired result (\ref{eq:P.mu.in}) immediately follows for the action
of $P^\mu$ on $|f_1,\dots,f_n;\,\In \rangle$.  The corresponding result for
more general non-product states $|\underline{f};\,\In\rangle$ follows, since
such states can be obtained as linear combinations of product states,
with possible limit operations. 
The same arguments and results apply to the out-states
$|g_1,\dots,g_n;\,\Out \rangle$ and $|\underline{g};\,\Out\rangle$.

\subsection{Conversion to momentum space}

We now return to the matrix element between out and in states in
(\ref{eq:f.i.coord.T}), and express it in terms of momentum-space
Green functions.  For this we use the definition (\ref{eq:tilde.G.N})
of momentum-space Green functions, and the last lines of Eqs.\
(\ref{eq:Af.dag.def.alt}) and (\ref{eq:Af.def.alt}), which give
momentum-space versions of the definitions of $A_g$ and $A_f^\dag$
operators.  The result is:
\begin{align}
\label{eq:f.i.1}
  \langle g_1,\dots,g_n;\,\Out | f_1,\dots,f_{n'};\,\In \rangle
  = {}&
  \frac{1}{c^n{c^*}^{n'}} \,
  \limT
  \int \prod_{k=1}^n \difftilde{q_k} \prod_{j=1}^{n'} \difftilde{p_j}
  \left[ \prod_{k=1}^n \tilde{g}_k(\3q_k) 
  \right]^* \,
  \prod_{j=1}^{n'} \tilde{f}_j(\3p_j)
  \int \prod_{k=1}^n \frac{\diff{q_k^0}}{2\pi} \prod_{j=1}^{n'} \frac{\diff{p_j^0}}{2\pi}
\times \nonumber \\ & \hspace*{-4cm} \times
   \prod_{k=1}^n \xleft[
          (E_{\3q_k}+q_k^0)
            \, e^{i(E_{\3q_k}-q_k^0)t_{k,+}}
            \, \tilde{F}^*(q_k^0-E_{\3q_k}, \Delta t_{k,+})
         \right]
   \prod_{j=1}^{n'} \xleft[
            (E_{\3p_j}+p_j^0)
            \, e^{-i(E_{\3p_j}-p_j^0)t_{j,-}}
            \, \tilde{F}(p_j^0-E_{\3p_j}, \Delta t_{j,-}) 
         \right]
\times \nonumber \\ & \hspace*{-4cm} \times
   \, G_{n+n'}(-q_1, \dots, -q_n, p_1, \dots, p_{n'})
.
\end{align}
As intended, the $\tilde{F}$ factors restrict the energy components of
the external momenta ($p_j$ and $q_k$) of the Green function to be
within order $1/\Delta t_{j,\pm}$ of their on-shell values, so that the
momenta are exactly on-shell in the limit $\Delta t_{j,\pm}\to\infty$.  The phase
factors involving $t_{j,-}$ and $t_{k,+}$ give rapid oscillations as
functions of the energy variables $p_j^0$ and $q_k^0$.  They would
give a strong suppression in the infinite-time limit were it not for
the mass-shell poles on external lines of the Green function.  It is
the combination of the oscillations and the poles that will result in
a non-zero limit.

The Green function $G_{n+n'}$ can be decomposed into a sum of terms
each with different numbers of connected components, with a
momentum-conservation delta function for each component.  \emph{To
  reduce the combinatorial complexity of the analysis of the different
  cases, we now restrict to the standard case of two incoming
  particles, $n'=2$.}

The possible cases for connected components of $G_{n+2}$ are
\begin{itemize}
\item A fully connected term.  This will give the expected $2\to n$
  scattering term.
\item When the number of outgoing particles is $n=2$, there are terms
  which connect each of the incoming lines to one of the outgoing
  lines, one with $p_1$ to $q_1$, $p_2$ to $q_2$, and one with $p_1$
  to $q_2$, $p_2$ to $q_1$.  These will give the no-scattering term,
  i.e., the contribution corresponding to the $\delta_{\beta\alpha}$ term in Eq.\
  (\ref{eq:S.1+iT}).
\item All the remaining cases have at least one component that
  connects only incoming to incoming lines, or only outgoing to
  outgoing lines, or one single incoming particle to two or more
  outgoing particles.  For these components, the energy conservation
  condition cannot be satisfied in the on-shell limit, given that the
  particles are asymptotic particles that are stable by definition,
  and that we have restricted attention to the case that all the
  particles are massive.  Hence all these disconnected terms result in
  a zero contribution to $\langle g_1,\dots,g_n;\,\Out | f_1,f_2;\,\In \rangle$.
  These restrictions are imposed by the $\tilde{F}$ functions in the
  $\Delta t_{j,\pm}\to\infty$ limit.
\end{itemize}
If we changed the number of incoming particles from 2 to a higher
value $n'>2$, there would be further possibilities.  Their treatment
merely involves a mechanical extension from the case of $n'=2$, but
with combinatorial complications.

\subsection{Fully connected term}
\label{sec:derive.conn}

To analyze the fully connected term in the Green function
$\tilde{G}_{n+2}$, we factor it as in Eq.\ (\ref{eq:G.Gamma}) into an
amputated part, a product of external propagators, and a delta
function for momentum conservation.  In the limit we use, the
$\tilde{F}$ factors restrict the energies to be within order $1/\Delta
t_{j,\pm}$ of the on-shell values, with the relevant instance of $\Delta
t_{j,\pm}$.

Hence we can replace the values of energies by on-shell energies in
every factor in (\ref{eq:f.i.1}) that is smooth as a function of the
energies.  That is, wherever possible the momenta
$p_1,p_2,q_1,\dots,q_n$ are replaced by on-shell values.  The factors
where this cannot be done have rapid variation near the on-shell
position.  These factors are the poles in the external propagators,
the factors of $e^{-i(E_{\3p_j}-p_j^0)t_-} \tilde{F}(p_j^0-E_{\3p_j},
\Delta t)$ for the incoming lines, the corresponding factors for the
outgoing line, and the delta function for energy conservation.

Were it not for the delta function, there would be independent
integrals over $n+2$ energy variables, and after the approximations,
each integral would have the form
\begin{equation}
\label{eq:h}
  h(t_0,\Delta t) = \int \frac{\diff{\delta E}}{2\pi} e^{-i\delta E \, t_0}
              \tilde{F}(\delta E,\Delta t) \frac{i}{\delta E+i\epsilon}.
\end{equation}
Here the variable of integration, $\delta E$, is chosen to be the deviation
of an energy from an on-shell value, i.e., $p_j^0-E_{\3p_j}$ or
$q_k^0-E_{\3q_k}$.  The $\delta E+i\epsilon$ denominator is from a propagator
pole.  The parameters are $t_0$, which is one of $-t_{j,-}$ or
$t_{k,+}$, and $\Delta t$, which is one of $\Delta t_{j,-}$ or $\Delta t_{j,+}$.
This integral goes to unity in the relevant infinite-time limit:
\begin{equation}
  \label{eq:h.lim}
  \lim_{\substack{ t_0 \to \infty\\ 
                   \Delta t \to \infty\\ 
                   \Delta t/t_0 \to 0 }}  
  h(t_0,\Delta t) = 1.
\end{equation}
This can be proved by shifting the contour of integration from the
real axis slightly into the lower-half plane, with an imaginary part
for $\delta E$ of order $-1/\Delta t$.  The deformation crosses the pole, and
the residue contribution gives unity.  On the deformed contour, the
deformation is small enough not to change the order of magnitude of
the $\tilde{F}$ factor.  But the exponential factor gives a
suppression of by a factor of order $e^{-t_0/\Delta t}$, which goes to zero
in the stated limit.  There remains the unit term from the pole's
residue.

However, the delta function constrains the energy integrals, which
therefore appear not to be independent.  Nevertheless, as we will see,
it turns out to be correct to replace the energies in the delta
function $\delta\xleft(\sum p_j^0-\sum q_k^0\right)$ by on-shell values, as in
$\delta\xleft(\sum E_{\3p_j}-\sum E_{\3q_k}\right)$.  This will immediately lead
to the reduction formula in the form (\ref{eq:LSZ.2}).  But because
the delta function has rapid variations in some directions, a more
detailed derivation is needed.

Let us define deviation variables
\begin{equation}
    \delta p_j^0 = p_j^0-E_{\3p_j}, \quad \delta q_k^0 = q_k^0-E_{\3q_k},
\end{equation}
and a quantity
\begin{multline}
\label{eq:H.def}
  H(\delta q_1^0,\dots,\delta q_n^0, \delta p_1^0,\delta p_2^0)
  =
  (c^*)^n c^2\, 
  \prod_{k=1}^n \difftilde{q_k} \int \prod_{j=1}^2 \difftilde{p_j}
  \left[ \prod_{k=1}^n \tilde{g}_k(\3q_k) 
  \right]^* \,
  \prod_{j=1}^2 \tilde{f}_j(\3p_j)
\times \\ \hspace*{1cm} \times
   (2\pi)^4\delta^{(3)}\xleft(\sum\3p_j-\sum\3q_k\right) \, \delta\xleft(X+\sum E_{\3p_j}-\sum E_{\3q_k}\right)
   \, \Gamma_{n+2}(-q_1, \dots, q_{n-1}, p_1, p_2)  
\end{multline}
that contains all the smooth dependence on the energy variables
together with the energy-conservation delta function.  Here $X=\sum\delta
p_j^0-\sum\delta q_k^0$, and the momenta in the amputated Green function are
set as follows: $p_j^0=\delta p_j^0+E_{\3p_j}$ and $q_k^0=\delta
q_k^0+E_{\3q_k}$.  When all the deviation variables are set to zero,
$H$ is of the form of the integral of wave functions with the S-matrix
that is given by the right-hand side of (\ref{eq:LSZ.2}). Our task is
to prove that this quantity does in fact equal the left-hand side of
Eq.\ (\ref{eq:f.i.1}), and hence that the reduction formula is
correct.

We use the change of variable to allow us to apply the
energy-conservation delta function to the 3-momentum variables instead
of the energy variables.  Then we use (\ref{eq:prop.residue}) for each
external propagator pole, and find that Eq.\ (\ref{eq:f.i.1}) gives
\begin{multline}
\label{eq:main.int}
  \langle g_1,\dots,g_n;\,\Out | f_1,f_2;\,\In \rangle_{\text{conn}}
  =
  \limT
  \int \prod_{k=1}^n \frac{\diff{q_k^0}}{2\pi} \prod_{j=1}^2 \frac{\diff{p_j^0}}{2\pi}
  \, H(\delta q_1^0,\dots,\delta q_n^0, \delta p_1^0,\delta p_2^0)
\times\\\times
   \prod_{k=1}^n \xleft[
            \frac{i}{\delta q_k^0+i\epsilon}
            \, e^{-i\delta q_k^0t_{k,+}}
            \, \tilde{F}^*(\delta q_k^0, \Delta t_{k,+})
         \right]
   \prod_{j=1}^2 \xleft[
            \frac{i}{\delta p_j^0+i\epsilon}
            \, e^{-i\delta p_j^0(-t_{j,-})}
            \, \tilde{F}(\delta p_j^0, \Delta t_{j,-}) 
         \right]
.
\end{multline}
Now $H$ is smooth as the energy-deviation variables
$\delta p_1^0,\dots,\delta q_n^0$, and hence $X$, go to zero.  We can therefore
set the energy deviation variables to zero in $H$. The $n+2$
energy-deviation variables are now independently integrated.  So we
apply Eq.\ (\ref{eq:h.lim}) to each of the integrals, to obtain
\begin{align}
  \langle g_1,\dots,g_n;\,\Out | f_1,f_2;\,\In \rangle_{\text{conn}}
  = {}&
  \int \prod_{k=1}^n \difftilde{q_k} \prod_{j=1}^2 \difftilde{p_j}
  \left[ \prod_{k=1}^n \tilde{g}_k(\3q_k) 
  \right]^* \,
  \prod_{j=1}^2 \tilde{f}_j(\3p_j)
  \, S_{\3q_1,\dots,\3q_n;\,\3{p}_1,\3{p}_2} ,
\end{align}
where $S_{\3q_1,\dots,\3q_n;\,\3{p}_1,\3{p}_2} = H(0,\dots,0)$.  This
is of the form of the defining equation (\ref{eq:g.f}) of the
S-matrix, with the S-matrix now proved to be given by the already
stated form in Eq.\ (\ref{eq:LSZ.2}), as regards the connected component.

This completes the proof of the reduction formula for the connected
component of the S-matrix.

\subsection{No-scattering term}
\label{sec:derive.no.scatt}

We now consider the contribution of disconnected parts of the Green
function to the $2\to2$ S-matrix.  This part of the Green function is
\begin{equation}
  \tilde{G}_{4,\text{no scatt.}}(-q_1,-q_2,p_1,p_2)
  =
    \prod_{j=1}^2 \xleft[ \hat{G}_2(p_j^2) \, (2\pi)^4\delta^{(4)}(p_j-q_j)
  \right]
  + \, \mbox{Term with $q_1$ and $q_2$ exchanged}.
\end{equation}
We treat each factor separately, and each gives a contribution to 
$\langle g_1,g_2;\,\Out | f_1,f_2;\,\In \rangle$ of the form
\begin{equation}
\label{eq:main.int.no.scatt}
    \frac{1}{|c|^2} \,
  \limT
  \int \difftilde{p} 
  \, \left[ \tilde{g}(\3p) \right]^* \,
  \tilde{f}(\3p)
  \int \frac{\diff{p^0}}{2\pi} 
            \frac{ (E_{\3p}+p^0)^2 }{ 2E_{\3p} }
            \, e^{-i(p^0-E_{\3p})(t_+-t_-)}
            \, \tilde{F}^*(p^0-E_{\3p}, \Delta t_+)
            \, \tilde{F}(p^0-E_{\3p}, \Delta t_-)
   \, \hat{G}_2(p^2) 
.
\end{equation}
Use of Eq.\ (\ref{eq:h.lim}) show that this equals
\begin{equation}
  \langle g|f\rangle = \int \difftilde{p} 
            \, \left[ \tilde{g}(\3p) \right]^* \, \tilde{f}(\3p).
\end{equation}
Hence the no-scattering term for $\langle g_1,g_2;\,\Out | f_1,f_2;\,\In \rangle$ is
\begin{equation}
  \langle g_1,g_2;\,\Out | f_1,f_2;\,\In \rangle_{\rm no scatt.}
  = \langle f_1|g_1\rangle \langle f_2|g_2\rangle + \langle f_1|g_2\rangle \langle f_2|g_1\rangle,
\end{equation}
which is exactly the expected non-scattering term in the S-matrix.

\end{widetext}

\subsection{More general cases}

The above derivations get the connected and disconnected components of
the S-matrix for the case of $n'=2$ incoming particles.  Exactly the
same principles apply to other cases ($n'=1$ and $n'>2$).

For general values of $n'$ and $n$, there is a correspondence between
the connected components of Green functions and the connected
components of the S-matrix.  Each kind of object is a sum over all
possibilities for a product over connected components.  Each connected
component has a delta function for 4-momentum conservation.  However,
only those connected components of Green functions that can obey the
constraint of conservation of on-shell 4-momenta give non-zero
contributions.  Other terms give zero, of which an example, already
referred to, is a component that connects only incoming lines to
incoming lines, but not to outgoing lines.

For each connected component, either a version of the proof and
formula for the $2\to n$ connected component applies, or the method of
Sec.\ \ref{sec:derive.no.scatt} applies to a component with one
incoming and one outgoing line.

Another important situation is for matrix elements of an operator or a
time-ordered product of operators between between in- and out-states,
of the kind shown in (\ref{eq:OME}).  The method of derivation of the
reduction formula works equally well here.  The only change in the
proof that is needed is to insert the factors of extra operators into
the vacuum matrix element on the right of Eq.\ (\ref{eq:f.i.coord.T}),
between the $\phi(y_k)$ and the $\phi(x_j)$.  The remaining manipulations
all go through unchanged.

\section{Verification of properties of creation and annihilation
  operators} 
\label{sec:verify}

In a sense, the reduction formula both for the S-matrix and for matrix
elements of operators between in- and out-states has provided a
convenient way of formulating a solution of a QFT.  Then, in
accordance with the principles laid out in Sec.\
\ref{sec:logical.structure}, it is necessary to verify that the
solution has the properties attributed to it that were the basis of
the derivations.  In particular, we need to show that the creation and
annihilation operators defined in Sec.\ \ref{sec:Afdag.def} actually
self-consistently obey the properties that the definitions were
intended to provide.  These properties include their commutation
relations, and also that the limits defining the operators can be
taken as strong limits.  Underlying all the derivations are
established properties of time-ordered Green functions and of their
Fourier transforms into momentum space.  The properties are certainly
valid to all orders of perturbation theory.

We must first ensure that the limits of the annihilation and creation
operators exist as strong limits.  This is the biggest difference
compared with the standard LSZ formulation. Then other derivations,
e.g., of the commutation relations, are routine, unlike the case when
only weak limits exist.

Readers particularly concerned about rigor would probably look for yet
further properties to verify.

\subsection{Strong limit for in-creation operator}

By definition, saying that the strong limit exists for the in-creation
operators means that
\begin{equation}
\label{eq:strong.state}
  \left\| A_f^\dagger(t;\Delta t) \, | f_1,\dots,f_{n'};\,\In\rangle
     \,-\, | f, f_1,\dots,f_{n'};\,\In \rangle \right\|^2
\end{equation}
goes to zero when the standard limit of infinite past time is taken.

Following the observations in App.\ \ref{sec:limits}, a useful method
for showing that the strong limit exists for $\limTIn A_f^\dagger(t,\Delta t)$
starts from applying the reduction formula to obtain matrix elements
with basis out states:
\begin{equation}
\label{eq:q.Afdag.f}
    \langle \3q_1,\dots,\3q_n;\,\Out| \, A_f^\dagger(t;\Delta t) \, | f_1,\dots,f_{n'};\,\In \rangle.
\end{equation}
The expected expected limit is
\begin{equation}
\label{eq:q.ff}
    \langle \3q_1,\dots,\3q_n;\,\Out| f, f_1,\dots,f_{n'};\,\In \rangle. 
\end{equation}
The quantity in (\ref{eq:strong.state}) is the same as 
\begin{widetext}
\begin{equation}
\label{eq:strong.sum}
  \sum_n \frac{1}{n!} \, \int \prod_{k=1}^n \difftilde{q_k} 
  \,\left| 
     \langle \3q_1,\dots,\3q_n;\,\Out| \, A_f^\dagger(t;\Delta t) \, | f_1,\dots,f_{n'};\,\In\rangle
     - \langle \3q_1,\dots,\3q_n;\,\Out| f, f_1,\dots,f_{n'};\,\In \rangle
\right|^2,
\end{equation}
and we will show that this goes to zero in the limit of infinite time in
the past.

Generally in the cases of interest here, the difference between having
a strong limit and a weak limit arises because of time-dependent phase
factors such as we found in (\ref{eq:afdag.1.3}).  If the
infinite-time limit depends on a suppression obtained by integrating
that phase over final-state momenta in a matrix element with a wave
function for a normalizable out-state, then the limit is weak.  If
instead the integral with a smooth wave function is not needed to get
the limit, then the infinite-time limit exists in the matrix element
with a basis out-state.  Any remaining phase factor cancels in the
absolute square of the matrix element in (\ref{eq:strong.sum}), and
the limit is strong.

Now from the momentum-space formula in (\ref{eq:Af.dag.def.alt}) for
$A_f^\dagger(t;\Delta t)$, we get
\begin{align}
\label{eq:Adag.ME.q}
    \langle \3q_1,\dots,\3q_n;\,\Out| \, A_f^\dagger(t;\Delta t) \, | f_1,\dots,f_{n'};\,\In \rangle
\hspace*{-3cm}&
\nonumber\\
    ={}&
    \frac{1}{c^*} \int \frac{\diff[4]{p}}{(2\pi)^4} 
            \, \tilde{f}(\3p)
            \, \tilde{F}(p^0-E_{\3p}, \Delta t)
            \, \frac{E_{\3p}+p^0}{2 E_{\3p}}
            \, e^{-i(E_{\3p}-p^0)t}
    \, \langle \3q_1,\dots,\3q_n;\,\Out| \, \tilde{\phi}(p) \, | f_1,\dots,f_{n'};\,\In \rangle
\nonumber\\
    ={}&
    \frac{1}{c*} 
    \int \prod_{j=1}^{n'} \difftilde{p_j} \prod_{j=1}^{n'} \tilde{f}_j(\3p_j)
    \int \frac{\diff[4]{p}}{(2\pi)^4} 
        \, \tilde{f}(\3p)
        \, \frac{E_{\3p}+p^0}{2 E_{\3p}}
        \, e^{-i(E_{\3p}-p^0)t}
        \, \tilde{F}(p^0-E_{\3p}, \Delta t)
\times \nonumber\\& \hspace*{7cm} \times 
    \, \langle \3q_1,\dots,\3q_n;\,\Out| \, \tilde{\phi}(p) \, | \3p_1,\dots,\3p_{n'};\,\In \rangle.
\end{align}
In previous sections, we have generally taken care to write matrix
elements with normalizable states, i.e., with wave packet states.  Now we
have a matrix element that has states of particles of definite momentum.
For our discussion, it is \emph{defined} to be the momentum-space quantity
that appears when the reduction formula is used to express $\langle
g_1,\dots,g_n;\,\Out| \, \tilde{\phi}(p) \, | f_1,\dots,f_{n'};\,\In \rangle$ in
terms of the momentum-space Green functions
$G_{n'+n+1}(-q_1,\dots,-q_n,p_1,\dots,p_{n'},p)$.

We decompose by connected components.  Some examples of the connectivity
of graphs for $\langle \3q_1,\dots,\3q_n;\,\Out| \, \tilde{\phi}(p) \,
|\3p_1,\dots,\3p_{n'};\,\In \rangle$ are shown in Fig.\ \ref{fig:Afdag.ME}.
There is one component where all of $\tilde{\phi}(p)$ and the lines for the
external particles are connected.  Added to this is a sum over terms each
of which is a product of two or more smaller connected components.  In
each term, one component contains the field $\tilde{\phi}(p)$, and the
remainder are equivalent to connected factors that also appear in the
S-matrix with some subset of the $n+n'$ particles used here.

Observe that the momentum $p$ is defined to flow \emph{into} the Green
function at the external vertex for the Fourier transformed field
$\tilde{\phi}(p)$, and that the Green function has a pole when
$p^2=m_{\rm phys}^2$.

\begin{figure*}
  \centering
  \setlength\tabcolsep{5mm}
  \begin{tabular}{ccc}
  \includegraphics[width=4cm]{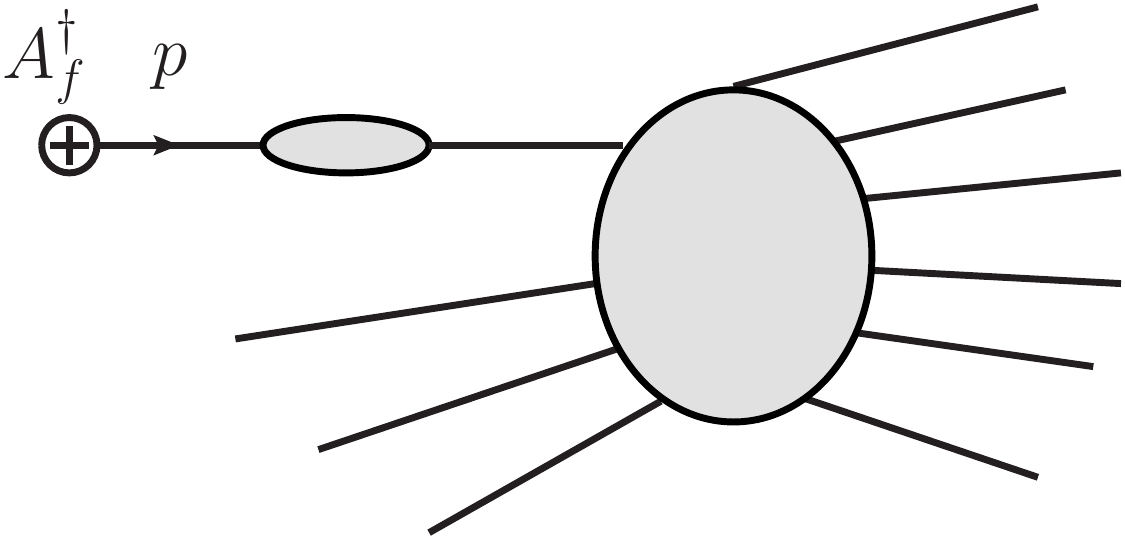} &
  \includegraphics[width=4cm]{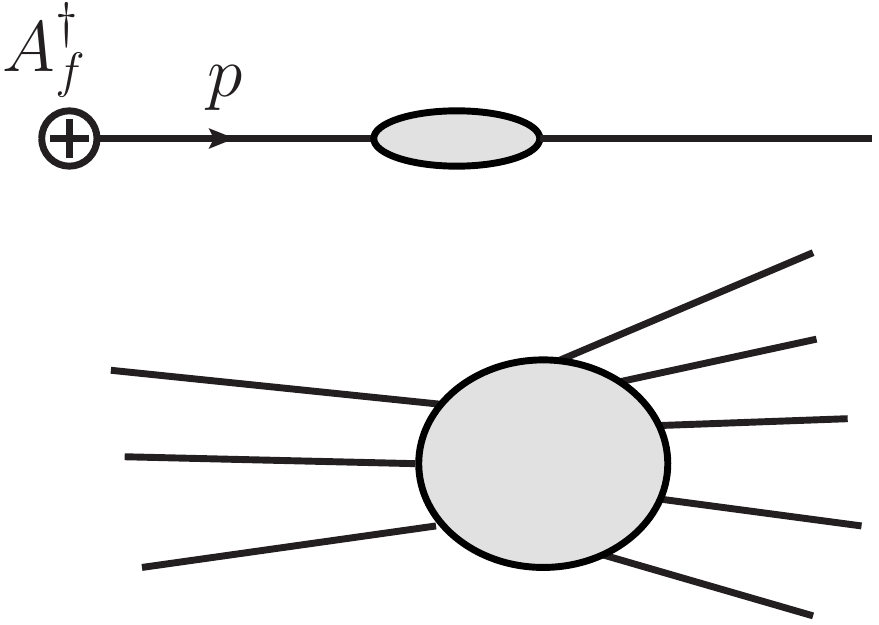} &
  \includegraphics[width=4cm]{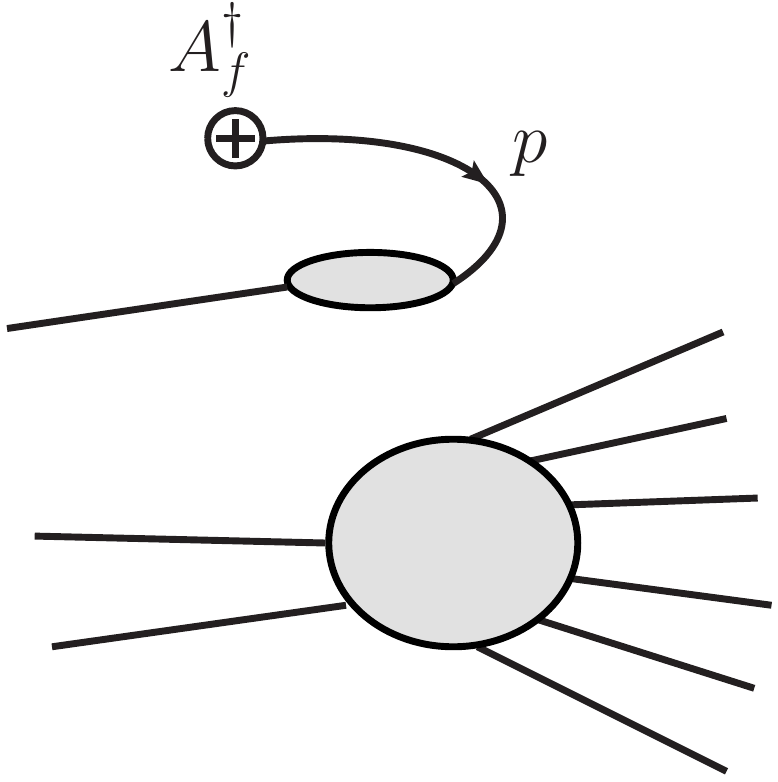} 
  \\
  (a) & (b) & (c)
  \end{tabular}
  \\*[10mm]
  \begin{tabular}{cc}
  \includegraphics[width=4cm]{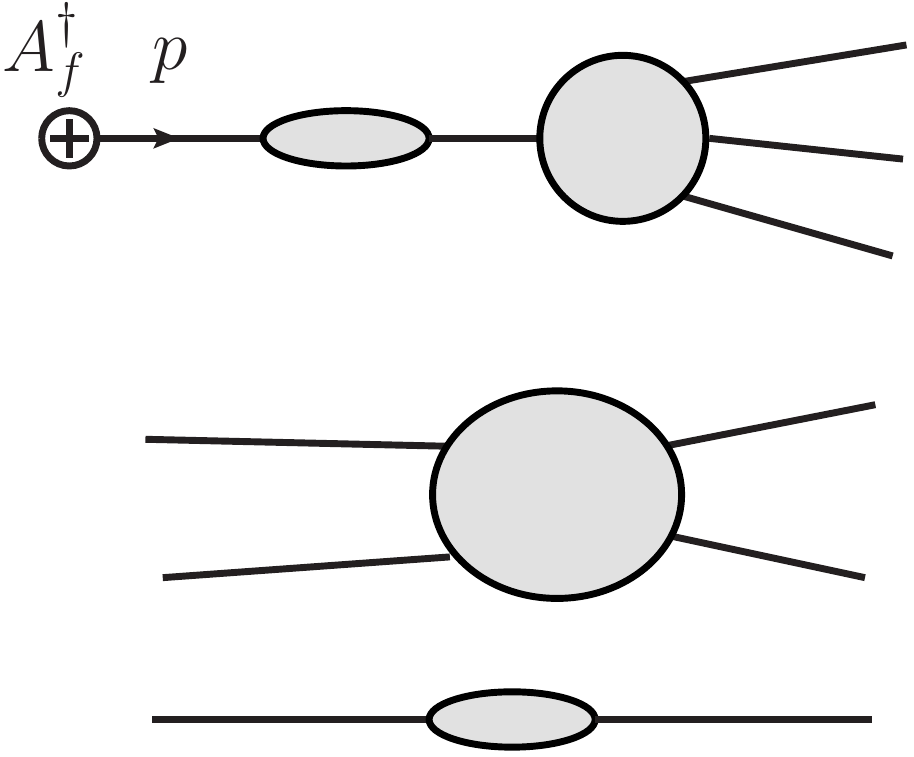} &
  \includegraphics[width=4cm]{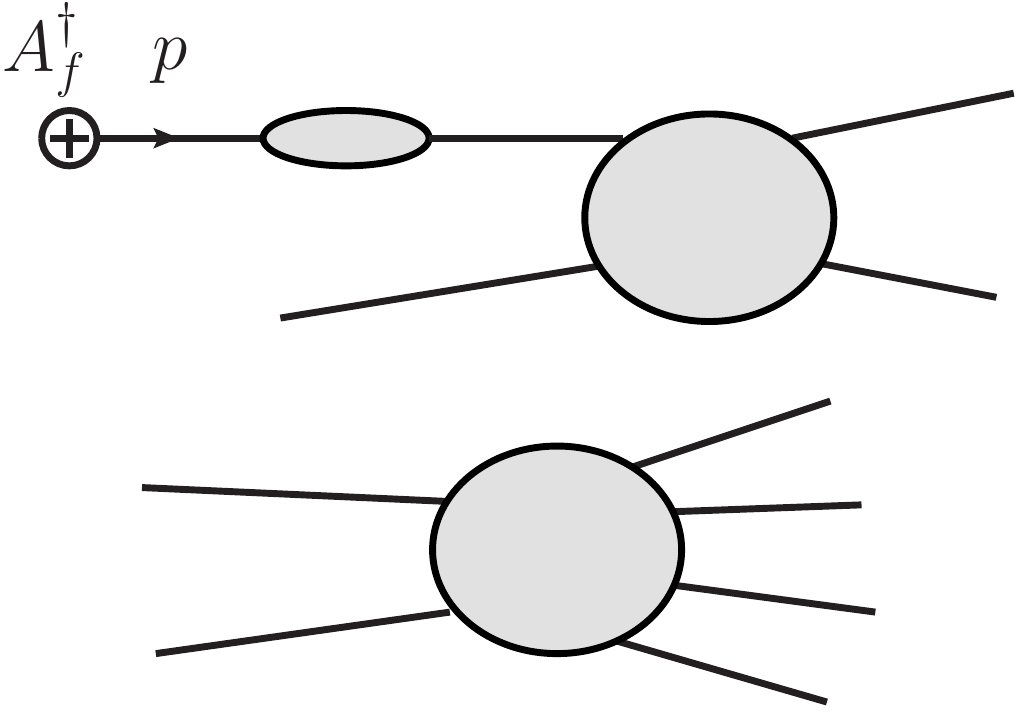} 
  \\
  (d) & (e)
  \end{tabular}
  \caption{Showing examples of connectivity structure for a matrix
    element of $A_f^\dagger(t;\Delta t)$ and the Green function corresponding to
    it, with $n'=3$ and $n=6$.}
    \label{fig:Afdag.ME}
\end{figure*}

First, consider a component connected to $\tilde{\phi}(p)$ that has one
or more particles in the initial state.  The $\tilde{F}$ factor
enforces that up to strongly suppressed contributions, $p$ is close to
mass shell; the suppression is the same for all final states, so it
will also apply to the sum and integral over states in
(\ref{eq:strong.sum}).  It also ensures that components without at
least two particles in the final-state are suppressed, by the
requirements for a momentum-conserving on-shell process.

The momentum conservation delta function removes the integral over
$p$, leaving an integral over the $p_j$.  Then the phase factor is
\begin{equation}
   e^{-i(E_{\3p}-p^0)t}
   =
   \exp\xleft[ -i t \left( E_{\sum\3q_k - \sum\3p_j}
                        - \sum E_{\3q_k} + \sum E_{\3p_j}
                   \right)
       \right].
\end{equation}
The argument in Sec.\ \ref{sec:derive.conn} shows that a small deformation
on the integral over the initial state momenta gives the relevant
connected component of (\ref{eq:q.ff}) as the limit.  There are bounds on
the errors that are uniform in the final state.

The remaining case is where there is a connected component with zero
particles in the initial state.  This corresponds to the example
calculations that we performed in Secs.\ \ref{sec:extra.particles} and
\ref{sec:Af.ME.simple} for the LSZ operators and for the modified
operators.  When there is more than one particle in the final state,
e.g., the first factor in Fig.\ \ref{fig:Afdag.ME}(d), we get a
strong suppression caused by the $\tilde{F}$ factor, unlike the case
for the LSZ operator. The suppression is uniform in the final state.

There remains the expected one-particle case, as in the first factor
in Fig.\ \ref{fig:Afdag.ME}(b). Momentum conservation between $p$
and the single initial particle ensures that the phase factor and
the $\tilde{F}$ factor are both unity.

The overall result is then a strong limit:
\begin{equation}
 A^\dag_{f;\,\In} \, | f_1,\dots,f_{n'};\,\In \rangle,
 =
 \limTIn A_f^\dagger(t;\Delta t) \, | f_1,\dots,f_{n'};\,\In \rangle,
 =
 | f, f_1,\dots,f_{n'};\,\In \rangle.
\end{equation}

A similar derivation applies to the final state for the creation
operators for outgoing particles.

\subsection{Strong limit for in-annihilation operator}

Next we consider the operator $A_f(t,\Delta t)$ in the infinite past, where
it is intended to annihilate one particle.  The equation for the
matrix element is obtained by a minor modification of (\ref{eq:Adag.ME.q}):
\begin{align}
\label{eq:A.ME.q}
    \langle \3q_1,\dots,\3q_n;\,\Out| \, A_f(t;\Delta t) \, | f_1,\dots,f_{n'};\,\In \rangle
\hspace*{-3cm}&
\nonumber\\
    ={}&
    \frac{1}{c} \int \frac{\diff[4]{q}}{(2\pi)^4} 
            \, \tilde{f}^*(\3q)
            \, \frac{E_{\3q}+q^0}{2 E_{\3p}}
            \, e^{i(E_{\3q}-q^0)t}
            \, \tilde{F}^*(q^0-E_{\3q}, \Delta t)
    \, \langle \3q_1,\dots,\3q_n;\,\Out| \, \tilde{\phi}(-q) \, | f_1,\dots,f_{n'};\,\In \rangle
\nonumber\\
    ={}&
    \frac{1}{c} 
    \int \prod_{j=1}^{n'} \difftilde{p_j} \prod_{j=1}^{n'} \tilde{f}_j(\3p_j)
    \int \frac{\diff[4]{q}}{(2\pi)^4} 
        \, \tilde{f}^*(\3q)
        \, \frac{E_{\3q}+q^0}{2 E_{\3q}}
        \, e^{i(E_{\3q}-q^0)t}
        \, \tilde{F}^*(q^0-E_{\3q}, \Delta t)
\times \nonumber\\& \hspace*{7cm} \times 
    \, \langle \3q_1,\dots,\3q_n;\,\Out| \, \tilde{\phi}(-q) \, | \3p_1,\dots,\3p_{n'};\,\In \rangle.
\end{align}
Here the momentum $q$ is defined to flow \emph{out} of the
corresponding Green function, as appropriate for an operator that is
intended to destroy one of the initial-state particles.  We will
highlight the changes with the calculation for the creation operator.
Unchanged is that $\tilde{F}$ strongly restricts $q$ to on-shell
momentum, but now with positive energy flowing out of the Green
function.  Examples of the connectivity structure are shown in Fig.\
\ref{fig:Af.ME}.

\begin{figure*}
  \centering
  \setlength\tabcolsep{5mm}
  \begin{tabular}{ccc}
  \includegraphics[width=4cm]{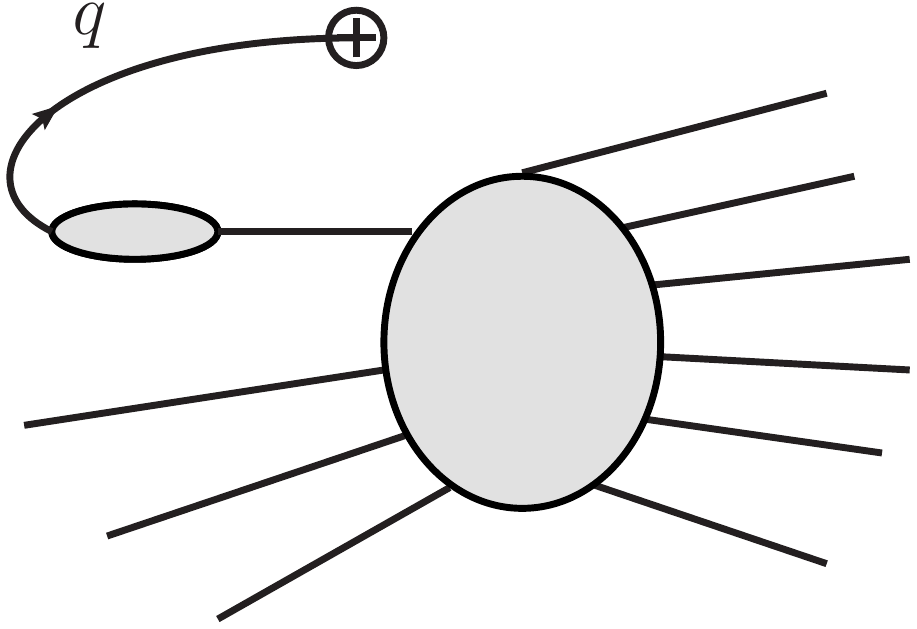} &
  \includegraphics[width=4cm]{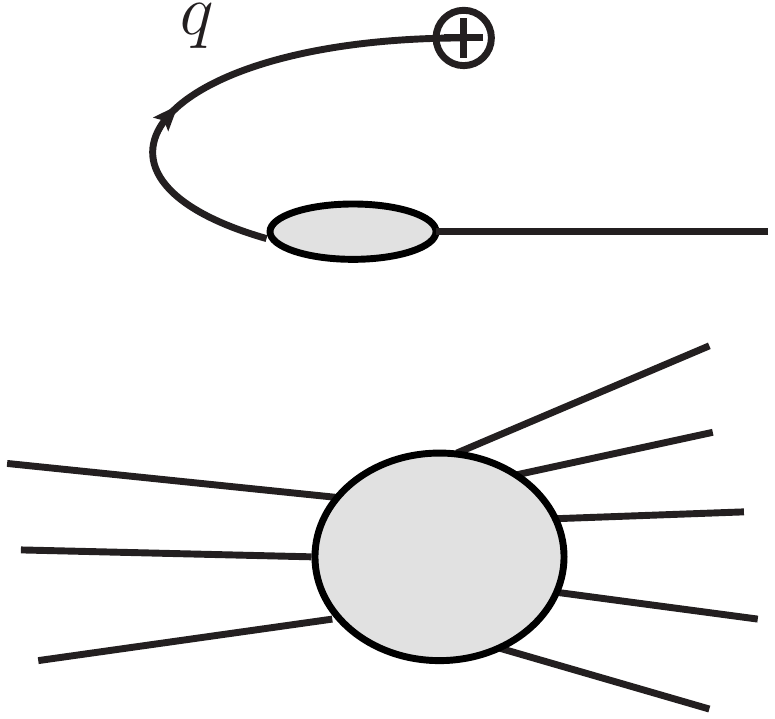} &
  \includegraphics[width=4cm]{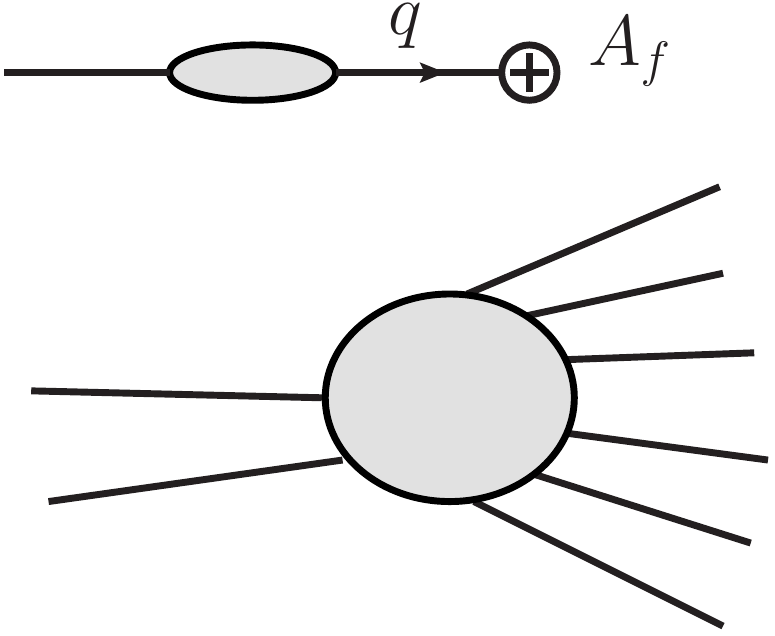} 
  \\
  (a) & (b) & (c)
  \end{tabular}
  \\*[10mm]
  \begin{tabular}{cc}
  \includegraphics[width=4cm]{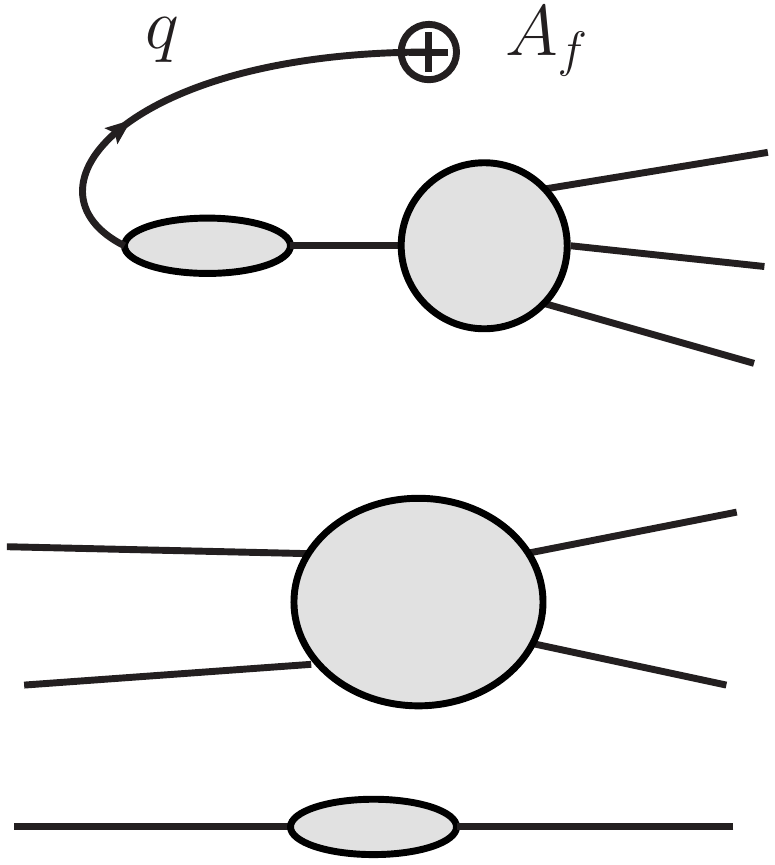} &
  \includegraphics[width=4cm]{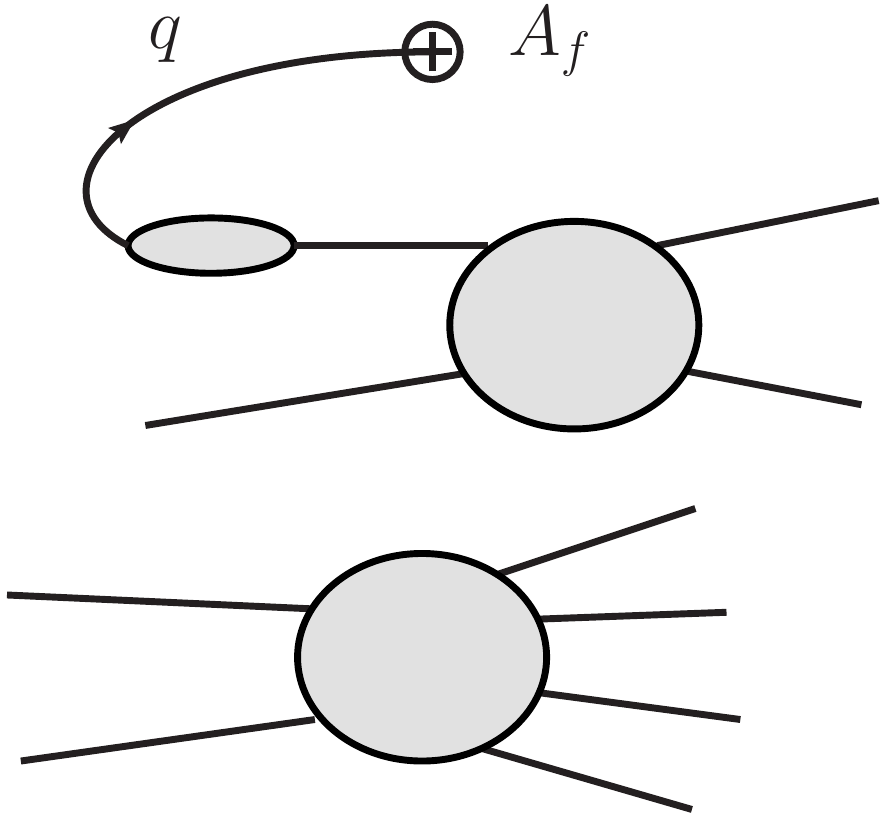} 
  \\
  (d) & (e)
  \end{tabular}
  \caption{Showing examples of connectivity structure for a matrix
    element of $A_f(t;\Delta t)$ and the Green function corresponding to
    it, with $n'=3$ and $n=6$. This is the same as Fig.\
    \ref{fig:Afdag.ME}, except that graphs are reoriented for the
    vertex for $A_f$ to emphasize its purpose as an annihilation
    operator.}
  \label{fig:Af.ME}
\end{figure*}

Consider first a connected component with at least one initial
particle and at least one final particle.

In the derivation of the reduction formula we used this annihilation
operator at large \emph{positive} time.  The contour deformation to
make $e^{i(E_{\3q}-q^0)t}$ suppressed on the deformed contour crossed
the pole in the propagator, and picked out the pole contribution.  But
now with the opposite sign of $t$, the contour deformation is in the
opposite direction, so the contribution is suppressed, with a
suppression uniform in the final state.  This result is illustrated by
the following limit of Eq.\ (\ref{eq:h}):
\begin{equation}
  \lim_{\substack{ t \to -\infty\\ 
                   \Delta t \to \infty\\ 
                   \Delta t/|t| \to 0 }}  
  \int \frac{\diff{\delta E}}{2\pi} e^{-i\delta E \, t}
              \tilde{F}(\delta E,\Delta t) \frac{i}{\delta E+i\epsilon}
  =0.
\end{equation}

All other connected factors have no particles in either the initial or
the final states.  The $\tilde{F}$ factor suppresses all of them with
one exception.  The exception is a component with one initial particle
and no final particle, i.e., of the form
\begin{equation}
  \langle0| \, A_f(t;\Delta t) \, | f_1 ;\,\In \rangle,
\end{equation}
illustrated by the first factor in Fig.\ \ref{fig:Af.ME}(c).

Because the single particle is on-shell, there is no longer a phase
factor. Instead we just get $\int\difftilde{p} \tilde{f}^*(\3p)
\tilde{f}_1(\3p)$, as in the calculations in Sec.\
\ref{sec:Af.ME.simple}. Observe that this is even independent of $t$
and $\Delta t$.

The overall result is that in the limit of infinite time in the past,
the operator annihilates one incoming particle, and that the limit is
a strong limit:
\begin{equation}
 A_{f;\,\In} \, | f_1,\dots,f_{n'};\,\In \rangle
 =
 \limTIn A_f(t;\Delta t) \, | f_1,\dots,f_{n'};\,\In \rangle
 =
 \sum_{j=1}^{n'}
 | f_1,\dots,f_{n'}, \text{with $f_j$ omitted};\,\In \rangle
 \int\difftilde{p} \tilde{f}^*(\3p) \tilde{f}_j(\3p).
\end{equation}

\end{widetext}

\subsection{Comparison with LSZ}

Let us now compare with the situation for the LSZ operators
$a_f^\dagger(t)$.  For the LSZ operators, the on-shell condition for a
created or annihilated particle, i.e., for $p$ or $q$, was obtained
from an integration over a rapidly oscillating phase factor identical
with the one in (\ref{eq:Adag.ME.q}) or (\ref{eq:A.ME.q}).  With the
new operators, the on-shell condition is imposed more robustly by the
$\tilde{F}$ factors.  (The comparison between the two cases is
assisted by observing that matrix elements of the LSZ operators can be
obtained from those for the new operators simply by omitting the
$\tilde{F}$ factor, or alternatively by setting $\Delta t=0$.)

In all but one case, the suppression in the LSZ case can be obtained
by integrating over the momenta in the initial state.  The one
exception is where we have a connected component with no initial-state
particles.  This is exactly the situation explored by an explicit
calculation in Sec.\ \ref{sec:3.afdag.0}, and more generally in Sec.\
\ref{sec:div.spectral}.  In this case, momentum conservation between
$p$ or $q$ and the external particles results in a phase factor that
depends only on final-state momenta.  Then an integral over
final-state momenta is needed to get a suppression from the
oscillations in the phase factor.  Without the integral, there is no
suppression, and hence we have a weak limit, but not a strong limit
for the operators: The phase factor cancels in the integral over final
states in (\ref{eq:strong.sum}).

\subsection{Commutation relations}

Given now that the strong limits exist for the annihilation and
creation operators, and that they have the expected action on
in-states, it is elementary to derive the commutators.

For $A_{f;\,\In}^\dagger$ and $A_{g;\,\In}^\dagger$ we have
\begin{equation}
  A_{f;\,\In}^\dagger A_{g;\,\In}^\dagger | f_1,\dots,f_{n'};\,\In \rangle
  = | f, g, f_1,\dots,f_{n'};\,\In \rangle .
\end{equation}
Now the states are symmetric under exchange of the labels (as follows
via the reduction formula from a corresponding property of Green
functions).  So we get the same result by applying the operators in
the reverse order.  Hence the operators commute:
\begin{equation}
  [ A_{f;\,\In}^\dagger, A_{g;\,\In}^\dagger] = 0.
\end{equation}

For the commutator of two annihilation operators, the derived result
for the action of an annihilation operator shows that two of them
commute.  

For one annihilation and one creation operator, we obtain the standard
commutator:
\begin{equation}
  [ A_{f;\,\In}, A_{g;\,\In}^\dagger]
  =
  \int\difftilde{p} \tilde{f}^*(\3p) \tilde{f}_j(\3p).
\end{equation}

\section{Generalizations}
\label{sec:generalizations}

Various extensions and generalizations of the derived results are
quite immediate:
\begin{enumerate}
\item The precise form of $F$ is irrelevant if it satisfies the same
  general specifications.
\item The same argument can be applied in any theory with any number
  of kinds of particle.  
\item The field used in the Green function for a particular particle
  can be any field that has a nonzero matrix element between the
  vacuum and a state of one of that kind of particle.  This generally
  means that the field should ``have quantum numbers corresponding to
  the particle''.  (E.g., for a proton we could use an operator with
  two anti-upquark fields and one anti-downquark field. These would be
  antiquark fields, because with the definition in \eqref{eq:1-0-ME},
  the fields need to be those that have the correct quantum numbers to
  create the particle.)
\item In this context the jargon is that the field is an
  ``interpolating field'' for the particle.
\item The choice of interpolating field is not unique.  Thus in $\phi^4$
  theory, $\phi^3$ or $\partial\phi/\partial x^\mu$ would work as interpolating fields
  instead of $\phi$, but at the expense of complication compared with the
  use of the elementary field.
\item The coefficient $c$ in the vacuum-to-one-particle matrix element
  generally depends on both the kind of particle and the field used.  It
  is in general complex; given a particular type of particle, the phase
  can be eliminated by convention only for one field.
\end{enumerate}
For another example of an interpolating field, consider a Schr\"odinger
field theory of an electron field and a proton field, with a Coulomb
interaction.  This theory has single particle states not only for
electrons and protons but also for every stable energy level of
hydrogen atom, and in fact for any stable ion and molecule. A possible
interpolating field for $s$-states of hydrogen would be
$\psi^\dagger_e(t,\3x)\psi^\dagger_p(t,\3x)$.  By choice of different time-dependent
wave functions in the derivation of the LSZ formula we can pick out
different energy levels of the atom for the particle that appears in
an S-matrix element.  If we want to deal with all energy levels, and
not just $s$-states, one could separate the electron and proton
fields: $\psi^\dagger_e(t,\3x+\3y)\psi^\dagger_p(t,\3x-\3y)$.

Particularly notable applications in strong interactions are for
pseudo-scalar mesons like the pion, since for the pion, certain
Noether currents for symmetries can be used as interpolating fields.
Interesting results can be found be applying Ward identities for the
symmetries to the Green functions used in the LSZ formula.  This gets
into the subject known as ``current algebra''.

\section{Relation to Haag-Ruelle}
\label{sec:HR.comparison}

In the formulation of Haag \cite{Haag:1958vt,Haag:1959ozr} and Ruelle
\cite{Ruelle:1962} the aim was to find a time-dependent operator that
when applied to the vacuum it creates a single particle.  They show
that in an infinite-time limit products of such operators applied to
the vacuum create general in- and out-states with desired wave
functions.  However they were concerned with general proofs rather
than providing, for example, fully explicit constructions of the
operators in a form useful for calculations in both coordinate and
momentum space.  Indeed Ruelle even states that Haag's results are
``less powerful than those of LSZ'': in contrast the formulation in
the present paper are intended to be more powerful than LSZ.

The formulation given in the present paper organizes the construction
of creation operators in a different way.  It starts from the LSZ
definition of a creation operator, which only has a spatial integral
over a product of field and wave function and no time integral.  This
is then modified by an average over time.  In contrast the Haag-Ruelle
formulation, the order of the averaging and the integration with a
wave function are reversed.  Moreover, the averaging is over
space-time, not just over time.

In the Haag-Ruelle method, the starting point is an ``almost local''
field defined by averaging $\phi$ with a test function $h$:
\begin{equation}
  \label{eq:avg.phi}
  \phi_h(x) = \int \diff[4]{y} \, h(x-y) \, \phi(y).
\end{equation}
Then a candidate creation operator is defined by applying the formula
for $a_f^\dag(t)$ to $\phi_h(x)$ instead of to $\phi(x)$:
\begin{equation}
\label{eq:afh.dag}
 a_{f,h}^\dagger(t) = -i \int \diff[3]{\3x} \, f(x) \derivB \phi_h(x) .
\end{equation}
Given the Fourier transform of $h$,
\begin{equation}
\label{eq:h.FT}
  h(x-y) = \int \frac{\diff[4]{p}}{(2\pi)^4} \, \tilde{h}(p) \, e^{\BDpos i p\cdot (x-y)},
\end{equation}
the momentum-space expression for $a_f^\dag(t)$ is
\begin{multline}
\label{eq:afh.dag.mom}
 a_{f,h}^\dagger(t)
        =   \int \difftilde{p} \tilde{f}(\3p)
            \int \frac{\diff{p^0}}{2\pi}
\times\\\times
            \tilde{\phi}(p)
            \, \tilde{h}(p)
            \, (E_{\3p}+p^0)
            \, e^{-i(E_{\3p}-p^0)t} .
\end{multline}

To enable this operator to create a single-particle state of specified
momentum content, but no multi-particle component, a restriction is
made on the support of $\tilde{h}(p)$.  Thus Duncan
\cite{Duncan:2012.book} required the support confined to a region
$am_{\rm phys}^2 < p^2 < bm_{\rm phys}^2$ with $0<a<1$ and $1<b<4$,
and that the function be non-zero on the whole of the one-particle
mass shell, where $p^0=\sqrt{\3p^2+m_{\rm phys}^2}$.  In addition,
$\tilde{h}(p)$ is zero when $p^0$ is negative.  Given the known matrix
element of the field between the vacuum and the one-particle states,
it follows that the state created by $a_{f,h}^\dagger(t)$ acting on the
vacuum is
\begin{equation}
\label{eq:HR.1part}
  a_{f,h}^\dagger(t) |0\rangle = \int \difftilde{p} \, \tilde{h}(E_{\3p},\3p)
  \, \tilde{f}(\3p) \, |\3p\rangle.
\end{equation}
As usual with test functions, $\tilde{h}(p)$ is both infinitely
differentiable and decreases faster than any power of $p$ when $p$
gets large.  Since therefore $\tilde{h}(E_{\3p},\3p)$ cannot be a
constant everywhere, the state differs from the one with wave function
$\tilde{f}(\3p)$.  Nevertheless, given any desired normalizable single
particle state, it can be created by suitably choosing
$\tilde{f}(\3p)$ in (\ref{eq:HR.1part}), i.e., by replacing
$\tilde{f}$ by $\tilde{f}(\3p)/\tilde{h}(E_{\3p},\3p)$.  (A slightly
different formulation with the same aim was given by Hepp
\cite{Hepp:1965.LSZ}.)

However what is not provided in this formalism is a simple formula
involving an integral of the coordinate-space field to give exactly the
one-particle state of a particular target wave function $|f\rangle$.

In contrast, the formulation in the present paper has reversed the
order of the averaging operation and the integration with the wave
function $f$, and has also made the averaging function a function of
time only.  For this to work, stronger dynamical requirements are
imposed on large momentum behavior of matrix elements than in the
Haag-Ruelle formulation.  These requirements are certainly valid in
the renormalizable QFTs that we are normally interested in, at least
as regards what can be seen in perturbation theory.  The Haag-Ruelle
method comes from a tradition that deliberately aims to derive general
properties of a relativistic QFT without appealing to more detailed
dynamical properties that generally are consequences of particular
QFTs.

A further difference compared with the Haag-Ruelle method is that the
operators $A_f^\dag(t,\Delta t)$ defined in the present paper, have a second
parameter $\Delta t$ that also has to be taken to infinity.  This change is
what enables explicit calculationally useful formulas to be given in
both coordinate and momentum space---see Eq.\
(\ref{eq:Af.dag.def.alt}).  The extra parameter has a useful
interpretation in terms of an energy uncertainty, and physically
should correspond to the experimental realities of making beams of
particles of almost exactly given momenta.  The explicit
coordinate-space formula for $A_f^\dag(t,\Delta t)$, together with the new 
proof of the reduction formula, are intended to be useful for further
applications, e.g., to the treatment of unstable particles, as in
Ref.\ \cite{Boyanovsky:2018swn}.

The strong commonalities between the Haag-Ruelle and the new methods
are that they perform some kind of averaging of the operators that
brings in an integral over time, and that the nature of the averaging
is arranged to restrict the momentum-space support of the creation
operator to correspond exactly to a single particle.  That is, the
coupling to multi-particle states is arranged to be zero by
construction of the definition (assisted by a limiting operation).


\begin{acknowledgments}
  I would like to thank the following for useful discussions:
  Dan Boyanovsky, Jainendra Jain, Radu Roiban, and Matt Schwartz.
\end{acknowledgments}


\appendix

\section{Strong v.\ weak limits, etc}
\label{sec:limits}

In treatments of scattering theory, we often discuss limits of
operators as a time parameter goes to infinity. It is useful to
distinguish three kinds of definition of a statement that
$\lim_{t\to\infty}A(t) = B$, where $A(t)$ and $B$ are linear operators on the
state space of a theory:
\begin{itemize}

\item A \emph{very strong limit} is where some measure of the
  difference between the operators $A(t)$ and $B$ themselves goes to
  zero:
  \begin{equation}
    \|A(t)-B\| \to 0.
  \end{equation}
  Here the measure of the size of $A(t)-B$ does not depend on the
  states acted on.  The name is coined here.  As will be pointed out
  below, limits of this kind typically do not exist for the operators
  used in the analysis of scattering.

\item A \emph{strong limit} is where the limit applies to states:
  \begin{equation}
    \left\| \left( A(t)-B \right) |f\rangle \right\| \to 0,
  \end{equation}
  for each state $|f\rangle$, but where the approach to the limit is
  permitted to depend on the state.

\item A \emph{weak limit} is where the limit is only for matrix
  elements
  \begin{equation}
    \left| \langle g| \left( A(t)-B \right) |f\rangle \right| \to 0,
  \end{equation}
  for each pair of normalizable states $|f\rangle$ and $\langle g|$, but where the
  approach to the limit is permitted to depend on the states.

\end{itemize}
The existence of a very strong limit implies the existence of the
strong and weak limits, and the existence of a strong limit implies
the existence of a weak limit.  But the reverse implications are
false.\footnote{For operators on a finite-dimensional space, all three
  concepts of a limit are equivalent.  The difference only appears
  when the operators act on an infinite-dimensional space, which is
  always the case in scattering problems.}

The existence of a very strong limit is equivalent to saying that the
approach of $\langle g|A(t)|f\rangle$ to $\langle g|B|f\rangle$ is uniform in both of the states
$\langle g|$ and $|f\rangle$.  The existence of a strong limit is equivalent to
uniformity in $|f\rangle$ only.

Generally, in applications such as to evolution operators, creation
operators, and the like in scattering theory, very strong limits do
not exist.  This is simply because we use theories that are invariant
under time translations.  Therefore (Heisenberg-picture) states may be
constructed that correspond to a physical scattering that occurs
arbitrarily far in the past or future.  How far in time one has to go
to get an approximation that corresponds to separated incoming or
outgoing particles depends on when the scattering(s) occur, and no
fixed time independent of the state suffices.  Thus very strong limits
do not exist in such cases.

Calculational methods (e.g., perturbation theory) lend themselves
naturally to the calculation of matrix elements rather than of the
operators themselves.  So it is useful to make definitions of the
different kinds of limit in terms of matrix elements, both with normalized
states and with states of particles of definite momenta.  Moreover the
matrix elements are typically between an in and an out state, i.e., in
using the definitions, we would normally replace $|f\rangle$ by
$|\underline{f};\,\In\rangle$, and $|g\rangle$ by $|\underline{g};\,\Out\rangle$, with
$\underline{f}$ and $\underline{g}$ denoting specifications of in- and
out-states in terms of their particle content, with the notation of Sec.\
\ref{sec:generalization}.

All the criteria for the different limits can be expressed in terms of
\begin{equation}
  \epsilon_{\rm ME}(t,f,g) \eqdef
  \frac{ \left| \langle g| (A(t)-B) |f\rangle \right| }
       { \||g\rangle\| ~ \||f\rangle\| }.
\end{equation}
Here there is a factor of the norms of the states in the denominator to
make the quantity invariant under scaling of the states.  

The definition that the weak limit exists is simply the statement that
$\epsilon_{\rm ME}(t,f,g)\to0$ as $t\to\infty$ for each $f$ and $g$.

Now the Cauchy-Schwartz inequality implies that 
\begin{equation}
  \left| \langle g| (A(t)-B) |f\rangle \right|
  \, \leq \,
  \||g\rangle\| ~ \| (A(t)-B)|f\rangle\|,
\end{equation}
with equality only when $|g\rangle$ is proportional to
$(A(t)-B)|f\rangle$.  So
\begin{equation}
  \epsilon_{\rm ME}(t,f,g) \leq 
  \frac{ \| (A(t)-B) |f\rangle \| }
       { \| |f\rangle \| }.
\end{equation}
This immediately shows that if the strong limit exists, so does the
weak limit.

Typically the proofs of existence of a limit are first made for $t\to\infty$ of
the matrix element with fixed states, with a finding that the weak limit
exists.  Then a possible approach to determining whether in addition the
strong limit exists is to find an upper bound to $\epsilon_{\rm ME}(t,f,g)$ as
$g$ is varied with $|f\rangle$ fixed.  So we define
\begin{align}
\label{eq:eps.state}
  \epsilon_{\rm state}(t,f) & \eqdef
  \sup_{\text{non-zero }|g\rangle}
  \epsilon_{\rm ME}(t,f,g)
\nonumber\\
  & =
  \sup_{\text{non-zero }|g\rangle}
  \frac{ | \langle g|(A(t)-B) |f\rangle }
        { \||g\rangle\| ~ \||f\rangle\| }.
\end{align}
(Here we use the notation for the supremum, i.e., the least upper
bound.)  We have already seen that the right-hand side is the same as
$\| (A(t)-B) |f\rangle \| / \| |f\rangle \|$.  So if we can bound $\epsilon_{\rm state}(t,f)$
on the basis of matrix element calculations and show that it goes to
zero as $t\to\infty$, then we know that the strong limit exists.

Now, in calculations and proofs we often express a matrix element in
terms of an integral over momenta --- see Secs.\ \ref{sec:w.v.s} and
\ref{sec:LSZ.derivation}, for example.  Then the $t$ dependence
appears as a phase, which commonly oscillates infinitely rapidly as
$t\to\infty$.  The limit then commonly involves strong cancellations because
of the oscillations, as in the weak limits of the LSZ operator
$a_f^\dagger(t)$.  Then for a given value of $t$, we may be able to change
the state $|g\rangle$ to another state $|g_{t,f}\rangle$ that has the same norm
and in which the wave functions are given phases that cancel the
first-mentioned phase.  See Sec.\ \ref{sec:w.v.s} for an example.

In this situation, since all possible states $|g\rangle$ are allowed in
calculating the bound in Eq.\ (\ref{eq:eps.state}), we therefore find
that $\epsilon_{\rm state}(t,f)$ is independent of $t$, and hence that the
strong limit does not exist.

Another approach is convenient when we specify $|f\rangle$ as an in-state,
is to express the norm of $(A(t)-B)|f;\,\In\rangle$ in terms of basis
out-states of particles of definite momenta:
\begin{multline}
\label{eq:mom.state}
  \| (A(t)-B)|f; \In\rangle \|^2
\\
  = \SumInt \diff{X}  \left| \langle X;\,\Out| (A(t)-B)|f; \In\rangle \right|^2.
\end{multline}
Here the integral over $X$ is a sum and integral over all out-basis
states, and includes the appropriate normalization factor.  Consider now
the situation (as with the LSZ operators) that the weak limit with
normalizable states relies on a cancellation of rapid oscillations of a
momentum-dependent phase factor, and that this needs an integral over wave
functions for both the initial and final states.  In the absolute value of
a matrix element with a momentum eigenstate, as in the right-hand side of
Eq.\ (\ref{eq:mom.state}), the phase is replaced by unity.  Then we no
longer have a suppression as $t\to\infty$.  A typical example of this situation
was shown in Sec.\ \ref{sec:3.afdag.0}.

Finally, we come to the possible existence of a very strong limit.
Let us define the size of the difference between the operators $A(t)$
and $B$ in terms of matrix elements by
\begin{align}
  \|A(t)-B\| &= \epsilon_{\rm op}(t)
\nonumber\\
  & \eqdef
  \sup_{\text{non-zero } |f\rangle, |g\rangle}
  \frac{ \left| \langle g| (A(t)-B)|f\rangle \right| }
       { \||g\rangle\| ~ \||f\rangle\| }
\nonumber\\
  &= 
  \sup_{\text{non-zero } |f\rangle, |g\rangle}
  \epsilon_{\rm ME}(t,f,g)
\nonumber\\
  &= 
  \sup_{\text{non-zero } |g\rangle}
  \epsilon_{\rm state}(t,f).
\end{align}
The variety of forms for this definition relates the size of
difference of operators to the formulas we used for the determining
the existence of weak and strong limits.

First these results show explicitly that if the very strong limit
exists, then so do the strong and weak limits.

We also see how it may happen that the strong and weak limits exist,
but the very strong limit can be shown not to exist.  Suppose, as is
usual in this context, that the $t$ dependence is confined to a phase
factor in momentum-space integral and that the integrand contains
factors of momentum-space wave functions. Then we may be able to
cancel the phase by changing the states to certain other states
$|f_t\rangle$ and $|g_t\rangle$ with appropriate $t$-dependent phases in their
momentum-space wave functions, and hence unchanged norms.  Such is the
case in the derivation of the reduction formula, where the
$t$-dependent phases in Eq.\ (\ref{eq:f.i.1}) can be canceled by
changing the wave functions.  Then $\langle g_t| (A(t)-B) |f_t\rangle$ is
independent of time, and so
\begin{align}
    \|A(t)-B\|
  &\geq 
  \frac{ \left| \langle g_t| (A(t)-B)|f_t\rangle \right| }
       { \||g_t\rangle\| \, \||f_t\rangle\| }
\nonumber\\
  &=
  \frac{ \left| \langle g| (A(0)-B)|f\rangle \right| }
       { \||g\rangle\| \, \||f\rangle\| },
\end{align}
which is nonzero. Then $\|A(t)-B\|$ cannot go to zero as $t\to\infty$, and
therefore in this case, the very strong limit does not exist.  As already
mentioned, the non-existence of a very strong limit is to be expected in a
time-translationally invariant theory; the approach to an infinite-time
limit is controlled by the time relative to a scattering event, and the
time of a scattering depends on the state considered.


\bibliography{jcc}

\providecommand{\noopsort}[1]{}
\begin{thebibliography}{33}%
\makeatletter
\providecommand \@ifxundefined [1]{%
 \@ifx{#1\undefined}
}%
\providecommand \@ifnum [1]{%
 \ifnum #1\expandafter \@firstoftwo
 \else \expandafter \@secondoftwo
 \fi
}%
\providecommand \@ifx [1]{%
 \ifx #1\expandafter \@firstoftwo
 \else \expandafter \@secondoftwo
 \fi
}%
\providecommand \natexlab [1]{#1}%
\providecommand \enquote  [1]{``#1''}%
\providecommand \bibnamefont  [1]{#1}%
\providecommand \bibfnamefont [1]{#1}%
\providecommand \citenamefont [1]{#1}%
\providecommand \href@noop [0]{\@secondoftwo}%
\providecommand \href [0]{\begingroup \@sanitize@url \@href}%
\providecommand \@href[1]{\@@startlink{#1}\@@href}%
\providecommand \@@href[1]{\endgroup#1\@@endlink}%
\providecommand \@sanitize@url [0]{\catcode `\\12\catcode `\$12\catcode
  `\&12\catcode `\#12\catcode `\^12\catcode `\_12\catcode `\%12\relax}%
\providecommand \@@startlink[1]{}%
\providecommand \@@endlink[0]{}%
\providecommand \url  [0]{\begingroup\@sanitize@url \@url }%
\providecommand \@url [1]{\endgroup\@href {#1}{\urlprefix }}%
\providecommand \urlprefix  [0]{URL }%
\providecommand \Eprint [0]{\href }%
\providecommand \doibase [0]{https://doi.org/}%
\providecommand \selectlanguage [0]{\@gobble}%
\providecommand \bibinfo  [0]{\@secondoftwo}%
\providecommand \bibfield  [0]{\@secondoftwo}%
\providecommand \translation [1]{[#1]}%
\providecommand \BibitemOpen [0]{}%
\providecommand \bibitemStop [0]{}%
\providecommand \bibitemNoStop [0]{.\EOS\space}%
\providecommand \EOS [0]{\spacefactor3000\relax}%
\providecommand \BibitemShut  [1]{\csname bibitem#1\endcsname}%
\let\auto@bib@innerbib\@empty
\bibitem [{\citenamefont {Lehmann}\ \emph {et~al.}(1955)\citenamefont
  {Lehmann}, \citenamefont {Symanzik},\ and\ \citenamefont
  {Zimmermann}}]{Lehmann:1954rq}%
  \BibitemOpen
  \bibfield  {author} {\bibinfo {author} {\bibfnamefont {H.}~\bibnamefont
  {Lehmann}}, \bibinfo {author} {\bibfnamefont {K.}~\bibnamefont {Symanzik}},\
  and\ \bibinfo {author} {\bibfnamefont {W.}~\bibnamefont {Zimmermann}},\
  }\bibfield  {title} {\bibinfo {title} {On the formulation of quantized field
  theories},\ }\href@noop {} {\bibfield  {journal} {\bibinfo  {journal} {Nuovo
  Cim.}\ }\textbf {\bibinfo {volume} {1}},\ \bibinfo {pages} {205} (\bibinfo
  {year} {1955})}\BibitemShut {NoStop}%
\bibitem [{\citenamefont {Duncan}(0012)}]{Duncan:2012.book}%
  \BibitemOpen
  \bibfield  {author} {\bibinfo {author} {\bibfnamefont {A.}~\bibnamefont
  {Duncan}},\ }\href@noop {} {\emph {\bibinfo {title} {The Conceptual Framework
  of Quantum Field Theory}}}\ (\bibinfo  {publisher} {Oxford University
  Press},\ \bibinfo {address} {Oxford},\ \bibinfo {year} {20012})\BibitemShut
  {NoStop}%
\bibitem [{\citenamefont {Goldberger}\ and\ \citenamefont
  {Watson}(1964)}]{Goldberger.Watson}%
  \BibitemOpen
  \bibfield  {author} {\bibinfo {author} {\bibfnamefont {M.~L.}\ \bibnamefont
  {Goldberger}}\ and\ \bibinfo {author} {\bibfnamefont {K.~M.}\ \bibnamefont
  {Watson}},\ }\href@noop {} {\emph {\bibinfo {title} {Collision Theory}}}\
  (\bibinfo  {publisher} {Wiley},\ \bibinfo {year} {1964})\BibitemShut
  {NoStop}%
\bibitem [{\citenamefont {Haag}(1958)}]{Haag:1958vt}%
  \BibitemOpen
  \bibfield  {author} {\bibinfo {author} {\bibfnamefont {R.}~\bibnamefont
  {Haag}},\ }\bibfield  {title} {\bibinfo {title} {{Quantum field theories with
  composite particles and asymptotic conditions}},\ }\href
  {https://doi.org/10.1103/PhysRev.112.669} {\bibfield  {journal} {\bibinfo
  {journal} {Phys. Rev.}\ }\textbf {\bibinfo {volume} {112}},\ \bibinfo {pages}
  {669} (\bibinfo {year} {1958})}\BibitemShut {NoStop}%
\bibitem [{\citenamefont {Haag}(1959)}]{Haag:1959ozr}%
  \BibitemOpen
  \bibfield  {author} {\bibinfo {author} {\bibfnamefont {R.}~\bibnamefont
  {Haag}},\ }\bibfield  {title} {\bibinfo {title} {{The framework of quantum
  field theory}},\ }\href {https://doi.org/10.1007/BF02724844} {\bibfield
  {journal} {\bibinfo  {journal} {Suppl. Nuovo Cim.}\ }\textbf {\bibinfo
  {volume} {14}},\ \bibinfo {pages} {131} (\bibinfo {year} {1959})}\BibitemShut
  {NoStop}%
\bibitem [{\citenamefont {Ruelle}(1962)}]{Ruelle:1962}%
  \BibitemOpen
  \bibfield  {author} {\bibinfo {author} {\bibfnamefont {D.}~\bibnamefont
  {Ruelle}},\ }\bibfield  {title} {\bibinfo {title} {On the asymptotic
  condition in quantum field theory},\ }\href@noop {} {\bibfield  {journal}
  {\bibinfo  {journal} {Helv. Phys. Acta}\ }\textbf {\bibinfo {volume} {35}},\
  \bibinfo {pages} {147} (\bibinfo {year} {1962})}\BibitemShut {NoStop}%
\bibitem [{\citenamefont {Hepp}(1965)}]{Hepp:1965.LSZ}%
  \BibitemOpen
  \bibfield  {author} {\bibinfo {author} {\bibfnamefont {K.}~\bibnamefont
  {Hepp}},\ }\bibfield  {title} {\bibinfo {title} {On the connection between
  the {LSZ} and {W}ightman quantum field theory},\ }\href@noop {} {\bibfield
  {journal} {\bibinfo  {journal} {Comm. Math. Phys.}\ }\textbf {\bibinfo
  {volume} {1}},\ \bibinfo {pages} {95} (\bibinfo {year} {1965})}\BibitemShut
  {NoStop}%
\bibitem [{\citenamefont {Akhmedov}(2019)}]{Akhmedov:2019iyt}%
  \BibitemOpen
  \bibfield  {author} {\bibinfo {author} {\bibfnamefont {E.}~\bibnamefont
  {Akhmedov}},\ }\bibfield  {title} {\bibinfo {title} {{Quantum mechanics
  aspects and subtleties of neutrino oscillations}}\ }(\bibinfo {year} {2019})\
  \Eprint {https://arxiv.org/abs/1901.05232} {arXiv:1901.05232 [hep-ph]}
  \BibitemShut {NoStop}%
\bibitem [{\citenamefont {Akhmedov}\ and\ \citenamefont
  {Kopp}(2010)}]{Akhmedov:2010ms}%
  \BibitemOpen
  \bibfield  {author} {\bibinfo {author} {\bibfnamefont {E.~K.}\ \bibnamefont
  {Akhmedov}}\ and\ \bibinfo {author} {\bibfnamefont {J.}~\bibnamefont
  {Kopp}},\ }\bibfield  {title} {\bibinfo {title} {{Neutrino oscillations:
  {Q}uantum mechanics vs.\ quantum field theory}},\ }\href
  {https://doi.org/10.1007/JHEP04(2010)008, 10.1007/JHEP10(2013)052} {\bibfield
   {journal} {\bibinfo  {journal} {JHEP}\ }\textbf {\bibinfo {volume} {04}},\
  \bibinfo {pages} {008}},\ \bibinfo {note} {[Erratum: JHEP10,052(2013)]},\
  \Eprint {https://arxiv.org/abs/1001.4815} {arXiv:1001.4815 [hep-ph]}
  \BibitemShut {NoStop}%
\bibitem [{\citenamefont {Coleman}(2011)}]{Coleman:2011xi}%
  \BibitemOpen
  \bibfield  {author} {\bibinfo {author} {\bibfnamefont {S.}~\bibnamefont
  {Coleman}},\ }\bibfield  {title} {\bibinfo {title} {{Notes from {S}idney
  {C}oleman's {P}hysics 253a: {Q}uantum {F}ield {T}heory}},\ }\Eprint
  {https://arxiv.org/abs/1110.5013} {arXiv:1110.5013 [physics.ed-ph]}
  \BibitemShut {NoStop}%
\bibitem [{\citenamefont {Chen}\ \emph {et~al.}(2018)\citenamefont {Chen},
  \citenamefont {Derbes}, \citenamefont {Griffiths}, \citenamefont {Hill},
  \citenamefont {Sohn},\ and\ \citenamefont {Ting}}]{Coleman.2018}%
  \BibitemOpen
  \bibfield  {author} {\bibinfo {author} {\bibfnamefont {B.~G.-g.}\
  \bibnamefont {Chen}}, \bibinfo {author} {\bibfnamefont {D.}~\bibnamefont
  {Derbes}}, \bibinfo {author} {\bibfnamefont {D.}~\bibnamefont {Griffiths}},
  \bibinfo {author} {\bibfnamefont {B.}~\bibnamefont {Hill}}, \bibinfo {author}
  {\bibfnamefont {R.}~\bibnamefont {Sohn}},\ and\ \bibinfo {author}
  {\bibfnamefont {Y.-S.}\ \bibnamefont {Ting}},\ }\href
  {https://doi.org/10.1142/9371} {\emph {\bibinfo {title} {Lectures of Sidney
  Coleman on Quantum Field Theory}}}\ (\bibinfo  {publisher} {World
  Scientific},\ \bibinfo {year} {2018})\ \Eprint
  {https://arxiv.org/abs/https://www.worldscientific.com/doi/pdf/10.1142/9371}
  {https://www.worldscientific.com/doi/pdf/10.1142/9371} \BibitemShut {NoStop}%
\bibitem [{\citenamefont {Boyanovsky}(2019)}]{Boyanovsky:2018swn}%
  \BibitemOpen
  \bibfield  {author} {\bibinfo {author} {\bibfnamefont {D.}~\bibnamefont
  {Boyanovsky}},\ }\bibfield  {title} {\bibinfo {title} {{Quantum decay in
  renormalizable field theories: quasiparticle formation, {Z}eno and
  anti-{Z}eno effects}},\ }\href {https://doi.org/10.1016/j.aop.2019.03.012}
  {\bibfield  {journal} {\bibinfo  {journal} {Annals Phys.}\ }\textbf {\bibinfo
  {volume} {405}},\ \bibinfo {pages} {176} (\bibinfo {year} {2019})},\ \Eprint
  {https://arxiv.org/abs/1810.01301} {arXiv:1810.01301 [hep-th]} \BibitemShut
  {NoStop}%
\bibitem [{\citenamefont {Kapec}\ \emph {et~al.}(2017)\citenamefont {Kapec},
  \citenamefont {Perry}, \citenamefont {Raclariu},\ and\ \citenamefont
  {Strominger}}]{Kapec:2017tkm}%
  \BibitemOpen
  \bibfield  {author} {\bibinfo {author} {\bibfnamefont {D.}~\bibnamefont
  {Kapec}}, \bibinfo {author} {\bibfnamefont {M.}~\bibnamefont {Perry}},
  \bibinfo {author} {\bibfnamefont {A.-M.}\ \bibnamefont {Raclariu}},\ and\
  \bibinfo {author} {\bibfnamefont {A.}~\bibnamefont {Strominger}},\ }\bibfield
   {title} {\bibinfo {title} {{Infrared Divergences in {QED}, Revisited}},\
  }\href {https://doi.org/10.1103/PhysRevD.96.085002} {\bibfield  {journal}
  {\bibinfo  {journal} {Phys. Rev.}\ }\textbf {\bibinfo {volume} {D96}},\
  \bibinfo {pages} {085002} (\bibinfo {year} {2017})},\ \Eprint
  {https://arxiv.org/abs/1705.04311} {arXiv:1705.04311 [hep-th]} \BibitemShut
  {NoStop}%
\bibitem [{\citenamefont {Hoare}\ \emph {et~al.}(2018)\citenamefont {Hoare},
  \citenamefont {Levine},\ and\ \citenamefont {Tseytlin}}]{Hoare:2018jim}%
  \BibitemOpen
  \bibfield  {author} {\bibinfo {author} {\bibfnamefont {B.}~\bibnamefont
  {Hoare}}, \bibinfo {author} {\bibfnamefont {N.}~\bibnamefont {Levine}},\ and\
  \bibinfo {author} {\bibfnamefont {A.~A.}\ \bibnamefont {Tseytlin}},\
  }\bibfield  {title} {\bibinfo {title} {{On the massless tree-level {S}-matrix
  in 2d sigma models}},\ }\Eprint {https://arxiv.org/abs/1812.02549}
  {arXiv:1812.02549 [hep-th]} \BibitemShut {NoStop}%
\bibitem [{\citenamefont {Zamolodchikov}\ and\ \citenamefont
  {Zamolodchikov}(1992)}]{Zamolodchikov:1992zr}%
  \BibitemOpen
  \bibfield  {author} {\bibinfo {author} {\bibfnamefont {A.~B.}\ \bibnamefont
  {Zamolodchikov}}\ and\ \bibinfo {author} {\bibfnamefont {A.~B.}\ \bibnamefont
  {Zamolodchikov}},\ }\bibfield  {title} {\bibinfo {title} {{Massless
  factorized scattering and sigma models with topological terms}},\ }\href
  {https://doi.org/10.1016/0550-3213(92)90136-Y} {\bibfield  {journal}
  {\bibinfo  {journal} {Nucl. Phys.}\ }\textbf {\bibinfo {volume} {B379}},\
  \bibinfo {pages} {602} (\bibinfo {year} {1992})}\BibitemShut {NoStop}%
\bibitem [{\citenamefont {Born}\ and\ \citenamefont
  {Jordan}(1925)}]{Born.Jordan}%
  \BibitemOpen
  \bibfield  {author} {\bibinfo {author} {\bibfnamefont {M.}~\bibnamefont
  {Born}}\ and\ \bibinfo {author} {\bibfnamefont {P.}~\bibnamefont {Jordan}},\
  }\bibfield  {title} {\bibinfo {title} {Zur {Q}uantenmechanik},\ }\href@noop
  {} {\bibfield  {journal} {\bibinfo  {journal} {Zeit. f. Phys.}\ }\textbf
  {\bibinfo {volume} {34}},\ \bibinfo {pages} {858} (\bibinfo {year} {1925})},\
  \bibinfo {note} {{E}nglish translation in Ref.\ \cite{van1967sources}, but
  without the last section (on the quantized electromagnetic
  field).}\BibitemShut {Stop}%
\bibitem [{\citenamefont {van Hove}(1951)}]{vanHove1951}%
  \BibitemOpen
  \bibfield  {author} {\bibinfo {author} {\bibfnamefont {L.}~\bibnamefont {van
  Hove}},\ }\bibfield  {title} {\bibinfo {title} {Sur l'op\'{e}rateur
  {H}amiltonien de deux champs quantifi\'{e}s en interaction},\ }\href@noop {}
  {\bibfield  {journal} {\bibinfo  {journal} {Bulletin de la Classe des
  Sciences Academie Royale de Belgique}\ }\textbf {\bibinfo {volume} {37}},\
  \bibinfo {pages} {1055} (\bibinfo {year} {1951})}\BibitemShut {NoStop}%
\bibitem [{\citenamefont {van Hove}(1952)}]{vanHove1952145}%
  \BibitemOpen
  \bibfield  {author} {\bibinfo {author} {\bibfnamefont {L.}~\bibnamefont {van
  Hove}},\ }\bibfield  {title} {\bibinfo {title} {Les difficult\'{e}s de
  divergences pour un mod\`{e}le particulier de champ quantifi\'{e}},\ }\href
  {https://doi.org/https://doi.org/10.1016/S0031-8914(52)80017-5} {\bibfield
  {journal} {\bibinfo  {journal} {Physica}\ }\textbf {\bibinfo {volume} {18}},\
  \bibinfo {pages} {145 } (\bibinfo {year} {1952})}\BibitemShut {NoStop}%
\bibitem [{\citenamefont {Haag}(1955)}]{Haag:1955ev}%
  \BibitemOpen
  \bibfield  {author} {\bibinfo {author} {\bibfnamefont {R.}~\bibnamefont
  {Haag}},\ }\bibfield  {title} {\bibinfo {title} {On quantum field theories},\
  }\href@noop {} {\bibfield  {journal} {\bibinfo  {journal} {Kong. Dan. Vid.
  Sel. Mat. Fys. Med.}\ }\textbf {\bibinfo {volume} {29}},\ \bibinfo {pages}
  {1} (\bibinfo {year} {1955})},\ \bibinfo {note} {[Phil. Mag.
  Ser.746,376(1955)]}\BibitemShut {NoStop}%
\bibitem [{\citenamefont {Streater}\ and\ \citenamefont
  {Wightman}(1964)}]{Streater:1964}%
  \BibitemOpen
  \bibfield  {author} {\bibinfo {author} {\bibfnamefont {R.~F.}\ \bibnamefont
  {Streater}}\ and\ \bibinfo {author} {\bibfnamefont {A.~S.}\ \bibnamefont
  {Wightman}},\ }\href@noop {} {\emph {\bibinfo {title} {{PCT}, spin and
  statistics, and all that}}}\ (\bibinfo  {publisher} {W. A. Benjamin},\
  \bibinfo {year} {1964})\BibitemShut {NoStop}%
\bibitem [{\citenamefont {Sterman}(1993)}]{Sterman:1994ce}%
  \BibitemOpen
  \bibfield  {author} {\bibinfo {author} {\bibfnamefont {G.}~\bibnamefont
  {Sterman}},\ }\href@noop {} {\emph {\bibinfo {title} {An Introduction to
  Quantum Field Theory}}}\ (\bibinfo  {publisher} {Cambridge University
  Press},\ \bibinfo {address} {Cambridge},\ \bibinfo {year} {1993})\BibitemShut
  {NoStop}%
\bibitem [{\citenamefont {Srednicki}(2007)}]{Srednicki:2007qs}%
  \BibitemOpen
  \bibfield  {author} {\bibinfo {author} {\bibfnamefont {M.}~\bibnamefont
  {Srednicki}},\ }\href@noop {} {\emph {\bibinfo {title} {Quantum Field
  Theory}}}\ (\bibinfo  {publisher} {Cambridge University Press},\ \bibinfo
  {address} {Cambridge},\ \bibinfo {year} {2007})\BibitemShut {NoStop}%
\bibitem [{\citenamefont {Peskin}\ and\ \citenamefont
  {Schroeder}(1995)}]{Peskin:1995ev}%
  \BibitemOpen
  \bibfield  {author} {\bibinfo {author} {\bibfnamefont {M.~E.}\ \bibnamefont
  {Peskin}}\ and\ \bibinfo {author} {\bibfnamefont {D.~V.}\ \bibnamefont
  {Schroeder}},\ }\href@noop {} {\emph {\bibinfo {title} {An Introduction to
  Quantum Field Theory}}}\ (\bibinfo  {publisher} {Addison-Wesley},\ \bibinfo
  {address} {Reading, MA},\ \bibinfo {year} {1995})\BibitemShut {NoStop}%
\bibitem [{\citenamefont {Jain}(2007)}]{Jain2007.book}%
  \BibitemOpen
  \bibfield  {author} {\bibinfo {author} {\bibfnamefont {J.~K.}\ \bibnamefont
  {Jain}},\ }\href@noop {} {\emph {\bibinfo {title} {Composite Fermions}}}\
  (\bibinfo  {publisher} {Cambridge University Press},\ \bibinfo {address}
  {Cambridge},\ \bibinfo {year} {2007})\BibitemShut {NoStop}%
\bibitem [{\citenamefont {Weinberg}(1995)}]{Weinberg:1995mt}%
  \BibitemOpen
  \bibfield  {author} {\bibinfo {author} {\bibfnamefont {S.}~\bibnamefont
  {Weinberg}},\ }\href@noop {} {\emph {\bibinfo {title} {The Quantum Theory of
  Fields, Vol.\ {I}, Foundations}}}\ (\bibinfo  {publisher} {Cambridge
  University Press},\ \bibinfo {address} {Cambridge},\ \bibinfo {year}
  {1995})\BibitemShut {NoStop}%
\bibitem [{\citenamefont {Taylor}(1972)}]{Taylor.scattering.1972}%
  \BibitemOpen
  \bibfield  {author} {\bibinfo {author} {\bibfnamefont {J.~R.}\ \bibnamefont
  {Taylor}},\ }\href@noop {} {\emph {\bibinfo {title} {Scattering Theory: The
  Quantum Theory of Nonrelativistic Collisions}}}\ (\bibinfo  {publisher}
  {Wiley},\ \bibinfo {year} {1972})\BibitemShut {NoStop}%
\bibitem [{\citenamefont {Itzykson}\ and\ \citenamefont
  {Zuber}(1980)}]{Itzykson:1980rh}%
  \BibitemOpen
  \bibfield  {author} {\bibinfo {author} {\bibfnamefont {C.}~\bibnamefont
  {Itzykson}}\ and\ \bibinfo {author} {\bibfnamefont {J.-B.}\ \bibnamefont
  {Zuber}},\ }\href@noop {} {\emph {\bibinfo {title} {Quantum Field Theory}}}\
  (\bibinfo  {publisher} {McGraw-Hill},\ \bibinfo {address} {New York},\
  \bibinfo {year} {1980})\BibitemShut {NoStop}%
\bibitem [{\citenamefont {K\"all\'en}(1952)}]{Kallen:1952zz}%
  \BibitemOpen
  \bibfield  {author} {\bibinfo {author} {\bibfnamefont {G.}~\bibnamefont
  {K\"all\'en}},\ }\bibfield  {title} {\bibinfo {title} {{On the definition of
  the Renormalization Constants in Quantum Electrodynamics}},\ }\href
  {https://doi.org/10.1007/978-3-319-00627-7_90} {\bibfield  {journal}
  {\bibinfo  {journal} {Helv. Phys. Acta}\ }\textbf {\bibinfo {volume} {25}},\
  \bibinfo {pages} {417} (\bibinfo {year} {1952})},\ \bibinfo {note}
  {[,509(1952)]}\BibitemShut {NoStop}%
\bibitem [{\citenamefont {Lehmann}(1954)}]{Lehmann:1954xi}%
  \BibitemOpen
  \bibfield  {author} {\bibinfo {author} {\bibfnamefont {H.}~\bibnamefont
  {Lehmann}},\ }\bibfield  {title} {\bibinfo {title} {{On the Properties of
  propagation functions and renormalization contants of quantized fields}},\
  }\href {https://doi.org/10.1007/BF02783624} {\bibfield  {journal} {\bibinfo
  {journal} {Nuovo Cim.}\ }\textbf {\bibinfo {volume} {11}},\ \bibinfo {pages}
  {342} (\bibinfo {year} {1954})}\BibitemShut {NoStop}%
\bibitem [{\citenamefont {Bogoliubov}\ and\ \citenamefont
  {Shirkov}(1959)}]{Bogoliubov.Shirkov}%
  \BibitemOpen
  \bibfield  {author} {\bibinfo {author} {\bibfnamefont {N.~N.}\ \bibnamefont
  {Bogoliubov}}\ and\ \bibinfo {author} {\bibfnamefont {D.~V.}\ \bibnamefont
  {Shirkov}},\ }\href@noop {} {\emph {\bibinfo {title} {Introduction to the
  Theory of Quantized Fields}}}\ (\bibinfo  {publisher} {Wiley-Interscience},\
  \bibinfo {address} {New York},\ \bibinfo {year} {1959})\BibitemShut {NoStop}%
\bibitem [{\citenamefont {Bjorken}\ and\ \citenamefont
  {Drell}(1965)}]{Bjorken:1965zz}%
  \BibitemOpen
  \bibfield  {author} {\bibinfo {author} {\bibfnamefont {J.~D.}\ \bibnamefont
  {Bjorken}}\ and\ \bibinfo {author} {\bibfnamefont {S.~D.}\ \bibnamefont
  {Drell}},\ }\href@noop {} {\emph {\bibinfo {title} {Relativistic quantum
  fields}}}\ (\bibinfo  {publisher} {McGraw-Hill},\ \bibinfo {year}
  {1965})\BibitemShut {NoStop}%
\bibitem [{\citenamefont {Schlieder}\ and\ \citenamefont
  {Seiler}(1972)}]{Schlieder:1972qr}%
  \BibitemOpen
  \bibfield  {author} {\bibinfo {author} {\bibfnamefont {S.}~\bibnamefont
  {Schlieder}}\ and\ \bibinfo {author} {\bibfnamefont {E.}~\bibnamefont
  {Seiler}},\ }\bibfield  {title} {\bibinfo {title} {Remarks on the null plane
  development of a relativistic quantum field theory},\ }\href
  {https://doi.org/10.1007/BF01877587} {\bibfield  {journal} {\bibinfo
  {journal} {Commun. Math. Phys.}\ }\textbf {\bibinfo {volume} {25}},\ \bibinfo
  {pages} {62} (\bibinfo {year} {1972})}\BibitemShut {NoStop}%
\bibitem [{\citenamefont {Van Der~Waerden}(1967)}]{van1967sources}%
  \BibitemOpen
  \bibfield  {author} {\bibinfo {author} {\bibfnamefont {B.}~\bibnamefont {Van
  Der~Waerden}},\ }\href@noop {} {\emph {\bibinfo {title} {Sources of Quantum
  Mechanics}}}\ (\bibinfo  {publisher} {North-Holland},\ \bibinfo {year}
  {1967})\BibitemShut {NoStop}%
\end{thebibliography}%

\end{document}